\documentclass[twocolumn]{aastex631}
\usepackage{xspace}

\newcommand{\fermi}{{\em Fermi}\xspace}
\newcommand{\fermiT}{{T$_{0}$}\xspace}

\newcommand{\kw}{{\em Konus}-Wind\xspace}
\newcommand{\keV}{{\rm ~keV}\xspace}

\newcommand{\s}{{\rm ~s}\xspace}
\newcommand{\swift}{{\em Swift}\xspace}

\newcommand{\tninty}{{T$_{90}$}\xspace}
\newcommand{\Ep}{E$_{\rm pt}$\xspace}
\newcommand{\sw}[1]{\texttt{#1}}
\graphicspath{{./}{figures/}}
\usepackage{color, colortbl}
\usepackage{amsmath}
\usepackage[figuresright]{rotating}
\usepackage{threeparttable}
\usepackage{tablefootnote}
\usepackage{multirow,longtable,supertabular}
\usepackage{xcolor}

\usepackage{placeins}
\usepackage{chngcntr}
\shorttitle{ULGRBs and GRB 221009A}
\shortauthors{A. K. Ror et al.}
\usepackage{float}
\graphicspath{{./}{figures/}}

\begin{document}

\title{Exploring Origin of Ultra-Long Gamma-ray Bursts: Lessons from GRB 221009A}

\author[0000-0003-3164-8056]{Amit Kumar Ror$^\dagger$}
\affiliation{Aryabhatta Research Institute of Observational Sciences (ARIES), Manora Peak, Nainital-263001, India}
\affiliation{Department of Applied Physics/Physics, Mahatma Jyotiba Phule Rohilkhand University, Bareilly-243006, India}
\author[0000-0002-0786-7307]{Rahul Gupta$^{*}$}
\affiliation{Aryabhatta Research Institute of Observational Sciences (ARIES), Manora Peak, Nainital-263001, India}
\affiliation{Astrophysics Science Division, NASA Goddard Space Flight Center, Mail Code 661, Greenbelt, MD 20771, USA}
\affiliation{NASA Postdoctoral Program Fellow}
\author[0000-0003-4905-7801]{Amar Aryan}
\affiliation{Aryabhatta Research Institute of Observational Sciences (ARIES), Manora Peak, Nainital-263001, India}
\affiliation{Graduate Institute of Astronomy, National Central University, 300 Jhongda Road, 32001 Jhongli, Taiwan}
\author[0000-0003-4905-7801]{Shashi Bhushan Pandey$^{**}$}
\affiliation{Aryabhatta Research Institute of Observational Sciences (ARIES), Manora Peak, Nainital-263001, India}
\author[0000-0003-4905-7801]{S. R. Oates}
\affiliation{Physics Department, Lancaster University, Bailrigg, Lancaster LA1 4YB, UK}
\author[0000-0003-4905-7801]{A. J. Castro-Tirado}
\affiliation{Instituto de Astrof\'isica de Andaluc\'ia (IAA-CSIC), Glorieta de la Astronom\'ia s/n, E-18008, Granada, Spain}
\affiliation{Departamento de Ingenier\'ia de Sistemas y Autom\'atica, Escuela de Ingenier\'ias, Universidad de M\'alaga, C\/. Dr. Ortiz Ramos s\/n, E-29071, M\'alaga, Spain}
\author[0000-0003-1186-2119]{Sudhir Kumar}
\affiliation{Department of Applied Physics/Physics, Mahatma Jyotiba Phule Rohilkhand University, Bareilly-243006, India}

\email{$^\dagger$amitror@aries.res.in}
\email{$^{*}$rahulbhu.c157@gmail.com}
\email{$^{**}$shashi@aries.res.in}

\begin{abstract}
The brightest Gamma-ray burst (GRB) ever, GRB 221009A, displays ultra-long GRB (ULGRB) characteristics, with a prompt emission duration exceeding 1000 s. To constrain the origin and central engine of this unique burst, we analyze its prompt and afterglow characteristics and compare them to the established set of similar GRBs. To achieve this, we statistically examine a nearly complete sample of \swift-detected GRBs with measured redshifts. Categorizing the sample to Bronze, Silver, and Gold by fitting a Gaussian function to the log-normal of \tninty duration distribution and considering three sub-samples respectively to 1, 2, and 3 times of the standard deviation to the mean value. GRB 221009A falls into the Gold sub-sample. Our analysis of prompt emission and afterglow characteristics aims to identify trends between the three burst groups. Notably, the Gold sub-sample (a higher likelihood of being ULGRB candidates) suggests a collapsar scenario with a hyper-accreting black hole as a potential central engine, while a few GRBs (GRB 060218, GRB 091024A, and GRB 100316D) in our Gold sub-sample favor a magnetar. Late-time near-IR (NIR) observations from 3.6m Devasthal Optical Telescope (DOT) rule out the presence of any bright supernova associated with GRB 221009A in the Gold sub-sample. To further constrain the physical properties of ULGRB progenitors, we employ the tool \sw{MESA} to simulate the evolution of low-metallicity massive stars with different initial rotations. The outcomes suggest that rotating ($\Omega \geq 0.2\,\Omega_{\rm c}$) massive stars could potentially be the progenitors of ULGRBs within the considered parameters and initial inputs to \sw{MESA}.
\end{abstract}

\section{Introduction} 
\label{sec:intro}
Gamma-ray bursts (GRBs) are characterized by intense and short-lived bursts of high-energy (a few keV to MeV) radiation. GRBs emit electromagnetic radiation in two phases. The first phase, known as ``prompt emission," typically persists for a duration ranging from a few milliseconds to several thousand seconds \citep{zhang, 2015AdAst2015E..22P}. The presence of bi-modality in the duration (\tninty\footnote{it is referred to the duration encompassing 5-95 \% of the fluence observed in soft gamma/hard X-ray channels.}) distribution of the prompt emission of GRBs has led to the classification of these events into two distinct categories \citep{1993ApJ...413L.101K}: Long-duration (\tninty $\geq$ 2 s) and Short-duration (\tninty $\leq$ 2 s) GRBs. Long GRBs (LGRBs) have been observed to originate from the demise of core collapse of massive stars \citep{1993ApJ...405..273W, 2003Natur.423..847H}. Conversely, short GRBs (SGRBs) have been attributed to mergers involving compact objects such as neutron stars - neutron stars (NS-NS) or a neutron star - black hole \citep{Perna_2002, Abbott_2017}. In addition to these two traditional classes, a unique and intriguing class known as Ultra-long GRBs (ULGRBs) has been suggested. These exceptional events defy the conventional timescales associated with standard GRBs, exhibiting durations (several thousand seconds) far beyond what is typically observed. However, the finding of \cite{2013ApJ...778...54V, 2014ApJ...781...13L, 2015ApJ...800...16B, 2015JHEAp...7...44L, 2018ApJ...859...48P} suggests that there is no precise boundary to separate LGRBs and ULGRBs. Even though not all the GRBs are considered ULGRBs based on their prompt emission duration, in some cases, combined duration in gamma-ray/hard X-ray and soft X-ray (flares or plateau) are utilized to separate between the two classes \citep{2014ApJ...787...66Z}. Unlike the duration of ULGRBs, the total fluence exhibited by these events is not an exception. This fluence stretched over a longer time scale, requiring a highly sensitive instrument for their detection \citep{2014ApJ...781...13L}. Further, some of the well-studied ULGRBs (GRB 060218 and GRB 100316D) are found to be intrinsically soft, posing energy constraints on the detecting instruments. The orbital constraints associated with space-based detectors also present challenges in capturing the complete emissions of ULGRBs \citep{2015JHEAp...7...44L}. \swift's remarkable sensitivity in soft energy channels and its unique observation strategy, both in the event rate and integrated image mode, has proven beneficial in detecting several ULGRBs \citep{2004ApJ...611.1005G}. However, despite the discovery of numerous well-classified ULGRBs such as GRB 060218, GRB 091024A, GRB 100316D, GRB 101225A, GRB 111209A, GRB 121027A, GRB 130925A, GRB 141121A, GRB 220627A, and many more over the years, our understanding of their progenitors, central engine, and the surrounding environments remains elusive \citep{2013ApJ...778...54V, 2014ApJ...781...13L, 2014MNRAS.444..250E, 2015ApJ...800...16B, 2015JHEAp...7...44L, 2018ApJ...859...48P, 2019MNRAS.486.2471G, 2023arXiv230710339D}.

Direct evidence regarding the progenitors of ULGRBs emerges from the observation of associated supernovae accompanying these long-lasting events. These observations strongly imply that the demise of massive stars (collapsar) may account for some ULGRBs \citep{2006Natur.442.1008C, 2011MNRAS.411.2792S, 2011Natur.480...72T, 2013ApJ...778...67N, 2015Natur.523..189G}. However, alternative explanations are also proposed as potential progenitors capable of launching long-lasting ultra-relativistic jets: (1) The tidal disruption of a white dwarf by a black hole has been proposed as a potential progenitor for ULGRBs \citep{2011Natur.480...69C}. According to \cite{Ioka_2016}, under specific circumstances, such an occurrence could give rise to the supernova-like features observed in the late afterglow light curve of ULGRBs. (2) A massive star (15-30 M$_{\odot}$) with low metallicity, possessing a rotation that culminates in its evolution into a blue supergiant (BSG), represents a potential progenitor for ULGRBs. BSG stars, characterized by significantly larger radii compared to Wolf-Rayet (WR) stars, can collapse into hyperaccreting black holes. This scenario offers a natural explanation for the unexpectedly prolonged durations observed in ULGRBs \citep{2018ApJ...859...48P}. (3) Another contender for the progenitor of ULGRBs is a highly magnetized millisecond pulsar, often referred to as a magnetar \citep{1992Natur.357..472U}. The energy released during the spin-down of a magnetar can play a significant role in the formation of a bipolar jet, and such mechanism holds promise in elucidating the long-lasting emission observed in ULGRBs \citep{2007MNRAS.380.1541B, 2009MNRAS.396.2038B}.

In recent years, significant progress in both observational technology and theoretical modeling has illuminated our understanding of ULGRB progenitors. The Modules for Experiments in Stellar Astrophysics (\sw{MESA}) code \citep{2011ApJS..192....3P, 2013ApJS..208....4P, 2015ApJS..220...15P, 2018ApJS..234...34P, 2019ApJS..243...10P, 2023ApJS..265...15J}, a highly robust tool for modeling stellar evolution, has played a pivotal role in these advancements. \cite{2018ApJ...859...48P} utilized \sw{MESA} to evolve stars with masses of 30 M$_{\odot}$ and 40 M$_{\odot}$ under varying initial rotation conditions. Their findings demonstrated that moderately rotating massive stars could culminate their evolution as BSG, which can successfully launch an ultra-relativistic jet to power ULGRBs. Moreover, \cite{2023arXiv230105401S} conducted a comprehensive exploration of the impact of initial mass, metallicity, and rotation on magnetar formation. This extensive study involved evolving 227 stellar models using \sw{MESA}. In this context, we leverage \sw{MESA} to distinguish between the progenitors of LGRBs and ULGRBs by evolving massive stars within the mass range of 15–30 M$_{\odot}$ while considering various initial rotation scenarios. The minimum mass limit for typical LGRB progenitor given by \cite{2007MNRAS.376.1285L} is 20\,M$_{\odot}$. However, modeling results of \cite{2018ApJ...859...48P} revealed BSG stars as the progenitors of ULGRBs. The standard mass of BSG stars is 15\,M$_{\odot}$ \citep{2018arXiv181207620D}. Therefore, we select 15\,M$_{\odot}$ as our starting point. The choice of minimum mass 15\,M$_{\odot}$ is also supported by \cite{2011ApJ...739L..55B}. The upper limit  of 30\,M$_{\odot}$ is motivated from \cite{2023arXiv230105401S, 2018ApJ...859...48P}. Therefore, we use a mass range of 15-30\,M$_{\odot}$ while evolving the massive star models in \sw{MESA} and all related analyses.

Efforts have also been made to account for the observed duration of ULGRBs by examining the properties of the surrounding medium rather than solely focusing on unique central engines or progenitors. As proposed by \cite{2014MNRAS.444..250E}, it is suggested that the circumburst environment of ULGRBs may distinguish them from LGRBs. ULGRBs could potentially be situated within exceedingly low-density surroundings, resulting in a deceleration of their ejecta at a slower rate compared to a denser medium. Until now, ULGRBs have shown diverse observed characteristics during the prompt emission and afterglow phases. For example, observed SNe emission associated with GRB 060218, GRB 100316D, and GRB 111209A \citep{2006Natur.442.1008C, 2011MNRAS.411.2792S, 2015Natur.523..189G}, the association of GRB 101225A, GRB 111209A, and GRB 121027A with active star-forming galaxies and exhibiting a mixed type of surrounding environment \citep{2014ApJ...781...13L}. Most ULGRBs exhibit early X-ray light curves featuring flares or plateau \citep{2014ApJ...787...66Z}, and in some cases, they show thermal components in early X-ray afterglow. However, all such properties are common in LGRBs or low-luminous GRBs. Consequently, \tninty stands out as the robust parameter distinguishing ULGRBs from the broader LGRB population. Therefore, considering \tninty as a separation criterion, we statistically examine a sub-sample of ULGRB candidates from the complete set of \swift detected bursts. This work investigates the underlying physical mechanism, possible progenitors, and central engine contributing to their unexpectedly long duration compared to LGRBs and SGRBs.

The Gamma-ray Burst Monitor (GBM, \citealt{Meegan}) on board {\em Fermi Gamma-ray Space Telescope} (hereafter \fermi) and the Burst Alert Telescope (BAT, \citealt{2005SSRv..120..143B}) on board \swift detected GRB 221009A, which stands out as the brightest (surpasses nearby monster GRBs such as GRB 130427A and GRB 190114C in terms of observed fluence and isotropic energy release) burst ever observed \citep{2023ApJ...952L..42L}. Remarkably, \fermi-GBM recorded emission from this burst for over 1000 s \citep{2023ApJ...952L..42L}, and \kw reported a soft tail emission that extended up to an astonishing 20,000 s \citep{2023arXiv230213383F}, thereby positioning it as a potential candidate as a ULGRB \citep{2023arXiv230213383F, 2023ApJ...946L..31B}. In this work, we thoroughly studied the characteristics of this burst and compared it to a larger sample of ULGRB candidates. Additionally, LHAASO and Carpet-2 missions have claimed the detection of photons with energies of 18 TeV \citep{2022GCN.32677....1H} and 250 TeV \citep{2022ATel15675....1F}, respectively. With this, GRB 221009A has become the first ULGRB candidate to belong to the class of very high energy (VHE, few hundred GeV to TeV) GRBs \citep{2023ApJ...942...34R}.

This paper is structured as follows: \S 2 presents our selection criteria for Bronze, Silver, and Gold sub-samples of ULGRBs candidates that we investigate. In \S 3, we analyze the multi-wavelength characteristics of GRB 221009A and compare them to those of a broader sample of ULGRBs. \S 4 describes the basic characteristics, possible progenitors, and central engines of the GRBs included in our sample. The simulations of massive stars with \sw{MESA} code are given in \S 5, and \S 6 provides the summary and conclusion.

\section{Sample Selection and methodology} 
\label{sec:selection_criteria}

\begin{figure*}[!ht]
\centering
\includegraphics[width=1.7\columnwidth]{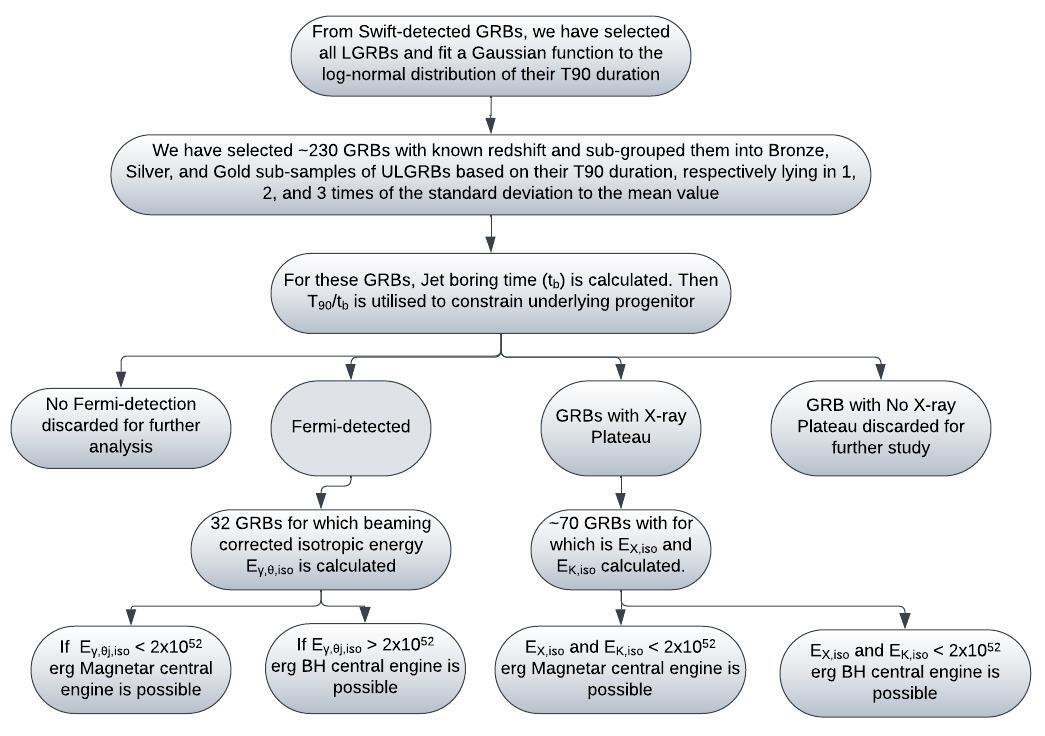} 
\caption{Description/Flow chart of the methods used to select the Bronze, Silver, and Gold sub-samples and to constrain their possible progenitors and central engines.}
\label{fig:selection_criteria}
\end{figure*}

Due to the limitations of duration-based classification, there is no exact boundary between LGRBs and ULGRBs. The different sensitivities of space-based gamma-ray detectors at different energies and orbital constraints can lead to the omission of significant amounts of prompt emission from some GRBs \citep{2015JHEAp...7...44L}. For instance, \kw observed three emission episodes for ULGRB GRB 091024A, while \swift only detected the first episode, resulting in a shorter \tninty duration \citep{2013ApJ...778...54V}. Consequently, it becomes imperative to establish a uniform sample selection methodology for conducting comprehensive analyses of ULGRBs. Motivated by this, we searched for possible candidates for ULGRBs in the complete sample of \swift-detected bursts\footnote{In this paper, we utilize \tninty as the criteria to discriminate between the various categories of GRBs and to compare their prompt and afterglow emission characteristics. We do not claim this is the only criterion for distinguishing ULGRB candidates from other SGRBs and LGRBs.}. However, our selection methodology extends beyond merely considering GRBs with durations exceeding a few thousand seconds. Our detailed sample selection approach and the methodology to constrain the possible progenitor and the central engine are described in Figure \ref{fig:selection_criteria}.

\begin{figure}[!ht]
\centering
\includegraphics[width=\columnwidth]{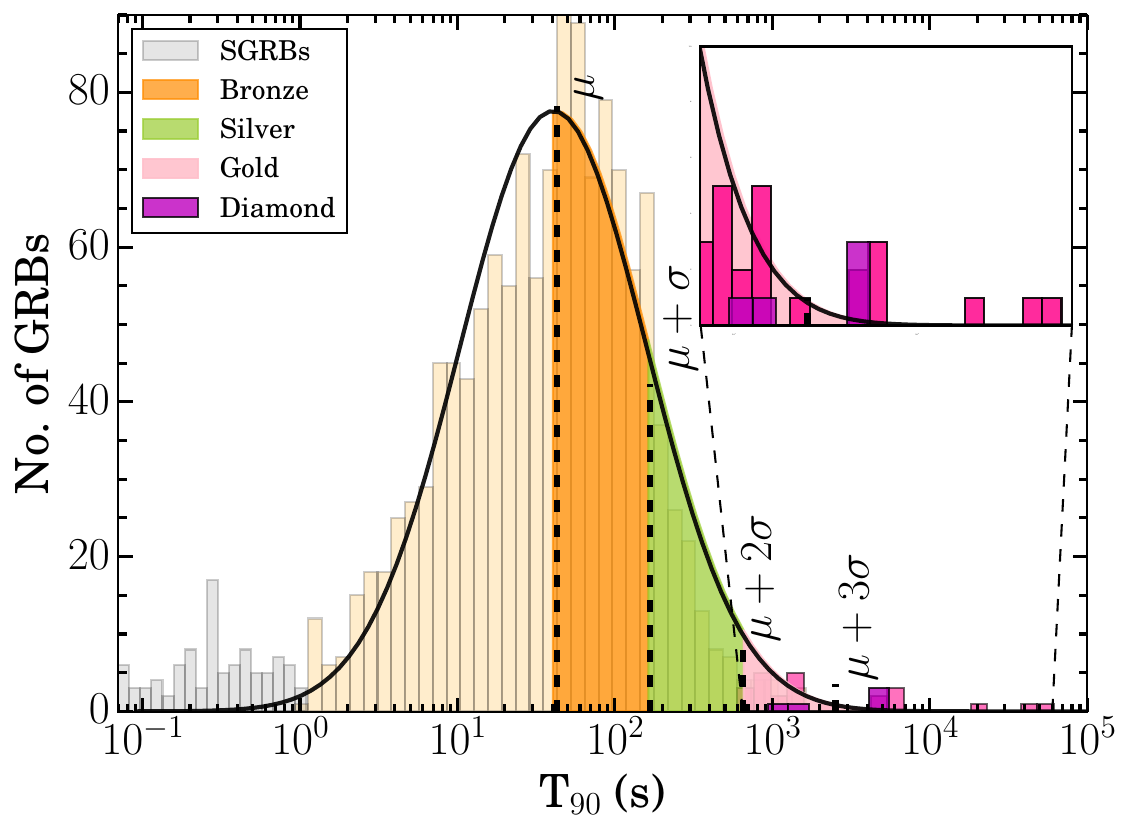}
\caption{Represents the \tninty distribution of \swift detected LGRBs. The orange, green, and pink shaded regions represent the Bronze, Silver, and Gold sub-samples of ULGRBs. The inset magnifies the region that represents the Gold sub-sample, and additional magenta bars represent the diamond sub-sample. The black dashed lines are plotted at $\mu$ = 43\,s, $\mu$+$\sigma$ = 167\,s, $\mu$+2$\sigma$ = 649\,s, and $\mu$+3$\sigma$ = 2519\,s.}
\label{fig:T90_dist}
\end{figure}

\begin{itemize}

\item Initially, we constructed a log-normal distribution representing the \tninty durations of a complete sample (from 2005 to 2023) of \swift-detected GRBs\footnote{\url{https://swift.gsfc.nasa.gov/results/batgrbcat/index_tables.html}}, which unveiled a bimodal pattern. For those GRBs listed as ULGRBs in \swift catalog\footnote{\url{https://swift.gsfc.nasa.gov/results/batgrbcat/summary_cflux/summary_GRBlist/list_ultra_long_GRB_comment.txt}}, we have used \tninty duration from the third \swift-BAT catalog \citep{2016ApJ...829....7L}. This is because bursts with longer durations are most likely to have emissions beyond the event data range (photons detected in T$_{\rm 0,BAT}$-250 s to T$_{\rm 0,BAT}$+950 s, where T$_{\rm 0,BAT}$ is the BAT trigger time). \cite{2016ApJ...829....7L} analyzed the BAT survey data and provided the complete duration for these bursts, combining the event and survey data. The observed bi-modality exhibited one peak associated with SGRBs and another with LGRBs. In our pursuit of ULGRB candidates, we deliberately omitted the peak associated with the SGRBs, as illustrated in Figure \ref{fig:T90_dist}.

\item We fitted a Gaussian function to the distribution of \swift-detected LGRBs. To create our ULGRB sample, we exclusively chose GRBs with \tninty durations greater than the mean of the distribution, denoted as $\mu$ (with $\mu$ = 43 s). There are $\sim$ 740 GRBs with \tninty $>$ 43 s. In order to compute the true energetics, we require these GRBs to have redshift, which reduces the sample down to 230 GRBs (see Table \ref{tab:collapsar}). We further subdivided the sample into Bronze, Silver, and Gold sub-samples through divisions in \tninty durations. With Bronze, Silver, and Gold bursts falling within the ranges of ($\mu$ - $\mu$+$\sigma$), ($\mu$+$\sigma$ - $\mu$+2$\sigma$), and ($\mu$+2$\sigma$ - $\mu$+3$\sigma$ or beyond), respectively.

\item Furthermore, to ensure comprehensiveness, we have incorporated well-studied instances of GRBs with \tninty $>$ 1000 s (that are not included in our Bronze, Silver, and Gold sub-samples) from the existing literature and put them in the diamond sub-sample. A detailed description of these ULGRB candidates is given in Table \ref{tab:lit_ulGRBs}.
\end{itemize}

\begin{figure}[!ht]
\centering
\includegraphics[width=\columnwidth]{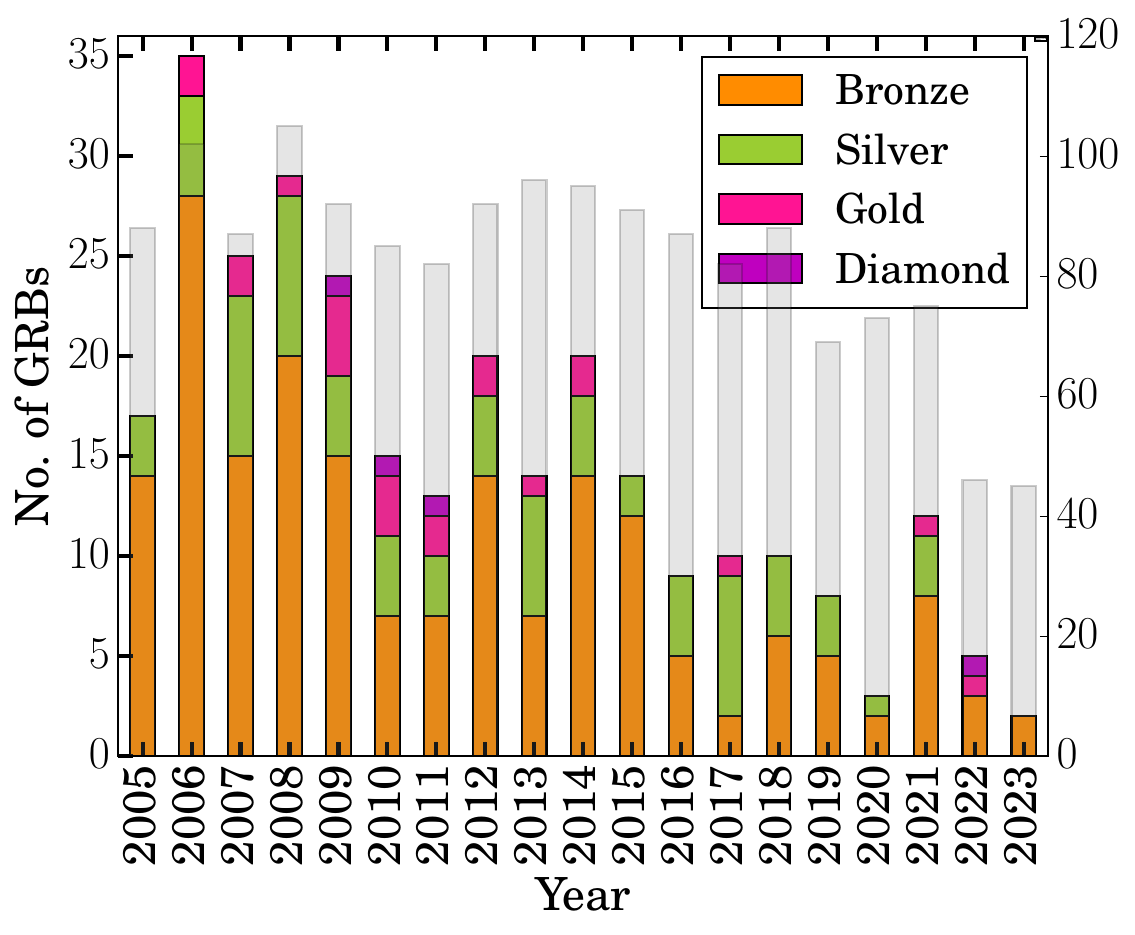}
\includegraphics[width=\columnwidth]{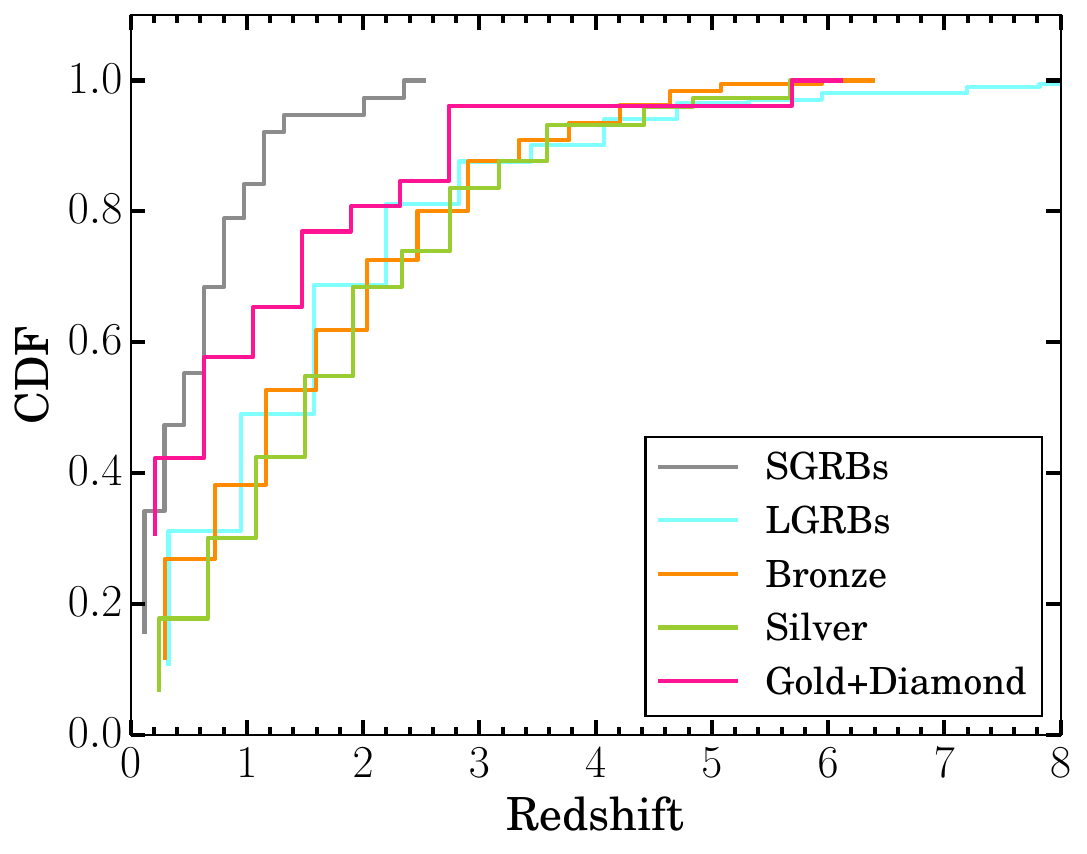} 
\caption{Upper panel represents the year-wise distribution of our Bronze, Silver, Gold, and Diamond samples of ULGRBs detected up to December 2023. The histogram plot of all \swift detected GRBs is shown in the background with gray color bars. The drop in long-duration GRBs (\tninty $>$ 43 s) over time might reflect the aging of BAT instruments, and the number of enabled detectors has decreased significantly over the years due to their permanently noisy behavior \citep{2022ApJ...927..157M}. On the other hand, \swift has conducted a significantly greater number of slews to observe more targets in recent years, resulting in shorter exposure times for each pointing. Given that many ULGRBs exhibit faint and longer emissions detecting them becomes more challenging with shorter exposure times. The bottom panel represents the cumulative distribution of redshift of the Bronze, Silver, and Gold+diamond sub-samples.}
\label{fig:CDF}
\end{figure}

The yearly distribution of our Bronze, Silver, and Gold sub-samples is depicted in the upper panel of Figure \ref{fig:CDF}. We have shown the cumulative distribution of redshift of the Bronze, Silver, and Gold sub-samples in the bottom panel of Figure \ref{fig:CDF}. We noted that the cumulative distribution of the Gold + Diamond sub-sample does appear at a lower redshift with respect to our Bronze and Silver sub-samples, mainly due to selection effects. The typical fluence observed from ULGRBs is not very different from LGRBs \citep{2015JHEAp...7...44L}. This observed fluence distributed over a longer time scale makes several ULGRBs faint. Therefore, these events are difficult to detect at higher redshift due to instrumental sensitivities or observational constraints. Consequently, detecting them at lower redshifts is more feasible, whereas only a few bright ULGRBs may be detectable at higher redshifts.

\section{Comparison among the characteristics of sample GRBs}
In this section, we compare the temporal and spectral characteristics of GRBs in our Bronze, Silver, and Gold sub-samples.

\subsection{Machine learning technique to differentiate between sub-samples}

We have used a machine learning tool, t-Distributed Stochastic Neighbor Embedding (t-SNE), developed by \cite{2023ApJ...951....4G}, to find differences between our selected sub-samples and other LGRBs and SGRBs detected by \swift-BAT till December 2023. t-SNE processes the high-energy light curve of GRBs and, based on similarities and dissimilarities between the light curves, places them in a two-dimensional map by forming a cluster of points where similar events lie close. The axes of this two-dimensional map do not have any significance. However, labeling each event with redshift and \tninty using different colors or markers allows us to observe their impact on grouping within the map. The clustering between two groups representing the bimodal distribution of GRBs (SGRBs and LGRBs) was observed by \cite{2020ApJ...896L..20J} using the t-SNE method. 

\begin{figure}[!ht]
\centering
\includegraphics[width=\columnwidth]{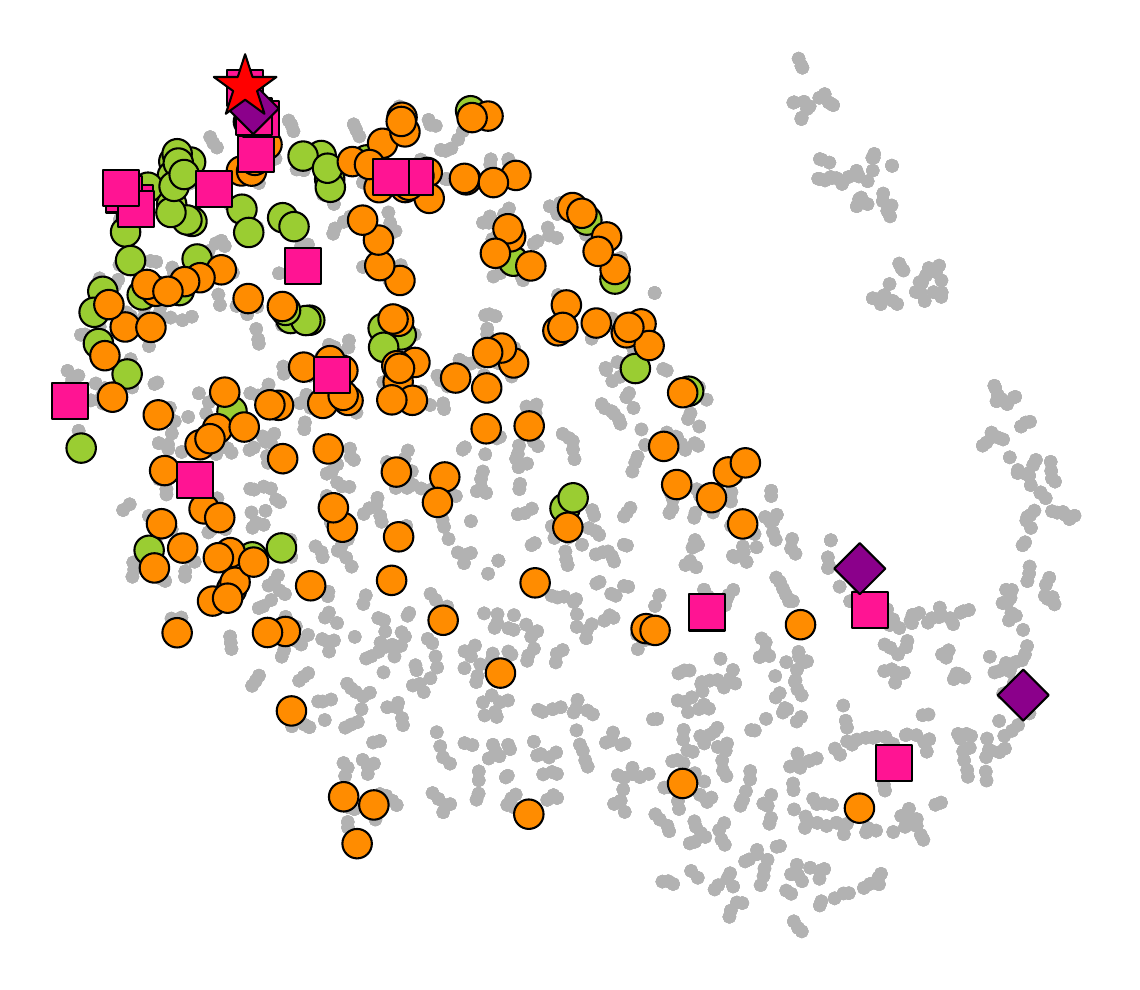} 
\caption{t-SNE distribution map for \swift-BAT GRBs 64 ms binned light curves are grouped into two main classes. Green and orange colored circles represent our Silver and Bronze sub-samples, respectively. Red and magenta squares represent the Gold and Diamond sub-samples. GRB 221009A is shown with a red star.}
\label{fig:HR_T90_tsne}
\end{figure}

To download and process the \swift-BAT data (GRBs detected up to December 2023) with t-SNE, we follow the procedures given in \cite{2023ApJ...951....4G}. t-SNE map of the \swift detected GRBs along with GRBs in the Gold, Silver, and Bronze sub-sample highlighted is shown in Figure \ref{fig:HR_T90_tsne}. In this figure, the Bronze, Silver, Gold, and Diamond sub-samples are represented by orange, green, red, and magenta colors, respectively. Since our Bronze, Silver, Gold, and Diamond sub-samples represent the light curves of different durations, we expect different clustering of these events by t-SNE. From the figure \ref{fig:HR_T90_tsne}, we noted that the Silver sub-sample GRBs mostly lie on the left of the map and gradually decrease toward the right. Similarly, the Gold and Diamond sub-sample also tends to cluster on the left of the map; however, five GRBs (GRB 070518 T$_{90}$ $\sim$ 5.5 s, GRB 090309A T$_{90}$ $\sim$ 3.0 s, GRB 090404 T$_{90}$ $\sim$ 82 s, GRB 091127 T$_{90}$ $\sim$ 7.42 s, and GRB 101024A T$_{90}$ $\sim$ 18.7 s) lying on the right are considered in our Gold/Diamond sub-sample based on the duration given in \cite{2016ApJ...829....7L}. GRB 221009A lies at the top left edge of the map, indicating the ultra-long nature of the burst. The detailed physical implications of the obtained results are given below:

Utilizing the t-SNE map as a tool to discern between different classes of GRBs based on their observed prompt emission light curves, we made several notable observations. Firstly, we observed that long and short GRBs are distinctly segregated into the bulk and tail regions of the t-SNE map, respectively. This clear grouping suggests that short GRBs exhibit prompt emission light curves that are fundamentally different from those of long GRBs, and both of these classes have diffident physical origins. However, upon closer examination, we found that the light curves of the Bronze sub-sample do not exhibit any discernible structural differences compared to long GRBs. Instead, they are uniformly distributed among the long GRBs. The uniform distribution indicates that the selected features used for the t-SNE analysis do not effectively discriminate between Bronze sub-sample and long GRBs, as expected. Furthermore, we noticed that GRBs in the Silver and Gold sub-samples predominantly cluster on the left side of the t-SNE map. This clustering suggests that the prompt emission light curve morphologies of these sub-classes may differ from those of long GRBs, and they might have different physical origins. However, it's worth noting that some GRBs in the Gold and Diamond sub-samples also appear on the right side of the map. This occurrence is primarily due to observation constraints; only a short portion of their light curve is utilized in the grouping. Our analysis revealed that the t-SNE grouping is primarily based on temporal features of observed light curves and does not adequately distinguish between different sub-classes based on the activity of the central engine. Consequently, relying solely on light curve morphology for the distinction between different GRB classes may be only partially appropriate. Therefore, in the subsequent sections, we utilize other methods to distinguish the characteristics of different sample sub-classes.

\subsection{High energy characteristics of sample GRBs}
\begin{figure}[!ht]
\centering
\includegraphics[width=\columnwidth]{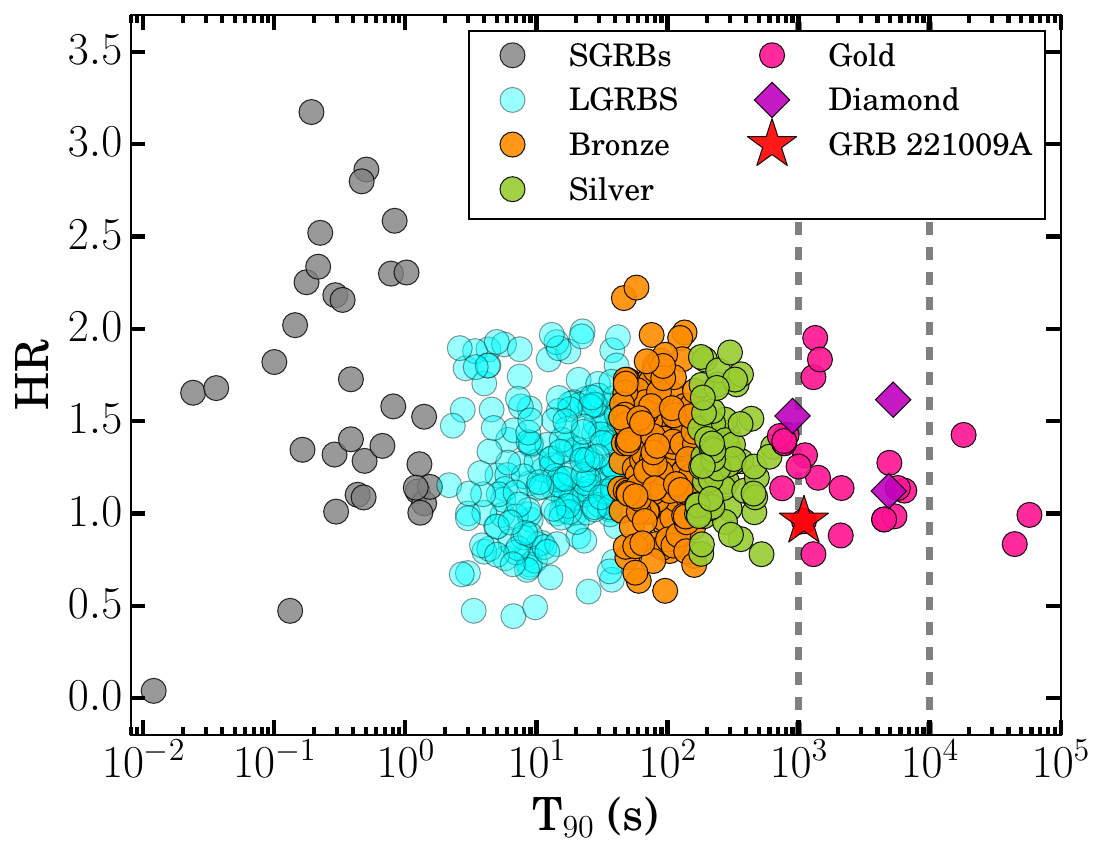} 
\caption{The hardness ratio in the \swift-BAT 50-100 \keV/25-50 \keV bands for our selected sub-samples is plotted with the burst's duration (\tninty). Two vertical gray lines at $\sim$ 10$^3$ and 10$^4$ s represent the proposed demarcation of T$_{90}$ duration between LGRBs and ULGRBs as published by \cite{2015ApJ...800...16B, 2015JHEAp...7...44L}, respectively.}
\label{fig:HR_T90}
\end{figure}

The prompt identification of ULGRBs is crucial for in-depth observational and theoretical investigations. We utilize the spectral characteristics of GRBs in Bronze, Silver, and Gold sub-samples and search for potential differences from other GRBs of well-studied sub-classes such as LGRBs with \tninty $<$ 43 s (as \tninty $>$ 43 s included in our Bronze sub-sample) and SGRBs. We calculated the hardness ratio (HR) for each GRB in our sample by comparing the fluence in the hard energy range (50-100\,\keV) to that in the soft energy range (25-50\,\keV). Figure \ref{fig:HR_T90} illustrates the distribution of HR as a function of \tninty for all the bursts. We noted that GRBs in the Gold sub-sample exhibit lower average hardness values. GRB 221009A also lies towards the softer ends of HR. 

\begin{figure}
\centering
\includegraphics[scale=0.36]{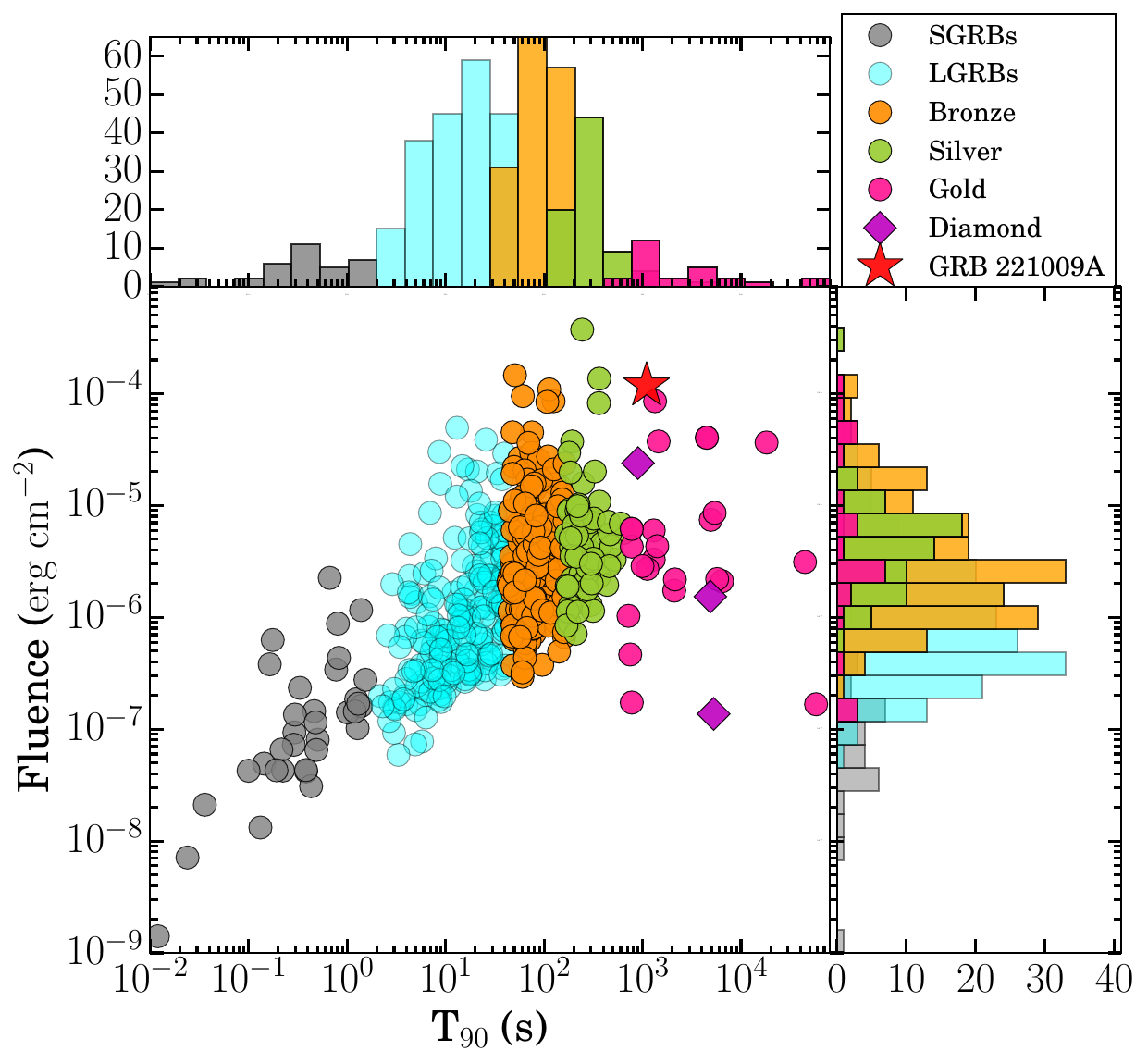} 
\includegraphics[scale=0.36]{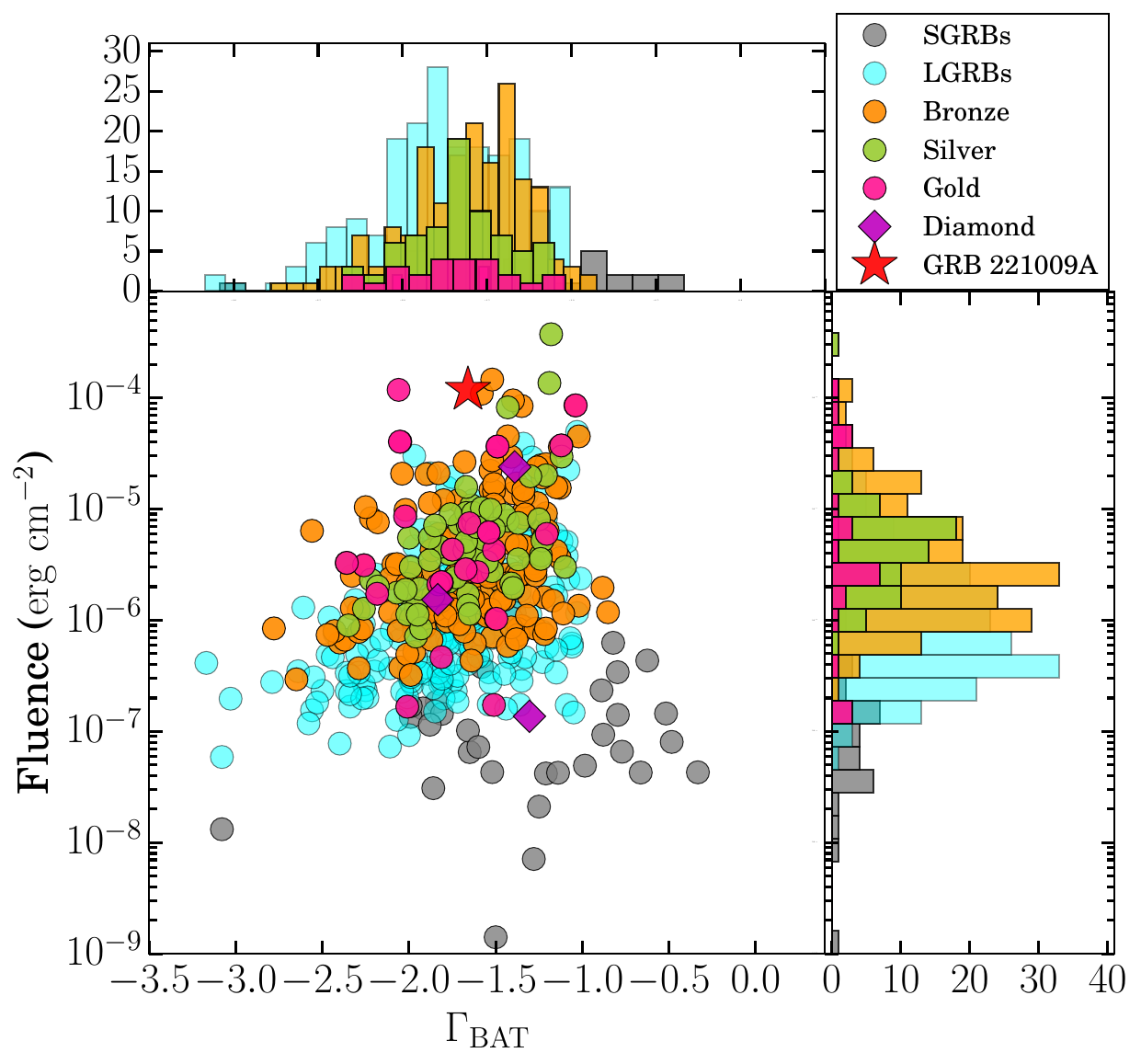} 
\includegraphics[scale=0.36]{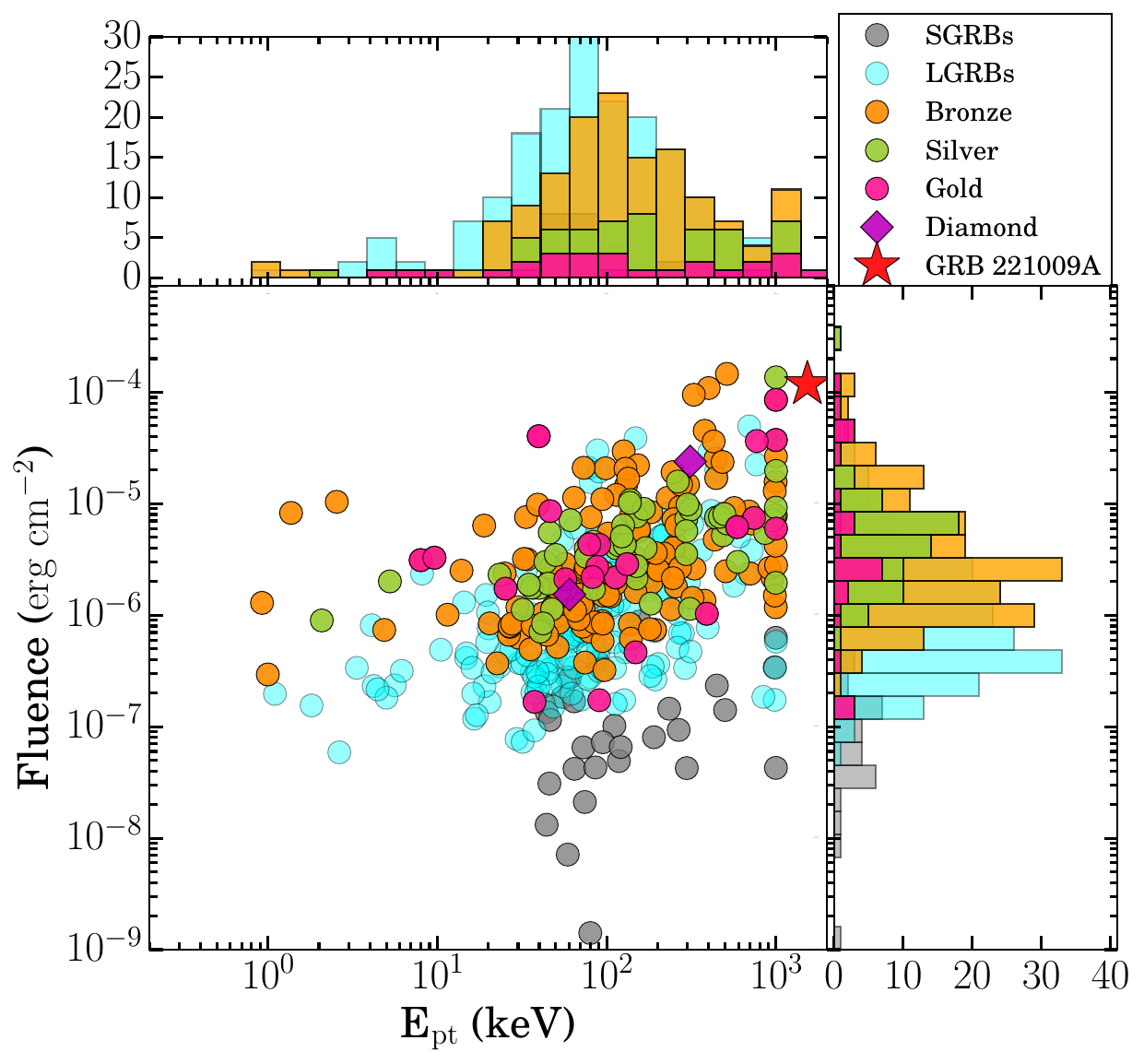} 
\caption{The upper panel represents fluence distribution in 15-150 \keV, along with the durations of GRBs in our Bronze, Silver, Gold, and Diamond sub-samples. The middle panel shows the spectral index ($\Gamma_{\rm BAT}$) obtained from fitting the time-integrated spectra of BAT with a power-law, along with the fluence obtained in the 15-150 \keV range. Similarly, the lower panel displays the distribution of \Ep along with the fluence obtained in the BAT energy range.}
\label{fig:gamma_Ep_flu}
\end{figure}

In the upper panel of Figure \ref{fig:gamma_Ep_flu}, we have shown the distribution of fluence in 15-150 \keV along with their durations. We found an increasing trend in the fluence with durations of the burst from SGRBs to GRBs in our Bronze and Silver sub-samples. However, Gold and diamond sub-samples do not seem to follow the trend. GRB 221009A is the brightest ever burst with observed fluence $\sim$ 0.09 erg cm$^{-2}$ (GBM, \citealt{2023ApJ...952L..42L}) and $\sim$ 10$^{-4}$ erg cm$^{-2}$ (BAT 15-150 keV, \citealt{eva07, eva09}). Since the Gold and diamond sub-samples show similar properties, from now on, we will present the combined properties of these two sub-samples.

In the middle panel and the lower panel of Figure \ref{fig:gamma_Ep_flu}, we have plotted the spectral parameters $\Gamma_{\rm BAT}$ and E$_{\rm pt}$, along with the fluence in the BAT 15-150 \keV range. The mean value of the spectral index $\Gamma_{\rm BAT}$ in the given energy range for SGRBs, LGRBs, Bronze, Silver, and Gold sub samples, respectively, are -1.36 $\pm$ 0.58, -1.75 $\pm$ 0.42, -1.63 $\pm$ 0.35, -1.66 $\pm$ 0.29, -1.65 $\pm$ 0.23, and -1.72 $\pm$ 0.34. Since the BAT detects GRBs in the soft energy range (limited spectral coverage), the spectral peak energy of many bright GRBs can surpass this range. Due to this, some GRBs in the lower panel of Figure \ref{fig:gamma_Ep_flu} show deviation from the distribution. We have calculated the mean values of the E$_{\rm pt}$ by removing the unphysical E$_{\rm pt}$ values, and the obtained mean values for SGRBs, LGRBs, Bronze, Silver, and Gold samples, respectively, are 99.93 $\pm$ 70.48, 92.69 $\pm$ 75.54, 111.54 $\pm$ 84.096, 119.15 $\pm$ 91.29, and 75.41 $\pm$ 53.16 KeV. Our observations indicate that the Gold and Silver samples demonstrate a softer spectrum (though consistent within error bars) compared to the Bronze sub-sample and LGRBs. At the same time, SGRBs exhibit the most hard spectral characteristics.

\subsection{Search for GeV emission using \fermi-LAT analysis}

\begin{figure}[!ht]
\centering
\includegraphics[width=\columnwidth]{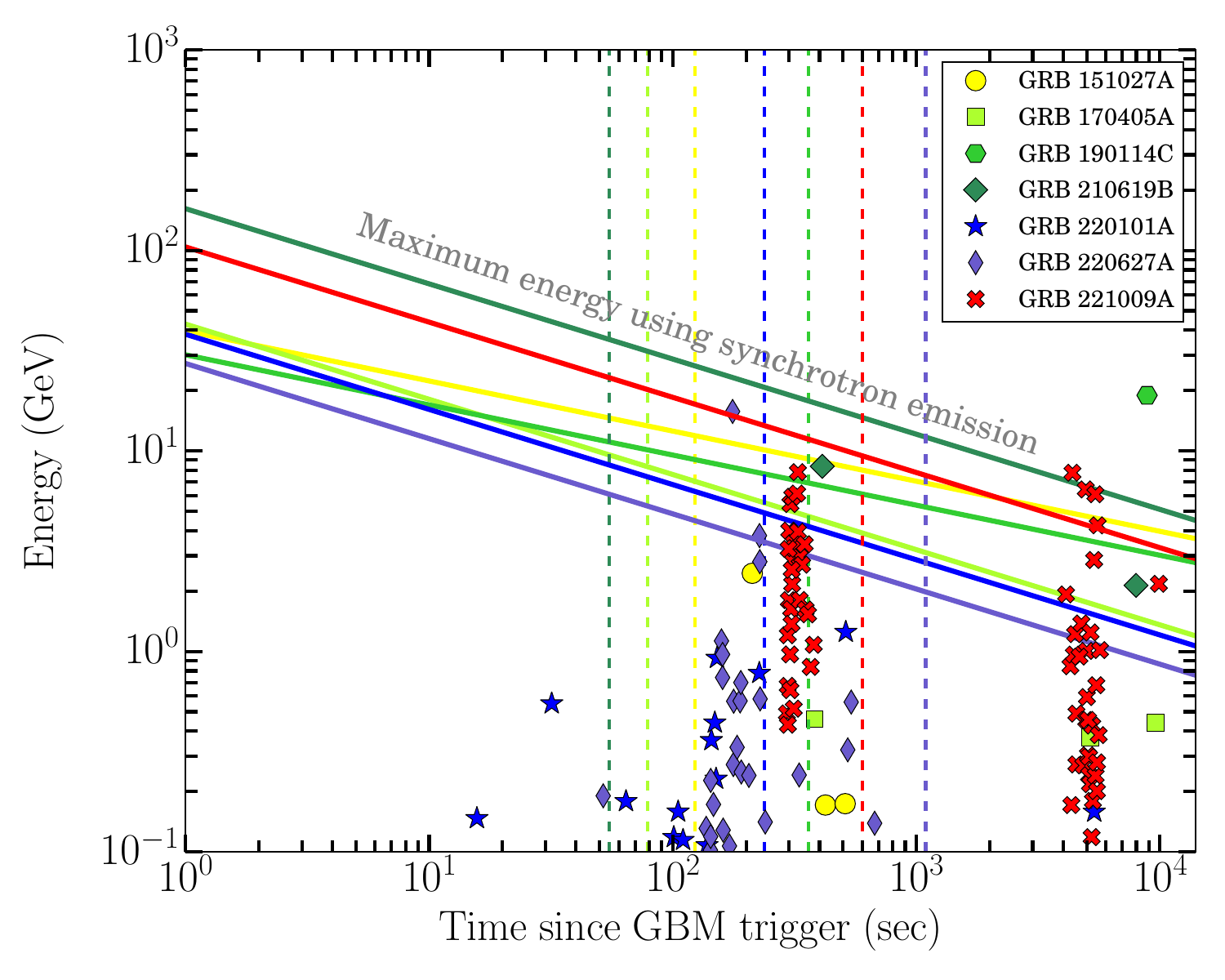} 
\caption{The \fermi-LAT observations of ULGRBs given in Table \ref{tab:lat_ulGRBs} with successful LAT detection. The various colored markers represent the photons with a probability of greater than 90\% associated with these bursts. The corresponding colored lines represent the maximum limit allowed for synchrotron emission for each GRB. The vertical dashed lines show the end epoch of \tninty prompt duration as observed by \fermi-GBM.}
\label{fig:lat_ulgrbs}
\end{figure}

The duration of prompt emission (\keV to MeV energy range) of ULGRBs is significantly longer, spanning 2-3 orders of magnitude compared to typical LGRBs \citep{2014ApJ...781...13L}. Nevertheless, owing to orbital constraints, Earth occultation, and limited sensitivity, it presents a challenge for instruments to capture all emissions throughout the entire duration of ULGRBs. To distinguish the high-energy emission (in the GeV energy range) of GRBs in the Bronze, Silver, and Gold sub-sample from other bursts, we conducted an analysis of \fermi-LAT (Large Area Telescope) data for GRBs that were simultaneously detected by both \swift-BAT and \fermi-GBM in our sample. We have listed the LAT boresight angle for these GRBs in Table \ref{tab:gbm_ulGRBs}. The analysis involved acquiring and examining LAT data using the \sw{gtburst} software within a temporal range of 0 to 10 ks post-detection. A region of interest around the burst (10 x 10 degrees) was defined while implementing a Zenith angle cut of 100$^{\circ}$ to reduce contamination from the Earth's limb. The \sw{P8R3\_SOURCE} instrument response file was utilized for the analysis. The \fermi LAT time-integrated spectra within the 100 MeV to 100 GeV range were fitted using a power-law model. To establish a detection threshold, Test Statistic (TS) was used, setting TS $>$ 15 for adequate LAT detection. A more detailed method of \fermi LAT data analysis is presented in \cite{2021MNRAS.505.4086G}. The likelihood of associating the photons with each burst is computed using the \sw{gtsrcprob}. Following this analysis method, we found that there are three Bronze (GRB 151027A \citealt{2018ApJ...869..151R}, GRB 170405A \citealt{2020ApJ...891..106A}, and GRB 210619B \citealt{2023MNRAS.519.3201C}), two Silver (GRB 190114C \citealt{2019ApJ...879L..26F}, GRB 220101A \citealt{2022ApJ...941...82M}), one Gold (GRB 221009A, \citealt{2023ApJ...952L..42L}), one diamond (GRB 220627A \citealt{2022ApJ...940L..36H, 2023arXiv230710339D}) bursts in our sample with confirmed LAT detection (see Table \ref{tab:lat_ulGRBs}). Figure \ref{fig:lat_ulgrbs} illustrates the number of high-energy photons detected with a probability greater than 90\%, plotted against time since the GBM trigger for all the seven bursts. In comparing LAT GeV light curves with the prompt emission duration of ULGRBs, the LAT emission persists beyond 10,000 s from the GBM trigger, indicating the presence of an extended high-energy emission for ULGRBs, typical to LGRBs. Furthermore, we also observe that for a few GRBs, the origin of LAT emission is consistent with prompt \keV-MeV emission \citep{2022ApJ...940L..36H, 2023arXiv230710339D, 2023ApJ...952L..42L}. This implies a shared internal region of emission encompassing the entire Fermi energy range for these bursts. However, the delayed and long GeV emission post prompt emission is expected to originate from the external shock model \citep{2009MNRAS.400L..75K, 2010MNRAS.409..226K, 2019ApJ...879L..26F, 2023MNRAS.519.3201C}. Further, to explore the radiation mechanism of GeV LAT emission, we computed the maximum photon energy emitted by the synchrotron radiation mechanism in an adiabatic external forward shock during the decelerating phase, assuming an ISM or Wind stellar external medium following \cite{2010ApJ...718L..63P} with number density values from literature \citep{2019ApJ...879L..26F, 2023MNRAS.519.3201C, 2023arXiv230710339D} or 1 cm$^{-3}$, if not available. It was observed that some late-time photons, with a source association probability exceeding 90\%, surpass the maximum synchrotron energy for GRB 190114C, GRB 220627A, and GRB 221009A. This observation suggests a non-synchrotron origin for these photons \citep{2019ApJ...879L..26F, 2023ApJ...952L..42L}. In the case of recently detected VHE GRBs, photons above the maximum synchrotron energy point towards a Synchrotron self-Compton origin for these GeV photons \citep{2019Natur.575..459M, 2019Natur.575..464A, 2019ApJ...879L..26F}.

\subsection{GRB 221009A in the context of ULGRBs}
\begin{figure*}
\centering
\includegraphics[scale=0.525]{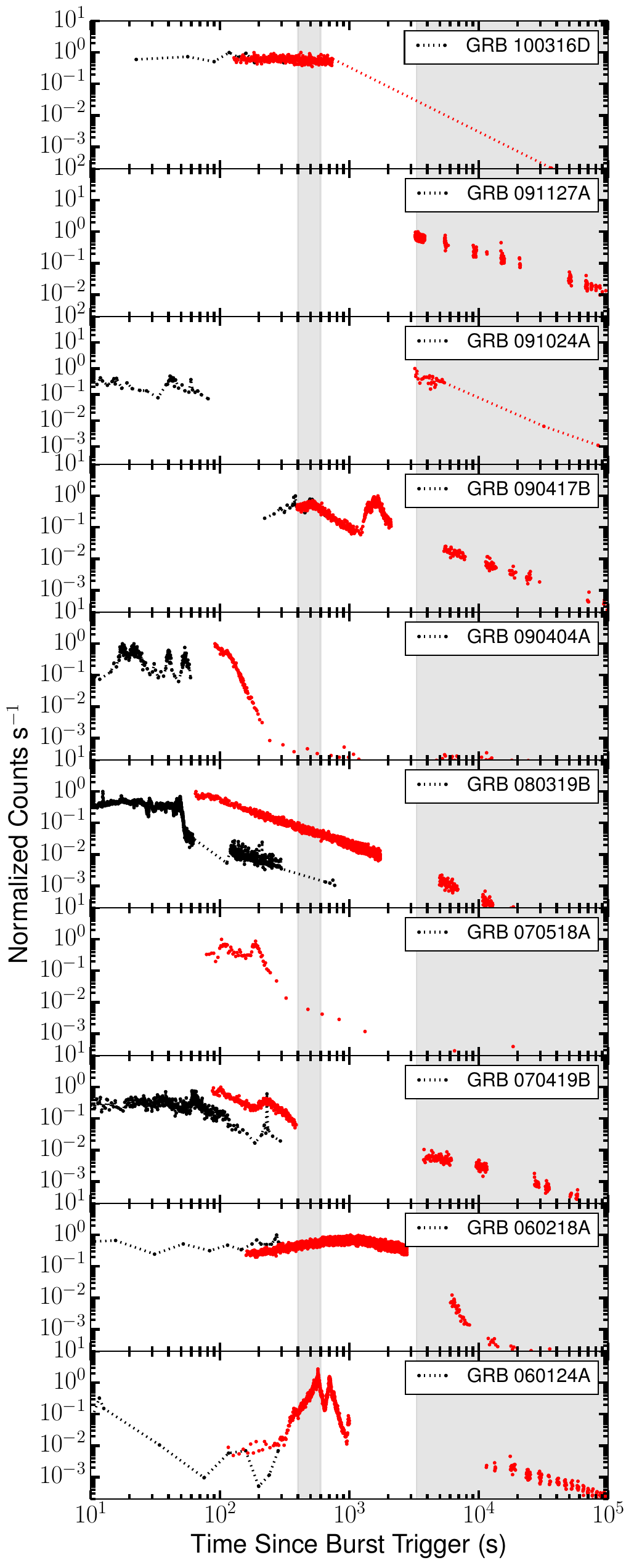}
\includegraphics[scale=0.525]{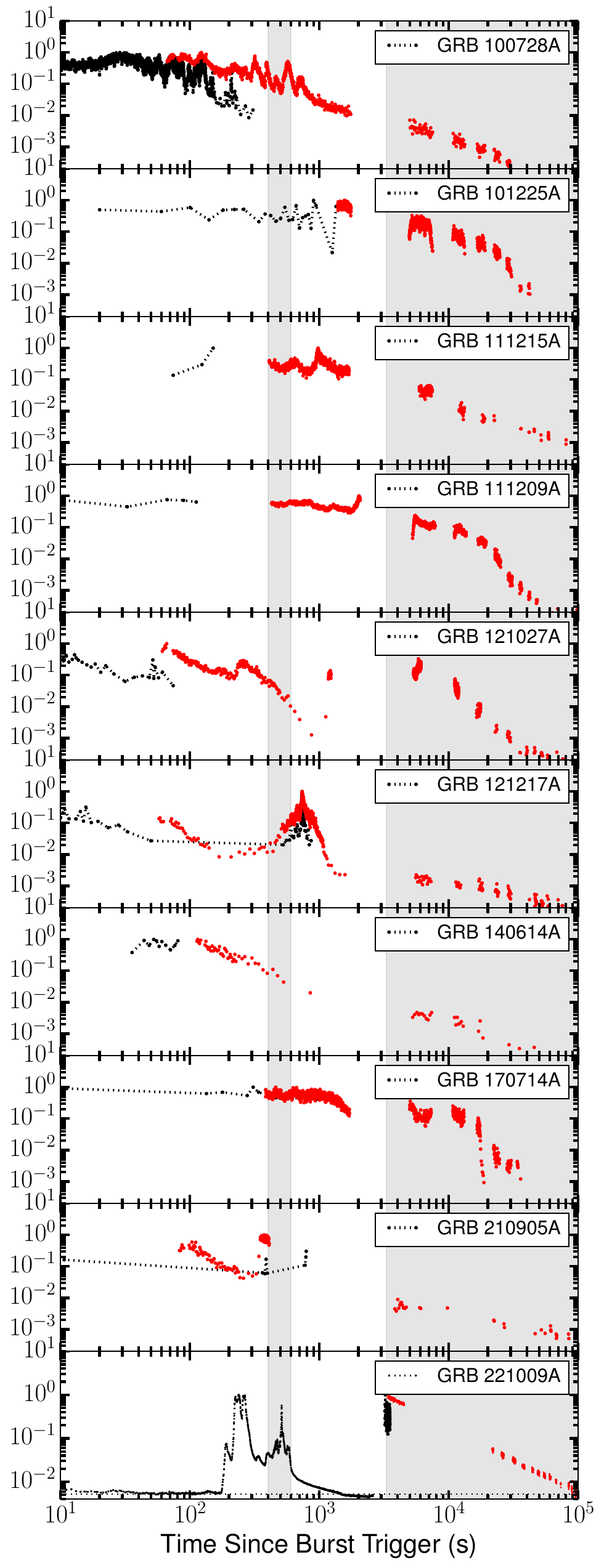}
\caption{A comparison of GRB 221009A with the GRBs included in our Gold sub-sample. The plot displays the normalized count rate light curve observed by BAT and XRT in the 0.3–10 keV range. However, for GRB 221009A, the GBM observation in NaI-7 was utilized for prompt emission, while BAT and XRT were employed within the same 0.3–10 keV range. A thin shaded region at 500 s represents the GBM flash, and a wider one starting at 3300 s covers XRT observations of GRB 221009A.}
\label{fig:ullc}
\end{figure*}

ULGRBs, characterized by their exceptionally long prompt emission durations, are comparatively uncommon in comparison to typical LGRBs and SGRBs. However, the situation changes significantly when considering the initial variability in the X-ray light curve \cite{2014ApJ...787...66Z}. Some of the well-studied candidates of ULGRBs detected in the \swift era are listed in our Gold and diamond sub-samples given in Table \ref{tab:collapsar} and \ref{tab:lit_ulGRBs} and in Figure \ref{fig:ullc}. It's worth noting that not all ULGRBs exhibit continuous emissions during the prompt phase, making it challenging to identify their ultra-long nature. Moreover, the duration of a GRB depends on the sensitivity of the detecting instruments and orbital constraints. For example, GRB 091024A has a weak pulse at 0 s, followed by brighter emissions at $\sim$ 600 s and $\sim$ 900 s post burst, respectively. However, \swift-BAT detected only the first pulse \citep{2013ApJ...778...54V}. On the other hand, GRB 220627A presents a double burst with the first pulse at 0 s and the second at 1000 s post-burst, separated by a quiescent gap of around 600 s \citep{2023arXiv230710339D}. GRB 060218 (SN 2006aj; \cite{2006Natur.442.1008C}) and GRB 100316D (SN 2010bh; \cite{2011MNRAS.411.2792S}) exhibit soft and long prompt emissions lasting for 2100 s and 1300 s, respectively, characteristics more akin to X-ray flashes than traditional GRBs. ULGRBs such as GRB 101225A (likely associated with a supernova \cite{2011Natur.480...72T}), GRB 111209A (SN 2011kl; \cite{2015Natur.523..189G}), GRB 130925A \citep{2014MNRAS.444..250E}, GRB 141121A \cite{2015ApJ...812..122C}, and GRB 170714A \cite{2018ApJ...854..104H} have been observed during the prompt emission with extremely long durations and accompanied by a highly variable initial X-ray light curve. In contrast, GRB 121027A \citep{2013arXiv1302.4876P}, despite having a prompt duration of only 80 s, is classified as a ULGRB due to its highly variable XRT light curve, extending up to 2000 s. Subsequently, from the BAT survey data, the duration of GRB 121027A was derived as 5730 s \citep{2016ApJ...829....7L}, placing it in our Gold sub-sample.

The prompt emission of GRB 221009A, as observed by \fermi-GBM, persisted above the background for more than 1000 s after the trigger, as depicted in the left panel of Figure \ref{fig:gbm_fit_param}. Notably, the prompt emission displayed a quiescent phase during which the central engine, although not entirely halted, produced multiple small pulses, maintaining the emission above the background levels. \cite{2023ApJ...952L..42L} demonstrated that after 600 s, the GBM detection smoothly transitioned to the afterglow emission. GBM observed this afterglow emission for up to 1500 s before it was occulted by the Earth. Similarly, \kw recorded GRB 221009A for more than 600 s. Moreover, \kw identified a subsequent tail emission persisting for approximately 20 ks \citep{2023arXiv230213383F}. In addition, authors have suggested that the duration of GRB 221009A is greater than 1000s and discussed the possibility of this burst being a ULGRB \citep{2023ApJ...946L..31B}. According to its reported \tninty duration, GRB 221009A satisfies the criteria of being ULGRB given by \cite{2015ApJ...800...16B} (\tninty $>$ 1000s) and also belongs to our Gold sub-sample.

However, no other low-energy X-ray or optical satellite was facing GRB 221009A to observe any soft flare during this time. \swift-BAT and XRT initiated observation of GRB 221009A afterglow at 3300 s after T$_{0}$, with XRT light curve decay with slope $\alpha_{x}$ = 1.66 $\pm$ 0.01. Figure \ref{fig:ullc} compares the temporal characteristics of GRB 221009A with those of other GRBs in our Gold sub-sample. Except for the GBM (NaI-7, 9-900 keV) observation of GRB 221009A, all the light curves are plotted in the temporal range (10 s-100 ks) and energy range of 0.3-10 keV for BAT (black) and XRT (red). Figure \ref{fig:ullc} shows that most of the GRBs in our Gold sample display either a plateau or flares during the early XRT light curve except for GRB 080319B and GRB 140614A, where a normal decay behavior can be seen throughout the afterglow phase. However, after 3300 s (time corresponding to the XRT trigger of GRB 221009A), the X-ray light curve for most of the GRBs decays following a simple power law, except for GRB 101225A, GRB 111209A and GRB 170714A, where the plateau extends for more than 10ks.

Our duration-based criteria place GRB 221009A in the Gold sub-sample, indicating that GRB 221009A is likely a potential ULGRB candidate detected by \fermi and \swift missions. To provide a comprehensive perspective, the prompt emission characteristics of GRB 221009A have been compared with those of other GRBs, as well as with GRBs from our Bronze, Silver, Gold, and Diamond sub-samples.

\section{Possible Origin of Extended Duration in ULGRBs: Methodology and Tools}
\label{origin}

In this section, we examine the possible progenitor, central engine, environments, and other key characteristics of our sub-samples using different methods \citep{2011ApJ...739L..55B, 2018ApJS..236...26L} and publicly available tools such as \sw{MESA}. Our detailed methodology to constrain the possible progenitor and the central engine is described in Figure \ref{fig:selection_criteria}.

\subsection{Supernova Connection with GRB 221009A ?} 

The emerging supernovae associated with nearby LGRBs are expected to cause a late red bump in the optical/NIR light curves and provide direct evidence of progenitors of GRBs \citep{Galama1998, 2003Natur.423..847H}. Previously studied nearby Very-High-Energy (VHE) detected GRBs, including GRB 190114C, GRB 190829A, and GRB 201015A, as well as ULGRB GRB 111209A, have revealed similar late bumps. These features are indicative of their associated supernovae and potential progenitor systems \citep{2006Natur.442.1008C, 2011MNRAS.411.2792S, 2015Natur.523..189G}. Both the close proximity and the long duration of GRB 221009A indicate the potential presence of a late optical bump in the afterglow light curve. Early spectroscopic observations taken using the 10.4\,m GTC telescope \citep{2022GCN.32800....1D} and subsequent photometric investigations \citep{2023arXiv230111170F} suggested the presence of an underlying supernova (initially dubbed SN 2022ixw) associated with GRB 221009A. However, findings from \cite{2023arXiv230203829S, 2023arXiv230204388L, 2023arXiv230207761L} neither support nor refute the presence of the underlying supernova emission associated with GRB 221009A. Further, the late-time ($\sim$ T$_{0}$+170 days) spectroscopic observation by the James Webb Space Telescope also favors the underlying supernova with observed spectral features similar to SN 1998bw \citep{2024NatAs.tmp...65B}.

\begin{figure}[!ht]
\centering
\includegraphics[scale=0.44]{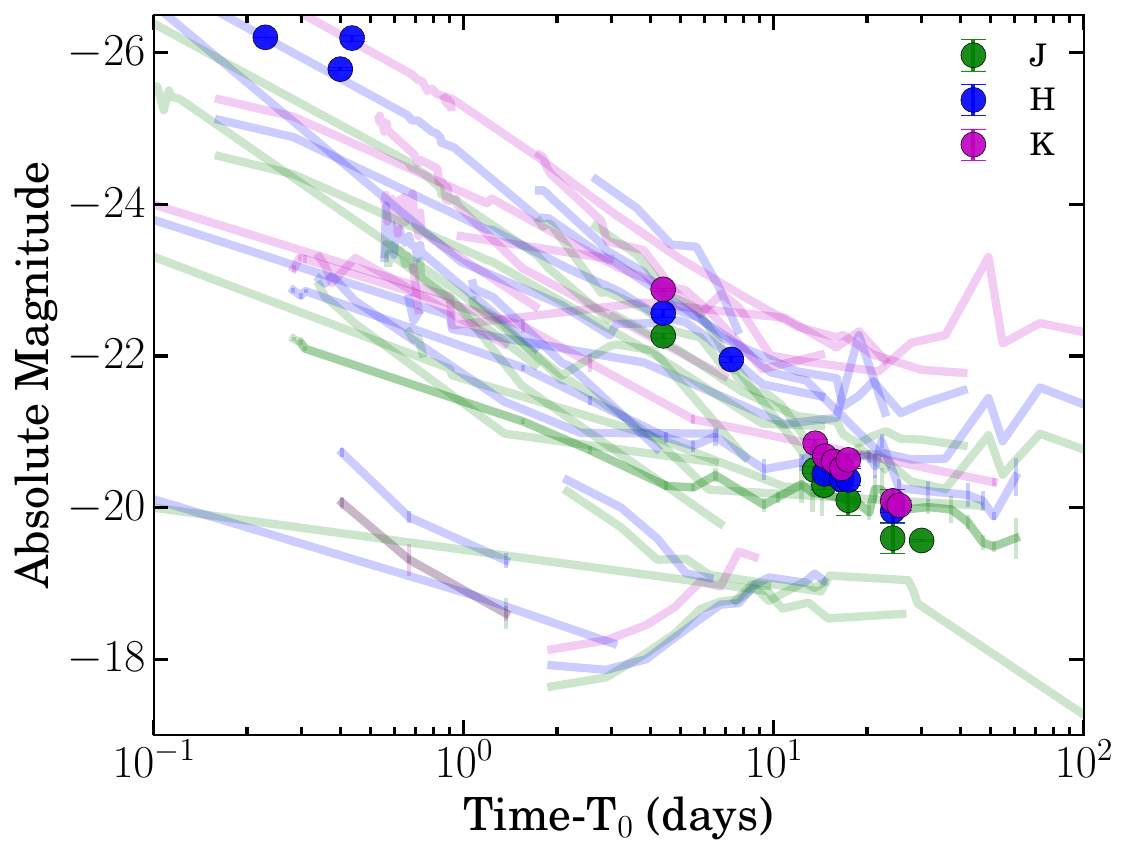}
\caption{The light curve of GRB 221009A in NIR bands (J, H, and K) is shown in colored circles. The data utilized for this plot is obtained from observations made with the 3.6m DOT \citep{2023arXiv230715585G} see also \citep{2016RMxAC..48...83P, 2018sn87.conf..149B}, published GCNs, and R. Sánchez-Ramírez et al. (2024, under review). Other GRBs connected to supernovae are also shown with colored lines in the background. The GRBs shown and corresponding references are: GRB 011121 \citep{2003ApJ...582..924G, 2002ApJ...572L..51P}, GRB 030329A \citep{2004AJ....127..252B}, GRB 060218A \citep{2007ApJ...663.1180K}, GRB 071112A \citep{2019A&A...622A.138K}, GRB 081007A \citep{2013ApJ...774..114J}, GRB 091127A \citep{2015A&A...577A..44O}, GRB 101219A \citep{2015A&A...577A..44O}, GRB 111228A \citep{2019A&A...622A.138K}, GRB 111209A \citep{2018A&A...617A.122K}, GRB 120422A \citep{2014A&A...566A.102S}, GRB 130215A \citep{2014A&A...568A..19C}, GRB 130427A \citep{2014ApJ...781...37P}, GRB 130702A \citep{2016ApJ...818...79T}, GRB 130831A \citep{2019A&A...622A.138K}, GRB 161219B \citep{2017A&A...605A.107C}, GRB 190114C \citep{2019Natur.575..459M, 2021RMxAC..53..113G}, GRB 190829A \citep{2021A&A...646A..50H}.}
\label{fig:SN_cmp}
\end{figure}

Figure \ref{fig:SN_cmp} illustrates the light curve of GRB 221009A obtained using our NIR observations along with data reported in GCNs and R. Sánchez-Ramírez et al. (2024, under review). Our observations revealed a consistent, smooth decay in the NIR light curve of GRB 221009A, distinct from the presence of bumps or flattening features as observed in the background light curves of supernova-connected GRBs. Hence, our NIR observations do not provide evidence for the presence of any bright supernova connected with GRB 221009A. However, faint supernova emission may be masked by the bright afterglow emission of GRB 221009A.

\subsection{Constraining the possible progenitor: Collapsar origin ?} \label{sec:collapsar}

Recent discoveries of GRB 200826A, GRB 211211A, and GRB 230307A have challenged our perception of the relation between \tninty and the origin of GRBs. GRB 200826A \citep{ahumada2021discovery} identified as an SGRB with a duration (\tninty) of 1.14 s, accompanied by an underlying supernova. Additionally, GRB 211211A \citep{2022arXiv220903363} and GRB 230307A \citep{2023arXiv230702098L}, with a duration of 50 and 35 s, respectively, are LGRBs originating from compact binary mergers. 

In this section, we determine the origin (collapsar or merger) of the bursts in our Bronze, Silver, and Gold sub-samples following \cite{2011ApJ...739L..55B} and determine their non-collapsar probability. The duration of the prompt emission of GRBs, represented by \tninty value, cannot be shorter than the time the engine remains active after the jet breakout. In most GRB models, these two durations are considered equal, denoted as \tninty = T$_{\rm Eng}$ - t$_{\rm b}$, where T$_{\rm Eng}$ is the duration for which engine is active, and t$_{\rm b}$ is the time taken by the jet to come out of the pre-existing envelope surrounding the progenitor star. It is unlikely that the engine will operate precisely long enough for the jet to break out of the star and then cease immediately afterward. This condition directly stems from the Collapsar model, implying that if ULGRBs originate from Collapsars, they must adhere to this criterion. To know the origin of GRBs in our sample, we calculated the jet opening angle ($\theta_{\rm j}$), which is then used to calculate t$_{\rm b}$. Then, we calculated the ratio T$_{\rm 90,z}$/t$_{\rm b}$ to constrain the possible progenitor of Bronze, Silver, and Gold sub-samples. We have used the following relation provided in \cite{2011ApJ...739L..55B} to calculate t$_{\rm b}$:

\begin{equation} \label{eqn:collapsar}
\rm t_{\rm b} (s) \sim{15 \epsilon_{\gamma}^{1/3} L_{\rm \gamma,iso,50}^{-1/3} \theta_{j,10^{\circ}}^{2/3} R_{11}^{2/3} M_{15\odot}^{1/3}}
\end{equation}

In this equation, L$_{\rm \gamma,iso,50}$  = $\frac{{\rm L}_{\gamma, iso}}{10^{50}}$ erg s$^{-1}$, where L$_{\gamma, iso}$ is the gamma-ray luminosity at the peak of the prompt light curve. $\theta_{\rm j, 10^{\circ}}$ = $\theta_{\rm j}/{10^{\circ}}$, where $\theta_{\rm j}$ is jet opening angle. R$_{11}$=R/10$^{11}$ cm, and M$_{15\odot}$= M/15 M$_{\odot}$, where R and M are the radius and the mass of the star, respectively. The calculated values of L$_{\rm \gamma,iso,50}$ for each burst included in our sample are given in Table \ref{tab:collapsar}. $\epsilon_{\gamma}$ is the radiative efficiency, fixed at 0.1 \citep{2011ApJ...739L..55B}. We calculated $\theta_{\rm j}$ using equation 4 of \cite{2021ApJ...908L...2S}. For the GRBs in our sample with clear evidence of a jet break, based on temporal and spectral indices, the time corresponding to the jet break (t$_{\rm j}$) is taken directly from the \swift-XRT webpage\footnote{\url{https://www.swift.ac.uk/xrt_live_cat/}}, otherwise, the last data point in the \swift-XRT light curve is assumed to be t$_{\rm j}$, providing a lower limit on $\theta_{\rm j}$. Initially, the mass (M) of the progenitor star is varied from 15\,M$_{\odot}$ to 30\,M$_{\odot}$, and the radius (R) is also accordingly varied. 

In this section, we initially determine the effect of mass and radius on t$_{b}$. We note that changing the mass from 15 to 30\,M$_{\odot}$ and keeping the radius fixed at 10$^{11}$ cm, there is only 20\% decrement in ratio T$_{\rm 90,z}$/t$_{\rm b}$ utilizing the equation \ref{eqn:collapsar}. Further, we vary the radius of star using the relation R = 1.33M$^{0.55}$ given by \cite{1991Ap&SS.181..313D} when mass is changed from 15 - 30 M$_{\odot}$. Thus utilizing equation \ref{eqn:collapsar} once again we estimate corresponding T$_{\rm 90,z}$/t$_{\rm b}$, we find that when both mass and radius are varied, there is a decrement of 38\% in ratio T$_{\rm 90,z}$/t$_{\rm b}$. We have shown above calculation in section \ref{tb_calculation}.

\begin{figure}
\centering
\includegraphics[scale=0.46]{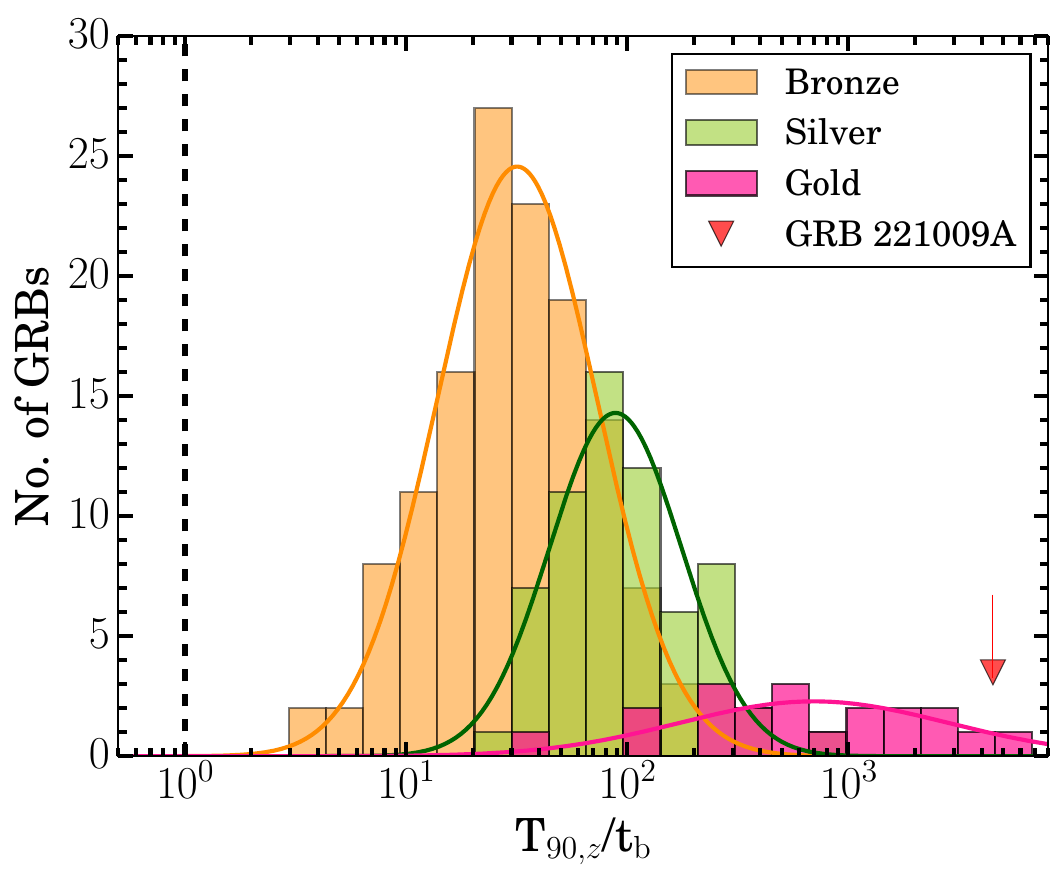} 
\caption{Distribution of the ratio of the rest frame T$_{\rm 90,z}$ and t$_{\rm b}$ calculated for Bronze, Silver, and Gold samples. The vertical dashed line indicates the rest frame \tninty = t$_{\rm b}$.}
\label{fig:T90_TB_ratio}
\end{figure}

Hence, there is no significant effect of M and R of the progenitor star in the calculation of T$_{\rm 90,z}$/t$_{\rm b}$ values in the considered range of mass and radius. Finally, we have used the progenitor star's mass and radius equal to 15M$_{\odot}$ and 10$^{11}$ cm, respectively \citep{2011ApJ...739L..55B}. The results of the distribution of T$_{\rm 90,z}$/t$_{\rm b}$ calculated for our Bronze, Silver, and Gold sub-samples are given in Figure \ref{fig:T90_TB_ratio} and Table \ref{tab:collapsar}. The vertical black dashed line in the plot indicates T$_{90}$ = t$_{\rm b}$, and all GRBs left of this line are considered to be of non-collapsar origin \citep{2011ApJ...739L..55B}. We noted that all GRBs in our sample lie to the right of the black dashed line, consistent with the collapsar origin. In addition, we have determined the probability of non-collapsar origin for our sample of GRBs by using equations 1 and 2 from \cite{2013ApJ...764..179B}. The probability values obtained are listed in Table \ref{tab:collapsar}. Negligible values of non-collapsar probabilities indicate the collapsar origin of the GRBs included in our Bronze, Silver, and Gold sub-samples.

\subsection{Constraining the central engine}  \label{central_engine}

Even after more than 50 years of GRB discovery, there is still much to learn about the central engines of GRBs. In this section, we constrain the possible central engines of our Bronze, Silver, and Gold sub-samples following the methodology presented in \citep{2018ApJS..236...26L, 2021ApJ...908L...2S}.

In the context of the compactness problem and the highly variable prompt emission light curve, a rapidly rotating compact object with an accretion disc is essential as a central engine for GRBs. Generally, two types of central engines are considered: a hyper-accreting black hole and a millisecond magnetar. While there is no direct observational evidence confirming the nature of the inner compact objects, certain features observed in the afterglow light curve provide some clues regarding their nature. A black hole central engine is the most important candidate for explaining the observed energy of GRBs. The power of a jet originating from a hyperaccreting black hole stems from two primary energy sources. First, the accretion energy present in the disk gives rise to neutrinos and antineutrinos. These particles annihilate each other, generating a collimated outflow \citep{1996ApJ...471..331Q}. Second, the rotational energy possessed by a Kerr black hole can be harnessed using magnetic fields through the process known as the Blandford-Znajek mechanism proposed by \cite{1977MNRAS.179..433B}. The energy released during the spin-down of a magnetar can also play a significant role in the formation of a bipolar jet \citep{2007MNRAS.380.1541B, 2009MNRAS.396.2038B}. However, the magnetar central engine relies on the fundamental concept that the maximum achievable rotational energy, approximately of the order of 10$^{52}$ erg, is possible to power jets from a millisecond magnetar. It's noteworthy that such a limit does not apply to a black hole central engine. Therefore, in this study, we leverage the maximum achievable rotational energy of a magnetar to investigate the potential central engine of GRBs. The rotational energy of a millisecond magnetar can be expressed as E$_{rot}$ = $\frac{1}{2}$ I $\Omega^2$, where I represent the moment of inertia of the magnetar with mass M$_{m}$ and radius R$_{m}$. The moment of inertia of a solid sphere is given by I = $\frac{2}{5}$ MR$^2$. The angular velocity $\Omega$ related to the period P$_{m}$ by the formula $\Omega$ = 2$\pi$/P$_{m}$. Considering the parameters M$_{m}$ = 1.4M${\odot}$, R$_{m}$ = 10 km, and P${m}$ = 1 ms as discussed in \citep{2014ApJ...785...74L}, the calculated value of rotational energy E$_{rot}$ is approximately 2.2 $\times$ 10$^{52}$ erg, which closely aligns with the value assumed for our analysis.\\

\begin{figure*}
\centering
\includegraphics[scale=0.48]{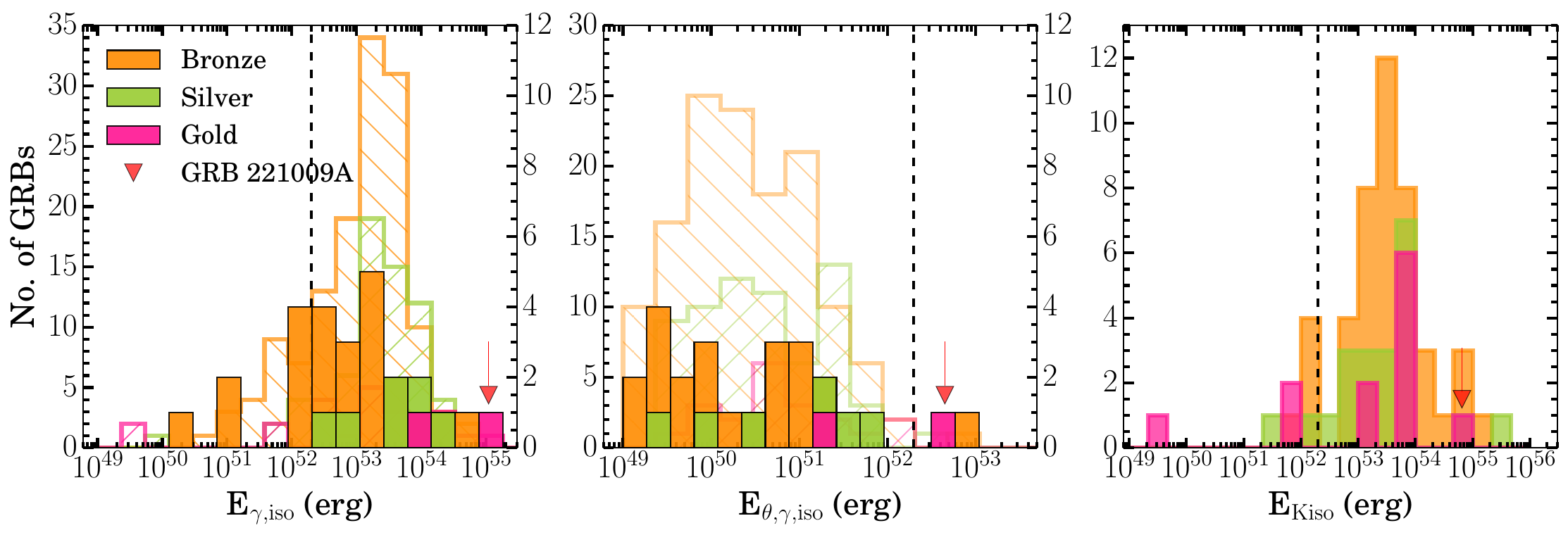} 
\caption{Histogram distribution of the isotropic energy (E$_{\rm \gamma, iso}$) (left), beaming corrected isotropic energy (E$_{\rm \theta_j, \gamma, iso}$) (middle) of \fermi (solid) and \swift (empty) detected bursts. Right: kinetic energy (E$_{\rm K, iso}$) of \swift detected bursts. Black dashed lines at 2 $\times$ 10$^{52}$ erg represent the maximum energy budget of a magnetar central engine.}
\label{fig:central_engine1}
\end{figure*}

For those GRBs in our sample simultaneously detected by \fermi-GBM and \swift-BAT (there are 32 such GRBs), we independently calculated the E$_{\rm \gamma,iso}$ using \fermi-GBM observations. For these GRBs, we have retrieved \fermi-GBM data from the official \fermi webpage\footnote{\url{https://heasarc.gsfc.nasa.gov/W3Browse/fermi/fermigbrst.html}}. Then, the time-integrated spectra for these GRBs are reduced by utilizing the latest version of the \sw{gtburst} software. Furthermore, the Multi-Mission Maximum Likelihood \sw{3ML} \citep{2015arXiv150708343V} framework is used for the spectral fitting of these time-integrated spectra. We fitted the \sw{Band} function to each spectrum, and the spectral parameters obtained are given in Table \ref{tab:gbm_ulGRBs}. The flux was calculated for each GRB in the energy range 10/(1+$z$) to 10000/(1+$z$), and using these values, we determined the isotropic gamma-ray energies E$_{\rm \gamma, iso}$ for each burst in the sample. Furthermore, we calculated the beaming corrected gamma-ray energy E$_{\rm \theta_j, \gamma, iso}$ = f$_{\rm b}$ $\times$ E$_{\rm \gamma, iso}$, where f$_{\rm b}$ = 1-cos($\theta_{\rm j}$) $\sim$ 1/2($\theta_{\rm j})^{2}$ is the beaming correction factor. If the beaming corrected energy is greater than the maximum energy budget of a magnetar (i.e., E$_{\rm \theta_j, \gamma, iso}$ $>$ 2 $\times$ 10$^{52}$), it rules out the possibility of a magnetar central engine \citep{2021ApJ...908L...2S}.
The histogram distributions of E$_{\rm \gamma, iso}$ and E$_{\rm \theta_j, \gamma, iso}$ for \fermi-GBM detected bursts are shown in left and middle panels of Figure \ref{fig:central_engine1}. We found only two GRBs (GRB 210619B and GRB 221009A) having E$_{\rm \theta_j, \gamma, iso}$ $>$ 2 $\times$ 10$^{52}$ erg for which a magnetar is excluded, while a black hole could be the possible central engine. For these cases, we have constrained the mass of the black hole using the equations 5 to 7 of \citep{2021ApJ...908L...2S} and obtained the black hole masses $\sim$3.4M$_{\odot}$ and $\sim$9.1M$_{\odot}$, respectively, for GRB 210619B and GRB 221009A.

For those GRBs in our sample only detected using \swift-BAT, we searched for the plateau in the \swift-XRT light curve and found 74 out of 230 GRBs exhibited at least one plateau. We retrieved the \swift-XRT spectra during the plateau phase and performed the spectral fitting. The methodology of \swift-XRT spectral fitting used to fit individual spectrum is given in the section \ref{sec:afterglow_analysis}. For these bursts, we have calculated the isotropic X-ray energy (E$_{\rm X,iso}$) corresponding to the plateau phase, total isotropic gamma-ray energy (E$_{\rm \gamma,iso}$), and the kinetic energy (E$_{\rm K,iso}$) to constrain the possible central engine of these \swift only detected burst. We have calculated E$_{\rm X,iso}$ released during the plateau phase using the relation:

\begin{equation} \label{eqn:EXiso}
E_{\rm X,iso} = \frac{4 \pi k D_{\rm L}^{2}}{\rm 1+z} \times {\rm F_{X}}
\end{equation}

Where D$_{\rm L}$ is the luminosity distance and F$_{\rm X}$ is X-ray fluence in the \swift-XRT energy range. $k$ = (1+z)$^{(\beta_{\rm X}-1)}$, where $\beta_{\rm X}$ is the spectral index obtained from the X-ray spectral fitting and z is redshift. Further, we have calculated E$_{\rm K,iso}$ of these GRBs using the relations given in \cite{2007ApJ...655..989Z}. E$_{\rm K,iso}$ of any GRB depends on the spectral regime and profile of the circumburst medium. We have considered the slow cooling regime. Even if the observed spectrum initially falls into the fast cooling regime (i.e., $\nu_{m}$ $>$ $\nu_{c}$), it's crucial to note that the evolution of $\nu_{m}$ ($\propto$ t$^{-3/2}$) is faster than the $\nu_{c}$ ($\propto$ t$^{-1/2}$ for ISM or t$^{1/2}$ for wind-like medium). Consequently, $\nu_{m}$ rapidly crosses $\nu_{c}$, resulting in the observed spectral shape predominantly lying within the slow cooling regime. Moreover, the X-ray afterglow modeling results presented by \cite{2015MNRAS.454.1073B} indicate that the X-ray emitting electrons typically lie in the slow cooling regime. Given these considerations, it is indeed relevant to consider the slow cooling regime in our analysis.\\

Based on the ISM or Wind-like surrounding medium and the location of break frequencies, the following three cases are possible:

(1) For the spectral regime $\nu_{\rm x}$ $>$ ($\nu_{\rm m}$, $\nu_{\rm c}$), the spectral indices are independent of the profile of the circumburst medium; we have used equation (8) of \cite{2018ApJS..236...26L} to calculate E$_{\rm K,iso}$. 
(2) For $\nu_{\rm x}$ $<$ $\nu_{\rm c}$ spectral regime and Wind-like surrounding media, the relation for E$_{\rm K,iso}$ is given by equations (10) of \cite{2018ApJS..236...26L}. 
(3) For $\nu_{\rm x}$ $<$ $\nu_{\rm c}$ spectral regime and ISM-like surrounding media, equation (11) from \cite{2018ApJS..236...26L} is utilized to determine (E$_{\rm K,iso}$). 
In these equations, $\nu F_{\nu}(\nu = 10^{18})$ is the energy flux at 10$^{18}$ Hz. $\epsilon_{e}$ (fixed at 0.1) and $\epsilon_{B}$ (fixed at 0.01) are the efficiencies of energy transfer to the electrons and magnetic field, respectively. $Y$ (fixed at 1) is the Compton parameter. The density parameter $n = 1$ is taken for an ISM-like surrounding medium, and A$_{*}$ is the density parameter for a wind-like surrounding medium. Initially, we used the closure relations by utilizing the temporal and spectral indices of the normal decay phase followed by the plateau phase of X-ray afterglow to constrain the spectral regime and the surrounding medium profile for each GRB. For the corresponding best possible spectral regime and the surrounding medium of each GRB, we calculated E$_{\rm K,iso}$ values. The histogram distribution of E$_{\rm K,iso}$ is shown in the right panel of Figure \ref{fig:central_engine1}. The calculated values of E$_{\rm X,iso}$ and E$_{\rm K,iso}$ are listed in Table \ref{tab:collapsar}.

\begin{figure}
\centering
\includegraphics[scale=0.44]{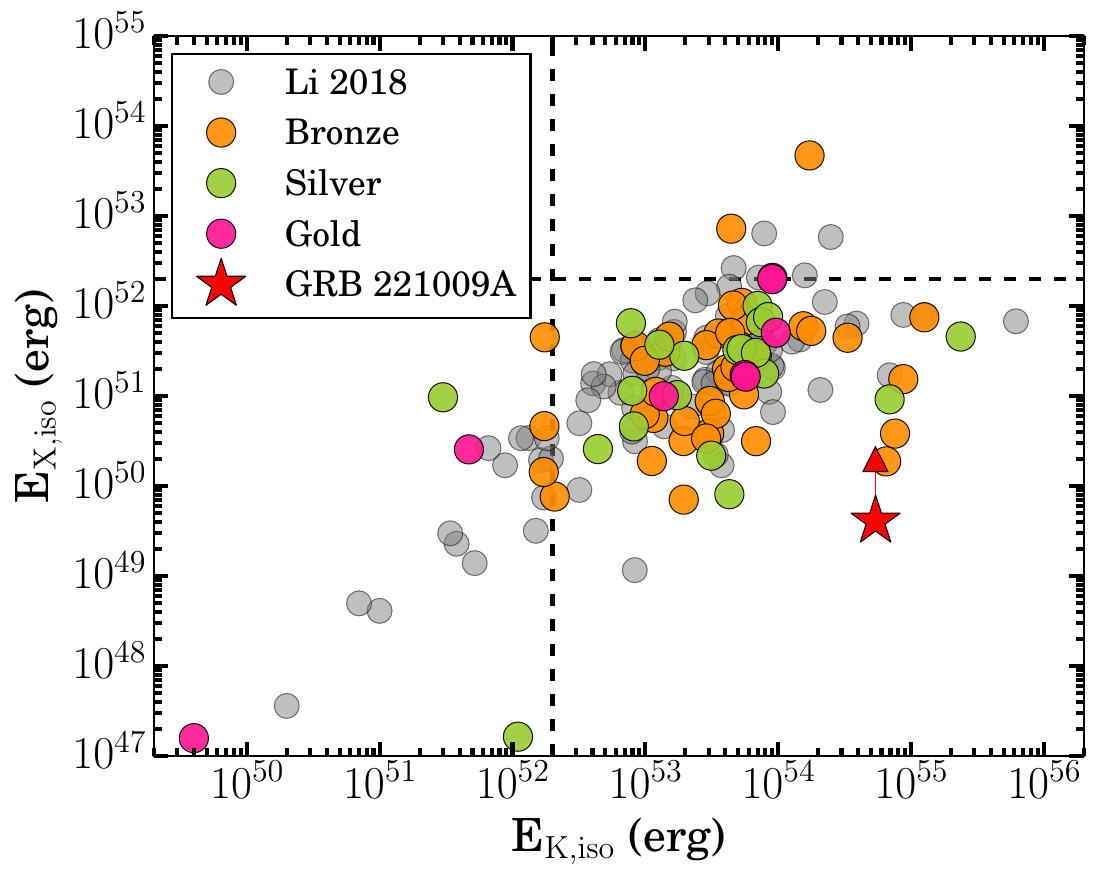}
\includegraphics[scale=0.45]{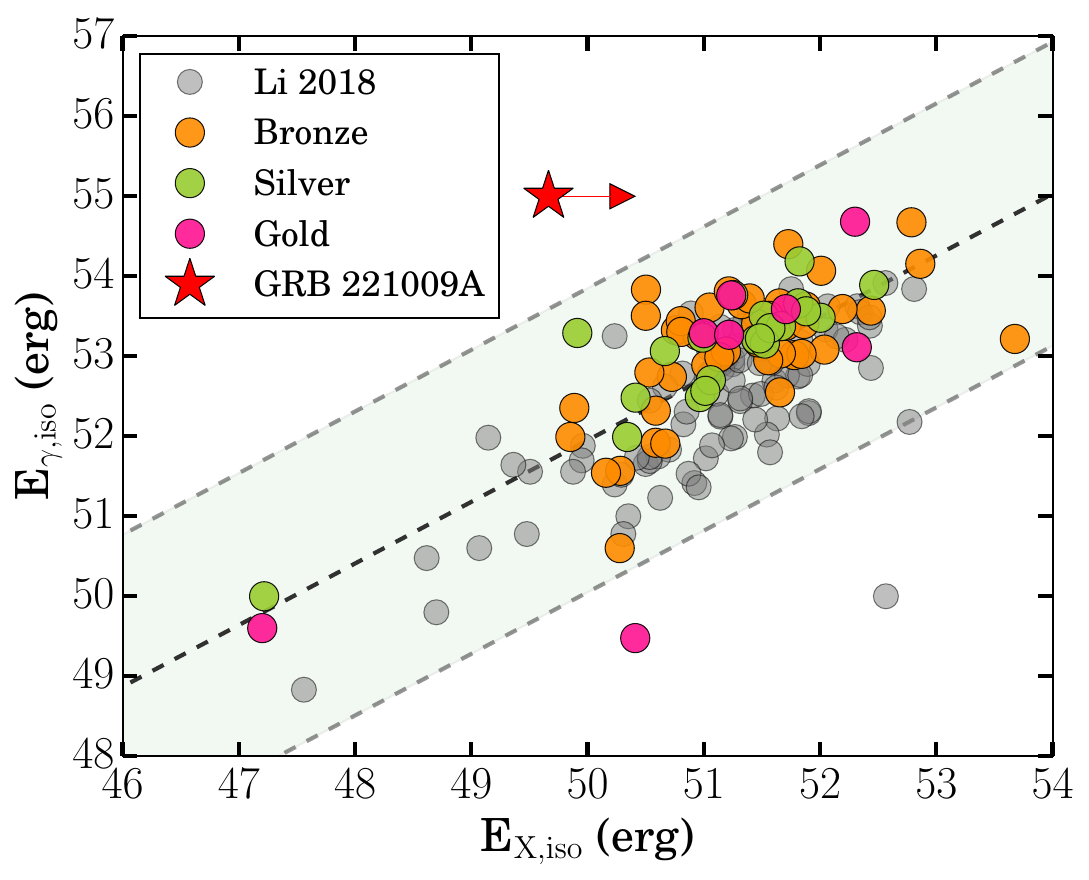}
\caption{Upper panel: Distribution of E$_{\rm X, iso}$ vs. E$_{\rm K,iso}$ calculated for GRBs with plateau in the XRT light curve in our Bronze, Silver, and Gold sub-samples. Similarly, gray circles represent the GRBs taken from \cite{2018ApJS..236...26L}. Black dashed lines at 2 $\times$ 10$^{52}$ erg represent the maximum energy budget of a magnetar central engine. Lower Panel: Distribution of E$_{\rm X,iso}$ vs. E$_{\rm \gamma, iso}$, the dashed lines represent the power-law fitted to the distribution along with 1-$\sigma$ uncertainty. GRB 221009A, as shown with a red star, does not have a plateau. The obtained lower limit of E$_{\rm X,iso}$ is calculated from the first data point of \swift-XRT observation.}
\label{fig:central_engine}
\end{figure}

The distribution of E$_{\rm X,iso}$ as a function of E$_{\rm K,iso}$, and E$_{\rm \gamma,iso}$ for our sample are shown in the upper and lower panels of Figure \ref{fig:central_engine}, respectively. The distribution of E$_{\rm K,iso}$ vs. E$_{\rm X,iso}$ shows that three GRBs (GRB 110213A, GRB 121027A, and GRB 140206A) have both E$_{\rm K,iso}$ $>$ 2 $\times$ 10$^{52}\,{\rm erg}$ and E$_{\rm X,iso}$ $>$ 2 $\times$ 10$^{52}\,{\rm erg}$, supporting the black hole as a possible central engine for these GRBs \citep{2018ApJS..236...26L}. Additionally, GRB 060218A, GRB 100316D, GRB 110808A, GRB 150915A, GRB 161108A, GRB 171205A, and GRB 180329B have both E$_{\rm K,iso}$ $<$ 2 $\times$ 10$^{52}\,{\rm erg}$ and E$_{\rm X,iso}$ $<$ 2 $\times$ 10$^{52}\,{\rm erg}$, for these GRBs a magnetar central engine is preferred. For the rest of the bursts of our sample, E$_{\rm K,iso}$ $>$ 2 $\times$ 10$^{52}\,{\rm erg}$ and E$_{\rm X,iso}$ $<$ 2 $\times$ 10$^{52}\,{\rm erg}$, a black hole central engine is poorly constrained \citep{2018ApJS..236...26L}. GRB 221009A does not have a plateau in the observed XRT lightcurve; considering the possibility of an early plateau, we calculated the lower limit of E$_{\rm X,iso}$ utilizing the first XRT data point (see Figure \ref{fig:central_engine}). The calculated lower limit also favors a black hole central engine for GRB 221009A. Further, we have also studied the distribution of E$_{\rm X,iso}$ as a function of E$_{\rm \gamma,iso}$ for our sample (see the lower panel of Figure \ref{fig:central_engine}). We calculated E$_{\rm X,iso}$ only during the plateau phase instead of the complete duration of X-ray afterglow and found a positive correlation \citep{2023ApJ...949L...4L}. We calculated the Pearson correlation and found a positive correlation with r = 0.73 and a p-value of $<$ 10$^{-4}$. We found a linear relation as log(E$_{\rm \gamma,iso}$) = (0.77 $\pm$ 0.05) $\times$ E$_{\rm X,iso}$ + 13.45 $\pm$ 2.08. We have also shown the data points given in \cite{2018ApJS..236...26L} along with our sample. 

\subsection{Environment of ULGRBs} 
\label{Ambient_medium}

\begin{figure}
\centering
\includegraphics[scale=0.40]{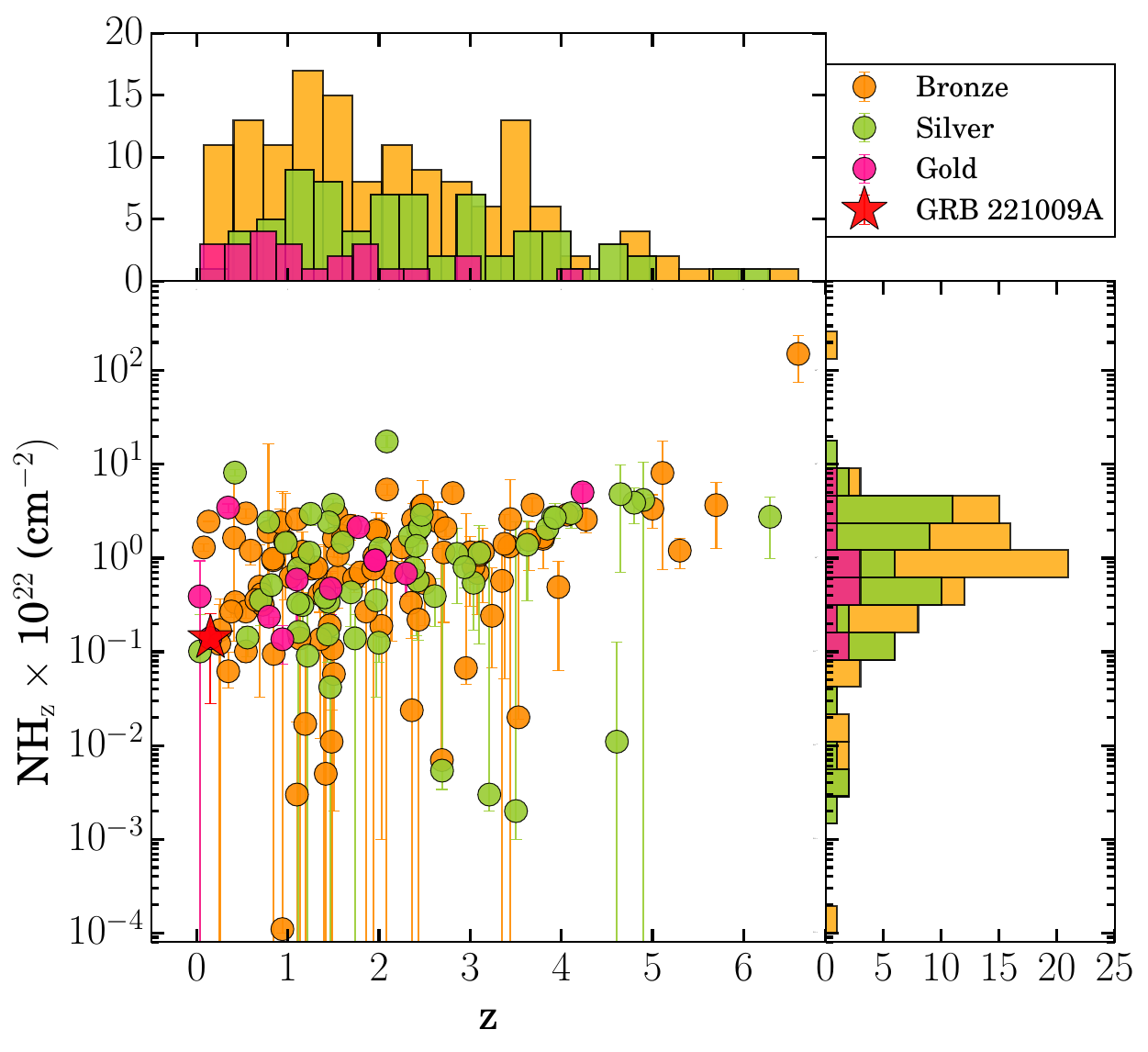}
\includegraphics[scale=0.50]{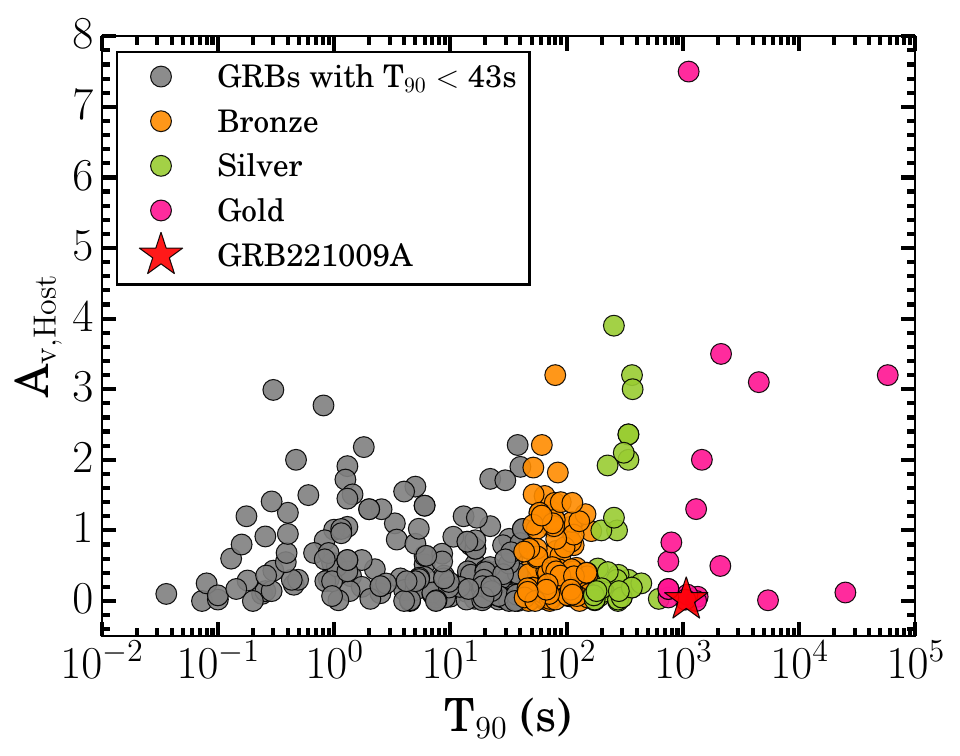}
\caption{Upper panel: represents the distribution of intrinsic X-ray column density (NH$_{\rm z}$) for our Bronze, Silver, and Gold sub-sample as a function of redshift. Lower Panel: Displaying the host extinction in the V-band (A$_{V, \text{host}}$) at the redshifts of the Bronze (orange), Silver (green), and Gold/ULGRBs (red) sub-samples, alongside additional GRBs (gray) documented in \cite{2010ApJ...720.1513K, 2010MNRAS.401.2773S, 2013ApJ...778..128P, 2014ApJS..213...15W, 2017MNRAS.467.1795L, 2018MNRAS.479.1542Z, 2022ApJ...940...57N, 2022ApJ...940...53S, 2023arXiv230207761L}.}
\label{fig:nhz}
\end{figure}

\cite{2014MNRAS.444..250E} suggested that ULGRBs may be distinguished from LGRBs due to their unique circumburst environment instead of different progenitor systems. The authors also proposed that ULGRBs occur in environments with extremely low densities, which can cause their ejecta to decelerate more slowly than they would in denser environments. To study the environmental properties of our sample, we used the intrinsic X-ray absorbing column density (NH$_{\rm z}$) as a parameter to estimate the amount of absorbing material along the line of sight, utilizing \swift-XRT data. The intrinsic column densities around a burst were calculated by fitting the X-ray afterglow spectra using \sw{XSPEC} software \citep{1996ASPC..101...17A}. We obtained our results by selecting a spectrum corresponding to the late time of the XRT light curve. This is particularly important because any early variation in the X-ray spectrum, such as a steep decay phase or flare, would reverberate the column density value and produce biased values \citep{2020MNRAS.495.2342D}. Each spectrum is then fitted by a power-law including the absorption components \sw{zphabs} and \sw{phabs}, respectively, due to the Galactic (NH$_{\rm Gal}$, fixed) and intrinsic host (NH$_{\rm z}$) at the redshift of the GRBs. The NH$_{\rm z}$ distribution for our sub-samples as a function of redshift is shown in Figure \ref{fig:nhz}. The mean values of NH$_{\rm z}$ for Bronze, Silver, and Gold sub-samples are 2.90 $\times$ 10$^{22}$, 1.83 $\times$ 10$^{22}$, and 1.25 $\times$ 10$^{22}$ cm$^{-2}$, respectively. Further, we check if there is any dependence of NH$_{\rm z}$ on redshift. There is an increasing trend of NH$_{\rm z}$ with redshift, as observed in the previous works \citep{2010MNRAS.402.2429C}. Although obtained from different methods, NH$_{\rm z}$ evolution with redshift in figure (1) of \citep{2019MNRAS.483.5380T} is nearly flat. Similarly, we conducted a comparative analysis of the optical host extinction (A$_{V, \text{host}}$) at the locations of GRBs within our Gold, Silver, and Bronze sub-samples along with GRBs given in \cite{2010ApJ...720.1513K, 2010MNRAS.401.2773S, 2011A&A...525A.109D, 2014ApJS..213...15W, 2017MNRAS.467.1795L, 2018MNRAS.479.1542Z, 2022ApJ...940...57N, 2022ApJ...940...53S}. The lower panel of Figure \ref{fig:nhz} illustrates that GRBs in SGRBs, LGRBs, Bronze, Silver, and Gold sub-samples display extinction characteristics consistent with each other. Four GRBs in the Gold sub-sample show a high value of A$_{V, \text{host}}$, possibly due to the dark nature of these bursts \citep{2010MNRAS.401.2005X, 2010ApJ...717..223H, 2015MNRAS.446.4116V}. However, the limited number of GRBs with measured host extinction properties in our sample poses a challenge in drawing definitive conclusions.

\section{Constraints on the ULGRB Progenitors Using MESA Simulation}
\label{Progenitor}

Late-time ($>$ 1 day) afterglow observations of low-redshift LGRBs have revealed that these events are sometimes accompanied by a special type of broad-line SNe-Ic \citep{Galama1998, 2003Natur.423..847H, 2022NewA...9701889K}, indicating the collapsar origin of LGRBs \citep{1993ApJ...405..273W}. SNe-Ic show no H and He lines in their spectral signatures, indicating extensive mixing of elements or violent mass loss in their progenitor stars. Simulations have shown that rapidly rotating massive stars with enhanced mixing rates can undergo quasi-chemical evolution \citep{2005A&A...443..643Y, 2006A&A...460..199Y}. Enhanced mixing ensures most of the H and He take part in the combustion due to the transport of these elements from the envelope to the core. The remaining H in the envelope can be removed by rotation-driven wind \citep{2005A&A...443..643Y, 2006A&A...460..199Y, 2018ApJ...858..115A}, that leads to the WR star as the final stage, which could be the progenitors of LGRBs. The progenitor of ULGRBs requires an additional condition: the free fall time of the envelope must be enough to feed the jet for a longer time scale than for typical LGRBs \citep{2018ApJ...859...48P}. There is significant interest in studying the evolution of massive stars that match the characteristics of ULGRB progenitors. 

\begin{figure*}
\centering 
\includegraphics[scale=0.50]{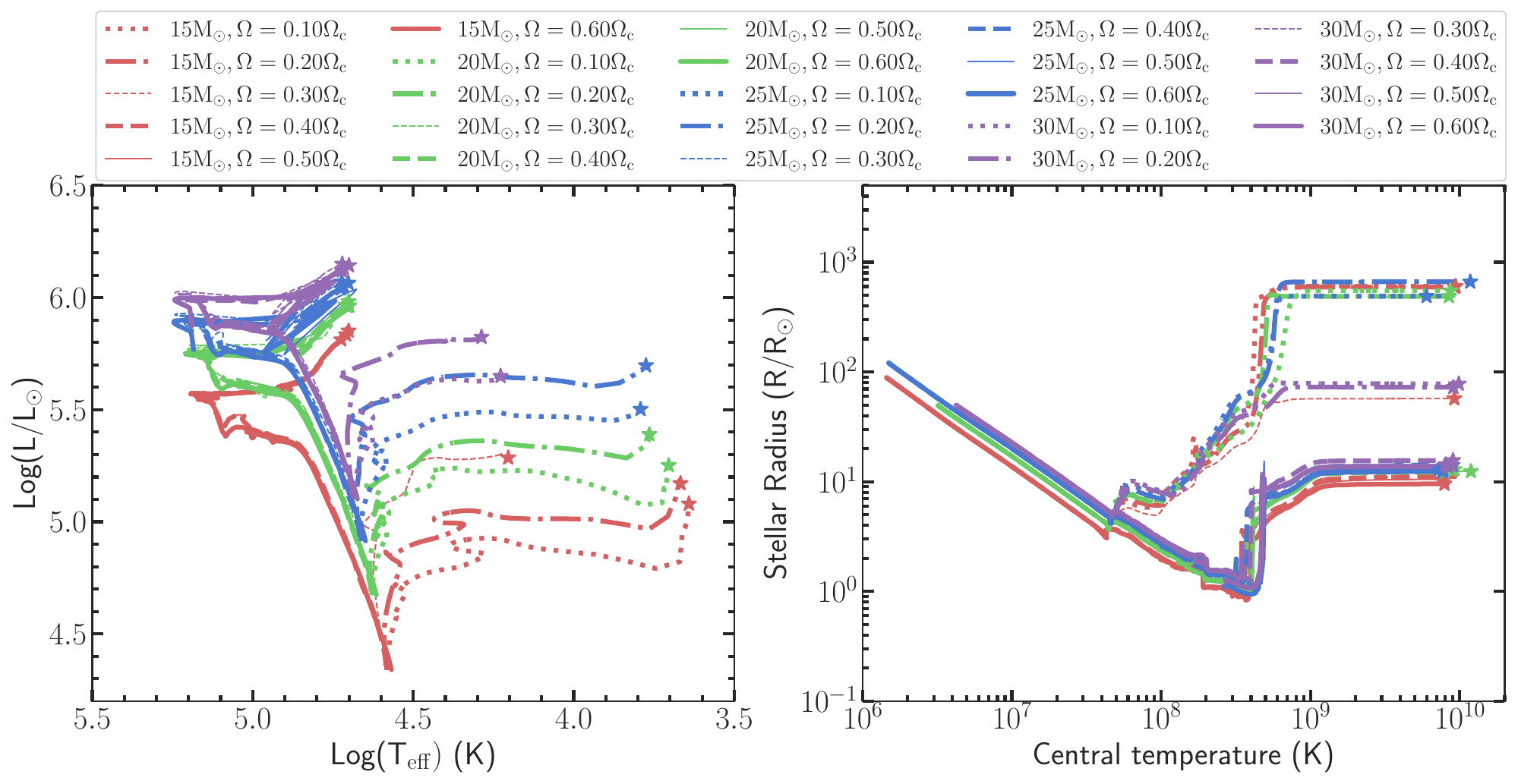}
\caption{Left: The evolution of our models having ZAMS masses of 15, 20, 25, and 30\,M$_{\odot}$ on the HR-diagram. For each model having a particular ZAMS mass, the initial rotation is varied in steps of 0.1\,$\Omega_{\rm c}$ up to 0.6\,$\Omega_{\rm c}$.  Right: The evolution of the stellar radius vs. central temperature curve of each model as the model progresses on the HR diagram. The stage of the onset of core collapse for each model is marked by $\star$.}
\label{fig:mesa}
\end{figure*}

The prolonged duration of ULGRBs in comparison to LGRBs suggests a distinct origin. In Section \ref{sec:collapsar}, our analysis rules out compact object mergers as the possible progenitor of ULGRBs and favors the collapsar scenario \citep{Galama1998, 2003Natur.423..847H}, similar to typical LGRBs (while certain LGRBs, such as GRB 211211A and GRB 230307A, have been identified as originating from compact object mergers). Moreover, the environment and prompt emission spectral properties of GRBs in our sample are consistent with those of typical LGRBs (see Figures \ref{fig:gamma_Ep_flu}, \ref{fig:nhz}, \& \ref{fig:prompt_param_ULGRBs}), implying similarities in jet composition and emission processes associated with ULGRBs and LGRBs, respectively. Therefore, it becomes important to distinguish the type of collapsing massive stars that can fuel the central engine 100-1000 times longer than collapsing typical WR stars for LGRBs \citep{1993ApJ...405..273W}.  It is assumed that due to the larger radius of BSG stars than WR stars, freely falling outer envelopes take longer time and, in turn, provide accretion for a longer duration to keep the central engine active to produce ULGRBs \citep{2018ApJ...859...48P}. Larger radii significantly increase the free fall time of the accreting material to the central engine or the active time for the central engine (T$_{\rm Engine}$). Consequently, \tninty = T$_{\rm Engine}$- t$_{\rm b}$ for the massive stellar object (15-30 M$_{\odot}$) can sufficiently account for the observed duration of ULGRBs. In addition, SN 2011kl associated with ULGRB 111209A differs from typical type Ic SNe \citep{2015Natur.523..189G} although the spectrum lacks H and He, it shows very little metal abundance. However, the missing H/He in the spectra of SNe associated with ULGRBs can also be due to the ionization of the ejecta due to the high energy emission from the central engine \citep{Ioka_2016}. 

Our analysis in Section 4.2 indicates that upon increasing mass within the selected mass range (15-30\,M$_{\odot}$), there is marginal change in the jet bore time t$_{\rm b}$ if the final collapsing star is a WR or BSG. This implies the GRB jet can bore through the WR and BSG stars. However, if the collapsing star is in RSG phase, the radius of the star could be several 100 to 1000\,R$_{\odot}$; in such a case, jet bore time could be very high ($\gtrsim$ 100s), and the emergence of the jet can not be possible through these stars. Such effect of larger pre-collapse radii of RSGs have been also observed by \citet[][]{2018ApJ...859...48P}. These results indicate that the WR and BSG stars could be the progenitors of long GRBs and the equation for the estimation of t$_{\rm b}$ is properly useful only for WR and BSG progenitors.

\subsection{Evolution of massive star with \sw{MESA}} 
\label{sec:mesa}

After studying the detailed prompt and afterglow properties of a number of ULGRBs in previous sections, we have performed the simulation of massive stars with different initial masses and rotations utilizing the state-of-the-art tool, \sw{MESA}, to refine our understanding of their progenitor. In this subsection, we provide the details of the 1D stellar evolution of possible progenitor models using \sw{MESA}. Beginning from their pre-main-sequence (PMS) stages, the models evolve up to the stage of the onset of core collapse. Considering various characteristics of the possible progenitors outlined in the previous studies \citep{2018ApJ...859...48P, 2018ApJ...858..115A, 2023arXiv230105401S}, we have chosen the initial conditions to simulate the evolution of massive stars. We obtain the final physical properties, including the radius, surface temperature, and luminosity of the collapsing stars, from \sw{MESA} as they enter the core collapse phase. These parameters are then used to constrain the free fall time of the collapsing star models. Further, we compare the derived free fall time with the observed T$_{90}$ duration of both LGRBs and ULGRBs, aiding in the understanding of the physical characteristics of stars capable of producing such GRBs.

In our study, to simulate the evolution of massive stars starting from the PMS until they reach the stage of the onset of core collapse, we employ \sw{MESA} version 23.05.1. Our primary objective is to utilize the final parameters of these massive stars at the stage of the onset of core collapse to estimate whether they can allow the formation and successful penetration of jets from the surrounding envelope to produce a GRB. Finally, we estimate the free fall time (t$_{\rm ff}$) to gain insights into how long the central engine can be fueled, which helps us to distinguish between LGRBs and ULGRBs \citep{2018ApJ...859...48P}. The variety of \sw{MESA} parameters in this study to evolve our models up to the stage on the onset of core collapse closely follow the \sw{MESA} settings of \citet[][]{2021MNRAS.505.2530A} and \citet[][]{2022MNRAS.517.1750A}. However, we discuss a few changes ahead. The stellar models in our study have Zero Age Main Sequence (ZAMS) masses of 15, 20, 25, and 30\,M$_{\odot}$. Starting from initial angular rotational velocity ($\Omega$) of 0.1\,$\Omega_{\rm c}$, the $\Omega$ for each model is varied up to 0.6\,$\Omega_{\rm c}$, where $\Omega_{\rm c}$ is the critical angular rotational velocity and is expressed as $\Omega_{\rm c}^2 = (1-L/L_{\rm edd})GM/R^3$, with $L_{\rm edd}$ representing the Eddington luminosity.
Further, for each model, we employ a metallicity (Z) of 2 $\times$ 10$^{-4}$, which is favored by host galaxy observations of LGRBs \citep{2003A&A...400..499L, 2011MNRAS.414.1263M, 2022JApA...43...82G}. 

We adopt the Ledoux criterion and model the convention utilizing the mixing length theory of \citet[][]{1965ApJ...142..841H} by fixing the mixing length parameter ($\alpha_{\rm MLT}$) to 2.0. The semiconvection coefficient $\alpha_{\rm sc}$ is fixed to 0.01 to introduce the effect of semiconvection by following \citet[][]{1985A&A...145..179L}. The thermohaline mixing in our models is modeled following \citet[][]{1980A&A....91..175K}. Incorporating the default \sw{MESA} settings for massive star evolution, the corresponding efficiency parameter for thermohaline mixing ($\alpha_{\rm th}$) is set to 2.0 and 0 for the phases before and after the core-He exhaustion, respectively. Convective overshooting in our models is modeled using the scheme mentioned by \citet[][]{2000A&A...360..952H}. The overshoot mixing parameters are fixed at f$_{\rm ov}$ = 0.005 and f$_{\rm 0}$ = 0.001. The choice of these values of f$_{\rm ov}$ and f$_{\rm 0}$ closely follow the settings of \citet[][]{2016ApJS..227...22F} and \citet[][]{2023MNRAS.521L..17A}. To incorporate the effects of wind, the `Dutch' wind scheme is employed with a wind scaling factor ($\eta_{\rm wind}$) of 0.5. The choices of these parameters are also primarily followed from prior studies, such as those given in \citet{2018ApJ...859...48P}, \citet[][]{2018ApJ...858..115A}, and \citet[][]{2023arXiv230105401S}. We have summarized a few of the initial parameters in Table \ref{tab:mesa_table}.\\

With the above-mentioned \sw{MESA} settings and initial parameters, we evolve all the models from PMS up to the stage of the onset of core collapse. The arrival of a model on ZAMS is marked at a stage where the ratio of the luminosity from nuclear reactions and the overall luminosity of the model becomes 0.4. Further, the beginning of the core collapse of the model is marked when the infall velocity of its Fe-core exceeds a limit of 1000\,km\,s$^{-1}$. The left panel of Figure \ref{fig:mesa} illustrates the evolutionary trajectory of the models in the current study on the HR diagram. Owing to the low initial metallicity, rotation, and a moderate wind scaling factor ($\eta_{\rm wind}$=0.5), most of the models terminate their evolution towards the relatively hotter end on the HR diagram, except the 15, 20 and 25\,M$_{\odot}$ models having angular rotational velocity $\leq$0.2\,$\Omega_{\rm c}$. These slowly rotating models with ZAMS mass (M$_{\rm ZAMS}$) of 15, 20, and 25\,M$_{\odot}$, end up their evolutions towards the cooler side of the HR diagram. These models also possess large final radii (R$_{\rm final}$) at their terminating stages, as indicated in the Right panel of Figure \ref{fig:mesa}. As listed in Table \ref{tab:mesa_table}, the final radii of these slowly rotating models exceed several 100\,R$_{\odot}$; thus indicating they terminate their evolution as massive RSGs. All the models in our simulations exceed the final radii of 10$^{11}$\,cm, which is a consistent result for ULGRB progenitors \citep[][]{2022MNRAS.510.4962G}. However, models terminating their evolution as RSGs cease to serve as the progenitors for the GRBs/ULGRBs since their enormous final radii (t$_{\rm b}$ $>$ 100s, Table \ref{tab:mesa_table}) do not allow successful penetration of the jet. Thus, the slowly rotating models with M$_{\rm ZAMS}$ of 15, 20, and 25\,M$_{\odot}$ are discarded as the progenitors of ULGRBs.

\begin{figure*}
    \centering
    \includegraphics[height=8cm,width=0.45\textwidth,angle=0]{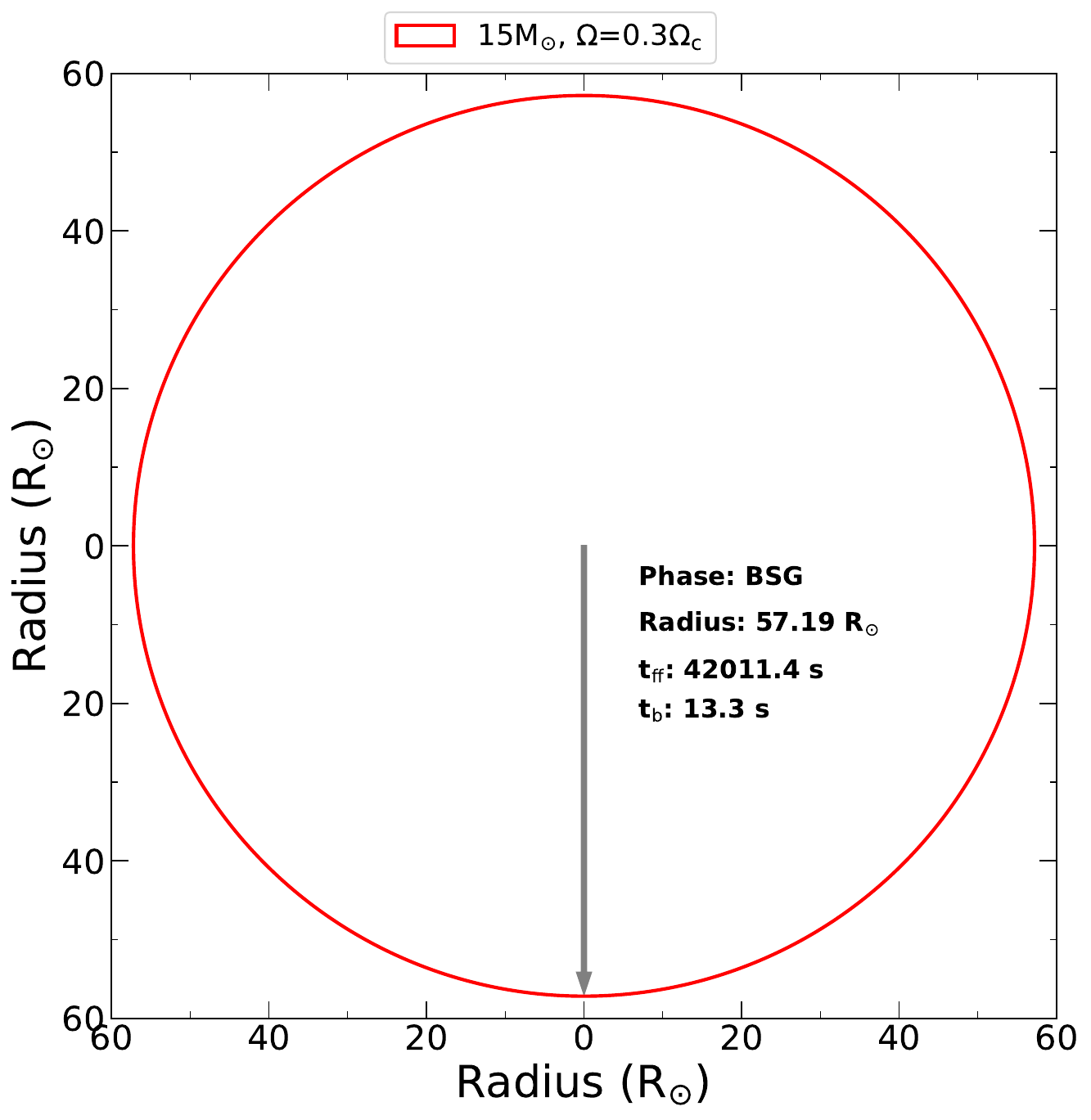}
    \includegraphics[height=8cm,width=0.45\textwidth,angle=0]{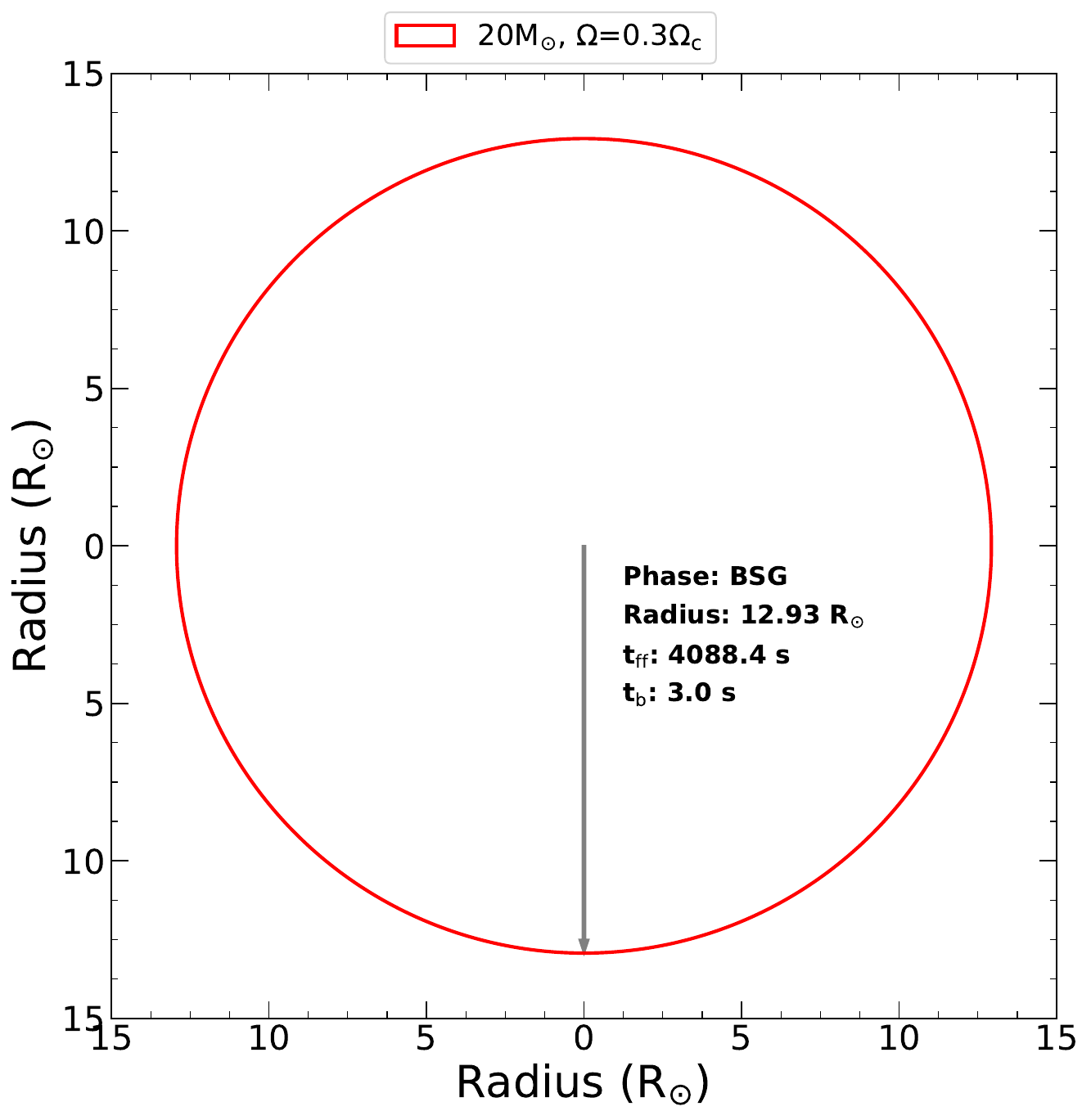}
    \includegraphics[height=8cm,width=0.45\textwidth,angle=0]{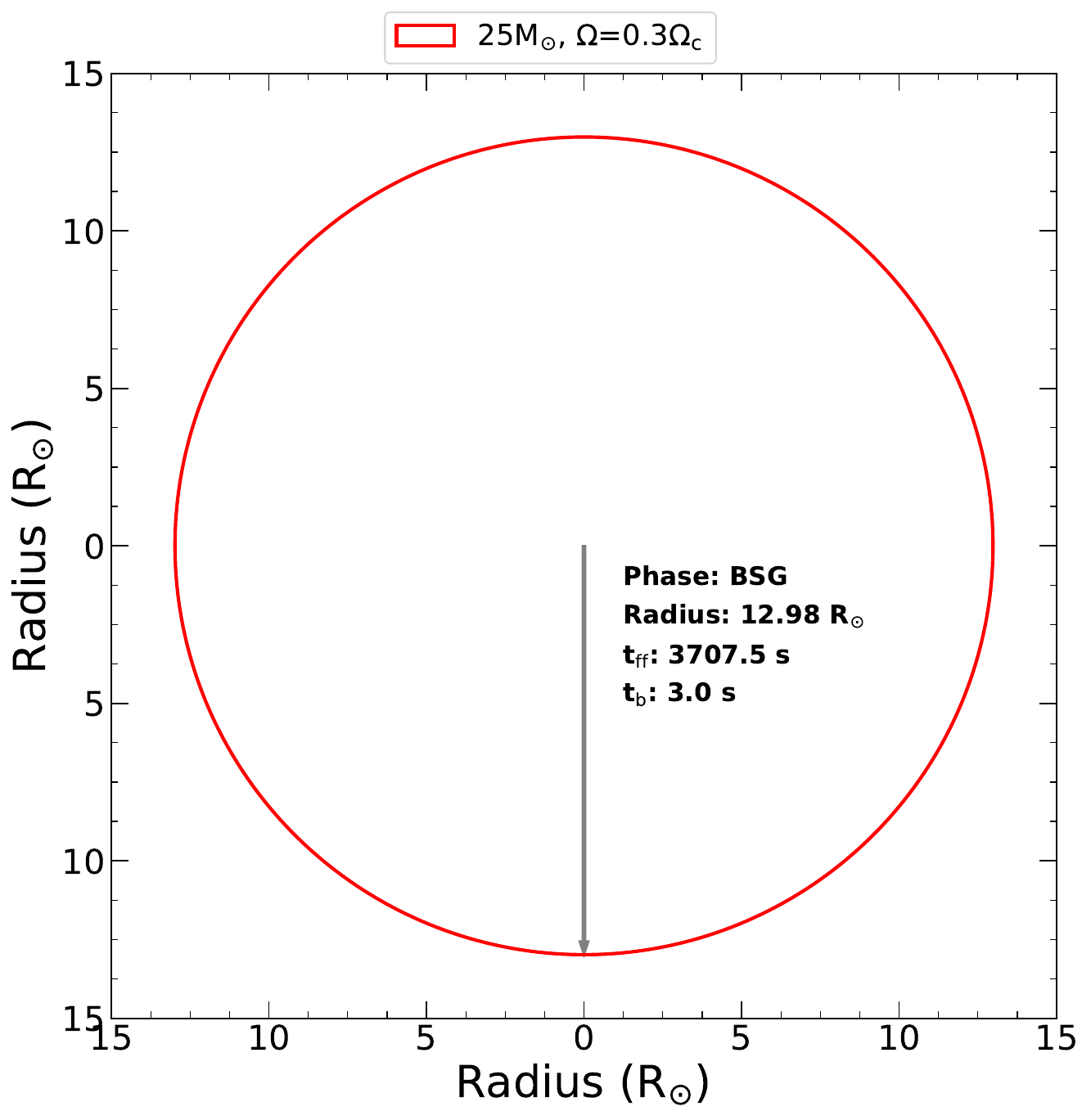}
    \includegraphics[height=8cm,width=0.45\textwidth,angle=0]{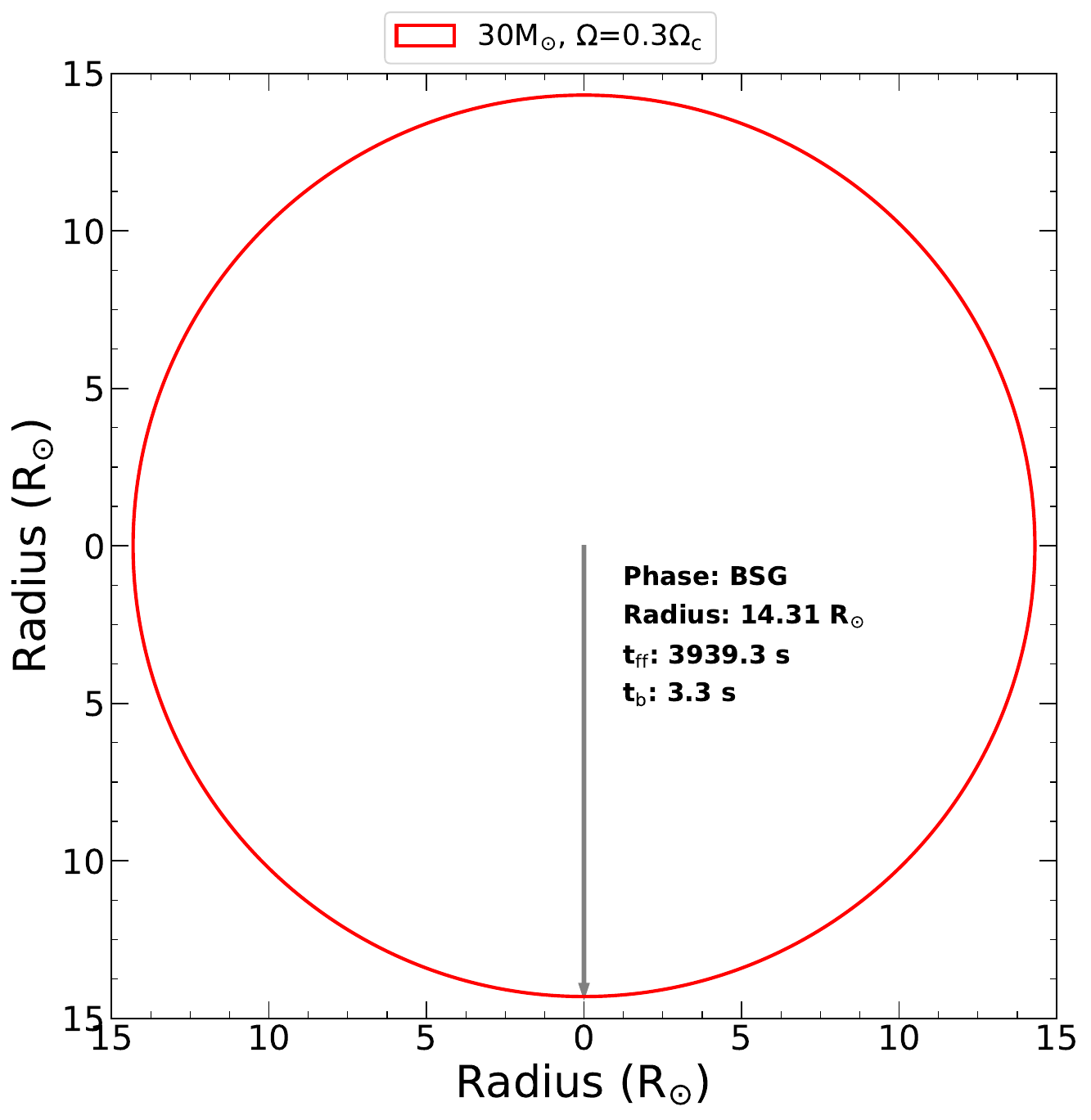}
   \caption {The radii of four models from the set of entire models, having M$_{\rm ZAMS}$ of 15, 20, 25, and 30\,M$_{\odot}$ respectively at the stage of the onset of core collapse. The t$_{\rm ff}$ and t$_{\rm b}$ for each model are also indicated. We have assumed a $\Gamma$ = 10 for each model while estimating t$_{\rm b}$.}
    \label{fig:radii_at_cc}
\end{figure*}

Utilizing the simulation parameters of the models at the stage of the onset of core collapse, we estimate the free fall timescales (t$_{\rm ff}$) by employing Equation 1 of \citet[][]{2018ApJ...859...48P}. The t$_{\rm ff}$ for each model is listed in Table~\ref{tab:mesa_table}. The estimation of t$_{\rm ff}$ is important to gain insights into how long the central engine can be fueled; thus, it can be compared with the T$_{90}$ duration of the GRBs. In a recent work by \citet[][]{2023ApJ...957...31S}, the authors propose a two-stage model for GRB 221009A and associate the precursor pulse with the weak jet arising due to the collapsed core. Thus, we estimate the bore-time (t$_{\rm b}$) of the weak jet for each of our models to get insight into the precursor pulse. We estimate t$_{\rm b}$ using a simple equation:
\begin{equation}
{\rm t_{\rm b}} = \frac{\rm R_{\rm final}}{({\rm u}\Gamma)}
\label{eq:3}
\end{equation}
In the above equation, u is the weak jet velocity corresponding to a Lorentz factor of $\Gamma$. The factor $\Gamma$ is divided in the denominator to account for the relativistic length contraction. While calculating the t$_{\rm b}$ using Equation~\ref{eq:3} above, we make a very simplified assumption that the weak jet moves with a constant $\Gamma$ of 10. The choice of $\Gamma$ = 10 is motivated from \citet[][]{2023ApJ...957...31S}, where the authors mention that at the time of break out, the $\Gamma$ is of the order of 10. With these assumptions, the estimated t$_{\rm b}$ for each model is listed in Table~\ref{tab:mesa_table}. Within the range of employed \sw{MESA} settings and initial parameters, in Figure~\ref{fig:radii_at_cc}, we have depicted the final stages of four models from the set of entire models, having M$_{\rm ZAMS}$ of 15, 20, 25, and 30\,M$_{\odot}$ and each one of them having $\Omega = 0.3\,\Omega_{\rm c}$. The final radius, t$_{\rm ff}$, and t$_{\rm b}$ are also indicated in the figure.

The above calculations are made for jet bore time (t$_{b}$) considering that the Lorentz factor within the envelope remains almost constant to its initial value ($\Gamma_{0}$ =10), independent of the mass and radius of the stars. However, in reality, the Lorentz factor of a fireball depends on the distance from the center of the star. Initially, during the expansion of the shell, i.e coasting phase, the Lorentz factor of the fireball remains almost constant to the initial value $\Gamma_{0}$ \citep{2019ApJ...883..187L}. After some time, known as deceleration time t$_{dec}$, the fireball enters in the self-similar expansion \citep{1976PhFl...19.1130B}, and the Lorentz factor of the jet evolves with the radius of star by the relation $\Gamma$ $\propto$ R$^{-3/2}$ (for T$_{90}$ $<$ t$_{dec}$) or R$^{-1/2}$ (for T$_{90}$ $>$ t$_{dec}$) \citep{2006RPPh...69.2259M}. This indicates that as the radius increases, the Lorentz factor of the fireball decreases, that eventually increases the corresponding t$_{\rm b}$. Thus, the quoted values of t$_{\rm b}$ in our work are obviously the lower limits of bore-time for the underlying Jets.

Now, we compare the t$_{\rm ff}$ estimated from our simulation parameters with a few actual T$_{\rm 90}$ duration of our Gold sample of ULGRBs mentioned in Table~\ref{tab:collapsar}. The model with M$_{\rm ZAMS}$ = 30\,M$_{\odot}$ and $\Omega = 0.4\,\Omega_{\rm c}$ has a t$_{\rm ff}$ of $\sim$ 4540\,s, which is close to the T$_{\rm 90}$ duration of GRB 070419B; the model with M$_{\rm ZAMS}$ = 30\,M$_{\odot}$ and $\Omega = 0.2\,\Omega_{\rm c}$ has a t$_{\rm ff}$ of $\sim$ 42500\,s, which is close to the T$_{\rm 90}$ duration of GRB 090404. Moreover, the t$_{\rm ff}$ obtained from our considered models are of similar order when compared to the actual T$_{\rm 90}$ durations of our Gold sample of ULGRBs.

\section{Summary and Conclusion}

The underlying physical mechanism, possible progenitor, and central engine of ULGRBs are still unclear. Previous findings have shown that ULGRBs, despite their exceptionally longer prompt emission duration, exhibit prompt and afterglow spectra, surrounding environment, and host properties similar to LGRBs. This paper aims to constrain the possible progenitor and central engine of GRB 221009A and other similar bursts exhibiting ULGRB characteristics based on their observed \tninty durations. In this context, we present a comprehensive search for ULGRB candidates using \swift detected GRBs (the most updated and complete sample). Specifically, we focus on GRBs with \tninty durations exceeding the mean value derived from a Gaussian distribution of \swift detected LGRBs with redshift measurements. Our sample incorporates a total of $\sim$ 230 GRBs. The selected GRBs are subsequently categorized into Bronze, Silver, and Gold sub-samples based on their \tninty duration, which lies in the 1, 2, and 3 $\sigma$ confidence levels of the distribution. For sample completeness, we included known cases of ULGRBs as the diamond sample. After sample selection, we performed the detailed prompt and afterglow analyses of GRB 221009A and the GRBs listed in our different sub-samples utilizing space-based data from \swift and \fermi satellites.

The prompt temporal and spectral examination of GRB 221009A revealed an ultra-long nature with a precursor activity, and for this precursor emission, the \sw{Band} function best describes the spectra while incorporating the thermal component into the \sw{Band} function only slightly improves the spectral fitting. Nevertheless, the evolution of $\alpha_{\rm pt}$ and \Ep during the precursor and main pulse of GRB 221009A suggests a potential synchrotron origin for the prompt emission \citep{2023ApJ...957...31S}. It is noteworthy that both $\alpha_{\rm pt}$ and \Ep exhibit flux-tracking evolution.

The distribution of spectral parameters obtained from the prompt and afterglow emission analysis of the Bronze, Silver, and Gold sub-samples is consistent with the broad sample of GRBs in the background (see Figures \ref{fig:gamma_Ep_flu} and \ref{fig:prompt_param_ULGRBs}). In the HR-\tninty space as plotted in Figure \ref{fig:HR_T90}, SGRBs are harder than LGRBs (a well-known feature of two classical families of GRBs), and the Bronze and Silver sub-samples are consistent with LGRBs, implying they might represent similar kinds of bursts as expected. However, GRBs in our Gold sub-sample show an overall soft spectral characteristic (including GRB 221009A). Again, in the fluence -\tninty space, GRBs included in SGRBs, LGRBs, Bronze, and Silver sub-samples are showing an increasing trend while the GRBs in Gold sub-samples are deviating from this trend. This might hint that the Gold sub-sample consists of GRBs with soft spectral characteristics and are relatively fainter than the other bursts, making them potential candidates for a new class of so-called ULGRBs. Furthermore, to constrain the origin of GRB 221009A and the GRBs in our Bronze, Silver, and Gold sub-samples, we have utilized the following methods:

First, we conducted a comparative analysis of the NIR light curve of GRB 221009A (including our observations taken using 3.6m DOT telescope) with that of other GRBs associated with supernovae. The NIR light curve of GRB 221009A exhibited a smooth decay, distinguishing it from other SN-detected GRBs where distinct bumps and flattening were observed. Our late-time near-IR observations obtained with 3.6m DOT and publicly available data rule out the presence of any prominent supernova associated with GRB 221009A. Subsequently, utilizing prompt and afterglow analyses, we attempted to constrain the progenitor of GRBs (collapsar or merger) within our sub-samples. GRB emission is accompanied by an ultra-relativistic jet that must bore through the pre-existing envelope surrounding the progenitor star. For a GRB to have a collapsar origin, the central engine powering the burst must remain active for a period longer than the jet bore-time. In other words, \tninty must be greater than t$_{\rm b}$. We first constrained the jet opening angle by utilizing the jet break time observed in the X-ray afterglow light curve, the isotropic energy release during the prompt emission, and other observed properties. Then using equation (\ref{eqn:collapsar}), we calculated the t$_{\rm b}$. The obtained values of t$_{\rm b}$ found much less than \tninty lead to the collapsar origin of all the bursts, including GRB 221009A. To further strengthen these results, we calculated the probabilities of non-collapsar origin for all GRBs included in our sample. The negligible values of non-collapsar probability again confirmed their collapsar origin. To further confirm our analysis results, we simulated the evolution of the low metallicity massive star having M$_{\rm ZAMS}$ of 15, 20, 25, and 30\,M$_{\odot}$ and different initial rotations utilizing \sw{MESA}. The bore-time obtained from the simulation closely matches our analysis results in Section \ref{sec:collapsar}. Subsequently, utilizing the simulation parameters of our models when they entered the core collapse phase, we estimated the free-fall time (t$_{\rm ff}$). Notably, a significantly extended final radius and t$_{\rm ff}$ observed in slowly rotating stars ($\Omega \leq 0.2,\Omega_{\rm c}$) that evolved to RSG contradict their potential to produce ultra-relativistic jets and their penetration through the surrounding envelope. For moderately rotating stars ($\Omega \geq 0.2,\Omega_{\rm c}$), the t$_{\rm ff}$ obtained from our simulated models closely matches the actual T$_{\rm 90}$ of a few ULGRBs from our Gold sample. These findings suggest that rotating ($\Omega \geq 0.2\,\Omega_{\rm c}$) massive stars could potentially be the progenitors of ULGRBs within the considered parameters and initial inputs to \sw{MESA}.

To constrain the central engine associated with the GRBs in our Gold, Silver, and Bronze sub-samples, we have utilized the following methods: (1) For \fermi-GBM detected bursts (32 GRBs), we calculated the isotropic gamma-ray energy E$_{\rm \gamma, iso}$ and beaming corrected energy E$_{\rm \theta_j, \gamma, iso}$ $\sim$ $\theta_{\rm j}^{2}$/2 $\times$ E$_{\rm \gamma, iso}$. For GRBs with E$_{\rm \theta_j, \gamma, iso}$ $<$ 2 $\times$ 10$^{52}$ erg, a magnetar can be the possible central engine for these bursts. For E$_{\rm \theta_j, \gamma, iso}$ $>$ 2 $\times$ 10$^{52}$ erg, a magnetar central engine is not possible due to its maximum energy constrain, and black hole central engine is favored. In this case, only two GRBs, GRB 210619B and GRB 221009A, can not be explained by a magnetar and require a black hole engine with black hole masses $\sim$3.4M$_{\odot}$ and $\sim$9.1M$_{\odot}$ respectively. (2) For 74 GRBs (47 Bronze, 21 Silver, 6 Gold) with the plateau in the \swift-XRT light curve, we calculated the isotropic X-ray energy E$_{\rm X, iso}$ released during the plateau phase as well as the kinetic energy E$_{\rm K, iso}$ of the burst. A magnetar central engine is favored by GRBs with E$_{\rm X, iso}$ $<$ 2 $\times$ 10$^{52}$ erg and E$_{\rm K, iso}$ $<$ 2 $\times$ 10$^{52}$ erg. For GRBs with E$_{\rm X, iso}$ $>$ 2 $\times$ 10$^{52}$ erg and E$_{\rm K, iso}$ $>$ 2 $\times$ 10$^{52}$ erg a black hole engine is preferred. A black hole central engine is poorly constrained for the rest of the GRBs. In this case, a hyper-accreting black hole is constrained as a potential central engine candidate for our Gold samples, and only a few GRBs (GRB 060218, GRB 100316D, and GRB 091024A) favor a magnetar.

In summary, utilizing \tninty as prevailing criteria, we present a method to search for ULGRB candidates. The observed properties of GRB 221009A (the brightest burst ever observed) are also discussed in this context. Further, we shed light on the origin and central engines of ULGRBs and the population of LGRBs. To achieve this, we statistically examine a nearly complete sample of \swift-detected GRBs and categorize them into Bronze, Silver, and Gold sub-samples. The properties of GRBs in the Bronze sub-sample do not show any difference from the LGRB population. Our Gold sub-sample indicates a higher likelihood of belonging to the ULGRB category. We successfully constrain the collapsar origin for all GRBs in our sample. Specifically, we found a hyper-accreting black hole central engine for GRB 221009A, featuring a black hole mass of $\sim$9.1M$_{\odot}$. Similarly, most GRBs in our Gold sub-sample favor a black hole central engine, except for three GRBs (GRB 060218, GRB 100316D, and GRB 091024A). In addition, the distribution of NH$_{\rm z}$ and A$_{v, Host}$ do not favor any particular kind of low-density environment for the GRBs in the Gold sub-sample, as suggested by \citep{2014MNRAS.444..250E}. Moreover, the striking similarities obtained in the observed parameters and simulation results from \sw{MESA} provide additional support for the low metallicity and rotating ($\Omega \geq 0.2\,\Omega_{\rm c}$) massive stars as progenitors for ULGRBs. It is also cautioned that except T$_{90}$ duration, the present analysis did not find any other robust criteria to distinguish between LGRBs and ULGRBs, and there could be other potential observed parameters to demarcate between the two populations of GRBs. Instead, we proposed a method to compare the properties of the GRB sub-samples with the increasing likelihood of being ULGRB candidates and understanding the nature of their progenitors. The upcoming Space-based multi-band astronomical Variable Objects Monitor (SVOM) mission is expected to detect more ULGRBs (at higher redshift) and provide insight into unusually long emissions from these bursts \citep{2020ExA....50...91D}.

\section*{Acknowledgments}
RG and SBP acknowledge the financial support of ISRO under AstroSat archival Data utilization program (DS$\_$2B-13013(2)/1/2021-Sec.2). SBP also acknowledges support from DST grant no. DST/ICD/BRICS/Call-5/CoNMuTraMO/2023(G) for the present work. RG is thankful to Dr. Amy Lien for updating the Swift BAT GRB catalog page on request and fruitful suggestions on the results. RG is also grateful to Dr. M. J. Moss for the discussion. RG was sponsored by the National Aeronautics and Space Administration (NASA) through a contract with ORAU. The views and conclusions contained in this document are those of the authors and should not be interpreted as representing the official policies, either expressed or implied, of the National Aeronautics and Space Administration (NASA) or the U.S. Government. The U.S. Government is authorized to reproduce and distribute reprints for Government purposes notwithstanding any copyright notation herein. AA acknowledges funds and assistance provided by the Council of Scientific \& Industrial Research (CSIR), India, under file no. 09/948(0003)/2020-EMR-I. AA also acknowledges the Yushan Young Fellow Program by the Ministry of Education, Taiwan, for financial support. AJCT acknowledges support from the Spanish Ministry project PID2020-118491GB-I00 and Junta de Andalucia grant P20\_010168. This research has used data obtained through the HEASARC Online Service, provided by the NASA-GSFC, in support of NASA High Energy Astrophysics Programs. We extend sincere thanks to all the observing and support staff of the 3.6m DOT to maintain and run the observational facilities at Devasthal Nainital.

\facilities{3.6m DOT, \swift, \fermi}

\software{\sw{MESA} \citep{2011ApJS..192....3P, 2013ApJS..208....4P, 2015ApJS..220...15P, 2018ApJS..234...34P, 2023ApJS..265...15J}, GBM-Tool \citep{GbmDataTools}, 3ML \citep{2015arXiv150708343V}, XSPEC \citep{1996ASPC..101...17A}, DAOPHOT-II \citep{1987PASP...99..191S}, IRAF \citep{1986SPIE..627..733T, 1993ASPC...52..173T}, Matplotlib \citep{2007CSE.....9...90H}}. {\bf The \sw{MESA} inlists and final models can be found on Zenodo at \dataset[]{https://zenodo.org/records/11119956}}.

\FloatBarrier
\bibliography{ref}{}
\bibliographystyle{aasjournal}

\appendix

\restartappendixnumbering

\section{Multi wavelength observations and analysis of GRB 221009A}

GRB 221009A is the brightest burst observed to date, with E$_{\gamma, iso}$ = 1 $\times$ 10$^{55}$ erg and L$_{\gamma, iso}$ = 9.91 $\times$ 10$^{53}$ \citep{2023ApJ...952L..42L}. The \swift team initially reported this burst as a detection of a new bright Galactic transient \citep{2022GCN.32632....1D}. However, the source was identified as an extremely bright burst based on the strong fading nature of the X-ray counterpart and the simultaneous detection/localization by \fermi-GBM (at 13:16:59.000 UT on October 9, 2022, hereafter \fermiT) and LAT \citep{2022GCN.32635....1K}. Due to the delay (about an hour) in the confirmation of the nature of the source post-GBM/BAT trigger, nearly all the ground-based telescopes missed the early emission. However, soon after the \swift and \fermi discovery report of extremely bright GRB 221009A, several space and ground-based telescopes (including 3.6m DOT facility) started a rigorous follow-up campaign across the electromagnetic band. In this section, we present the detailed analyses of space (\swift and \fermi) and ground-based observations of GRB 221009A.

\subsection{Prompt emission: Temporal and Spectral analysis}


\begin{figure}[!ht]
\centering
\includegraphics[scale=0.45]{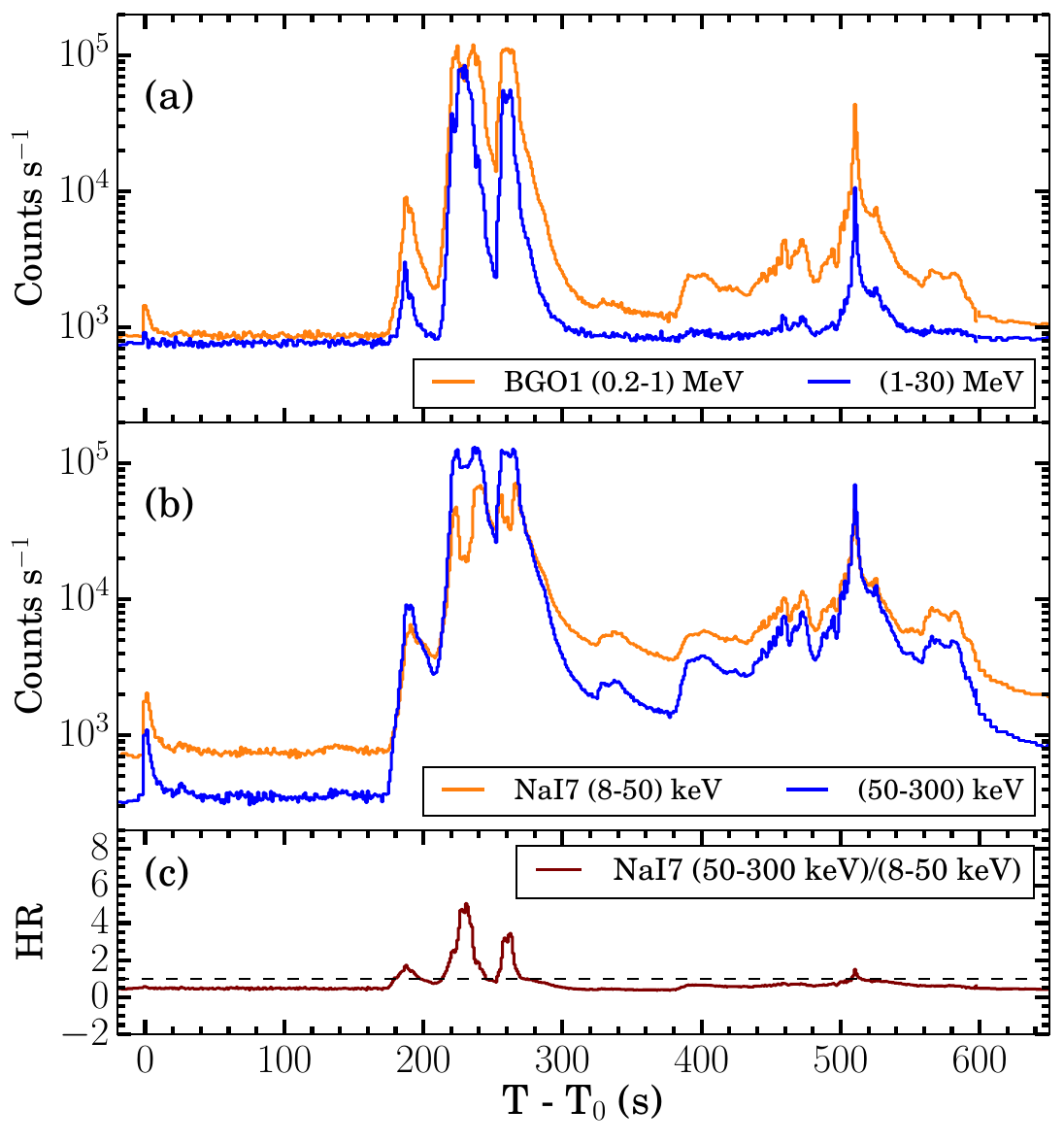} 
\caption{\fermi-GBM multi-channel light curve of GRB 221009A. Panels (a) and (b) display the count-rate light curve in the two energy channels of the BGO and NaI scintillation detectors, respectively. (c) depicts the hardness ratio in the two energy channels of the NaI scintillation detector. The specific energy channels used are indicated in the legends.}
\label{fig:multichannel_gbm}
\end{figure}

\begin{figure}
\centering
\includegraphics[height=6cm,width=0.44\textwidth]{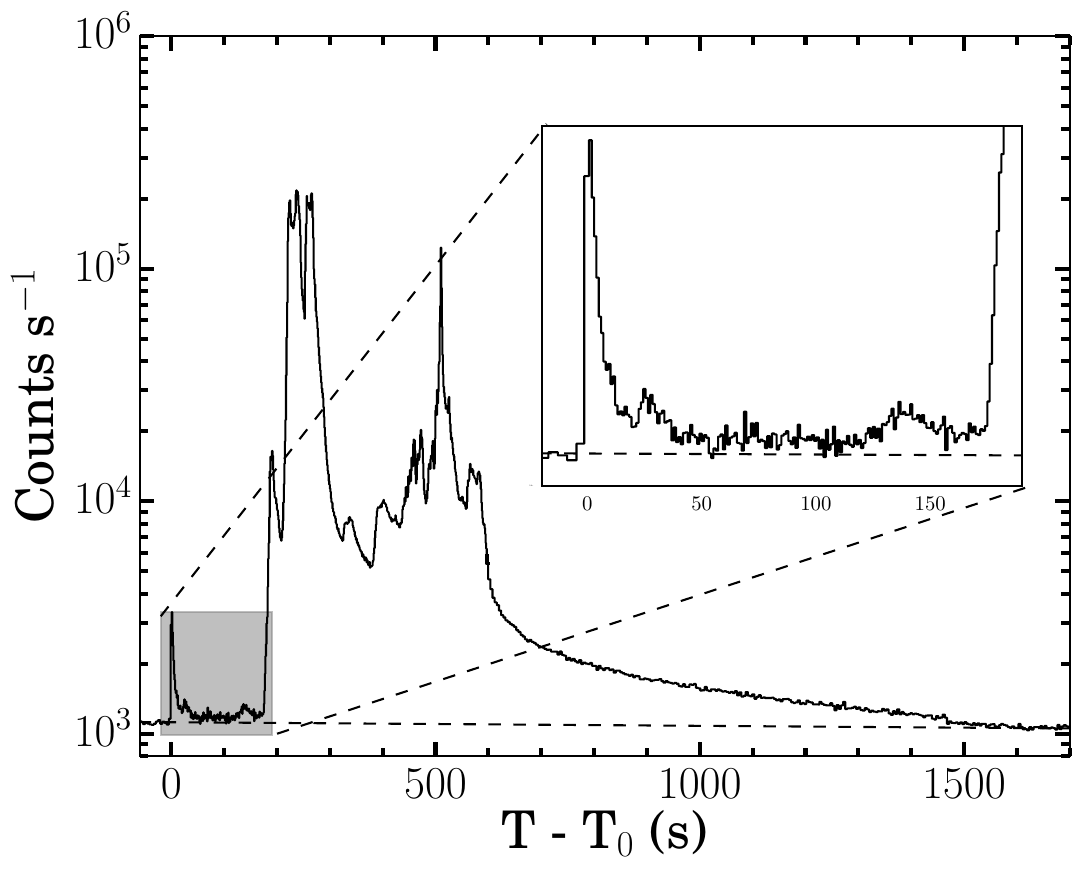}
\includegraphics[height=6cm,width=0.48\textwidth]{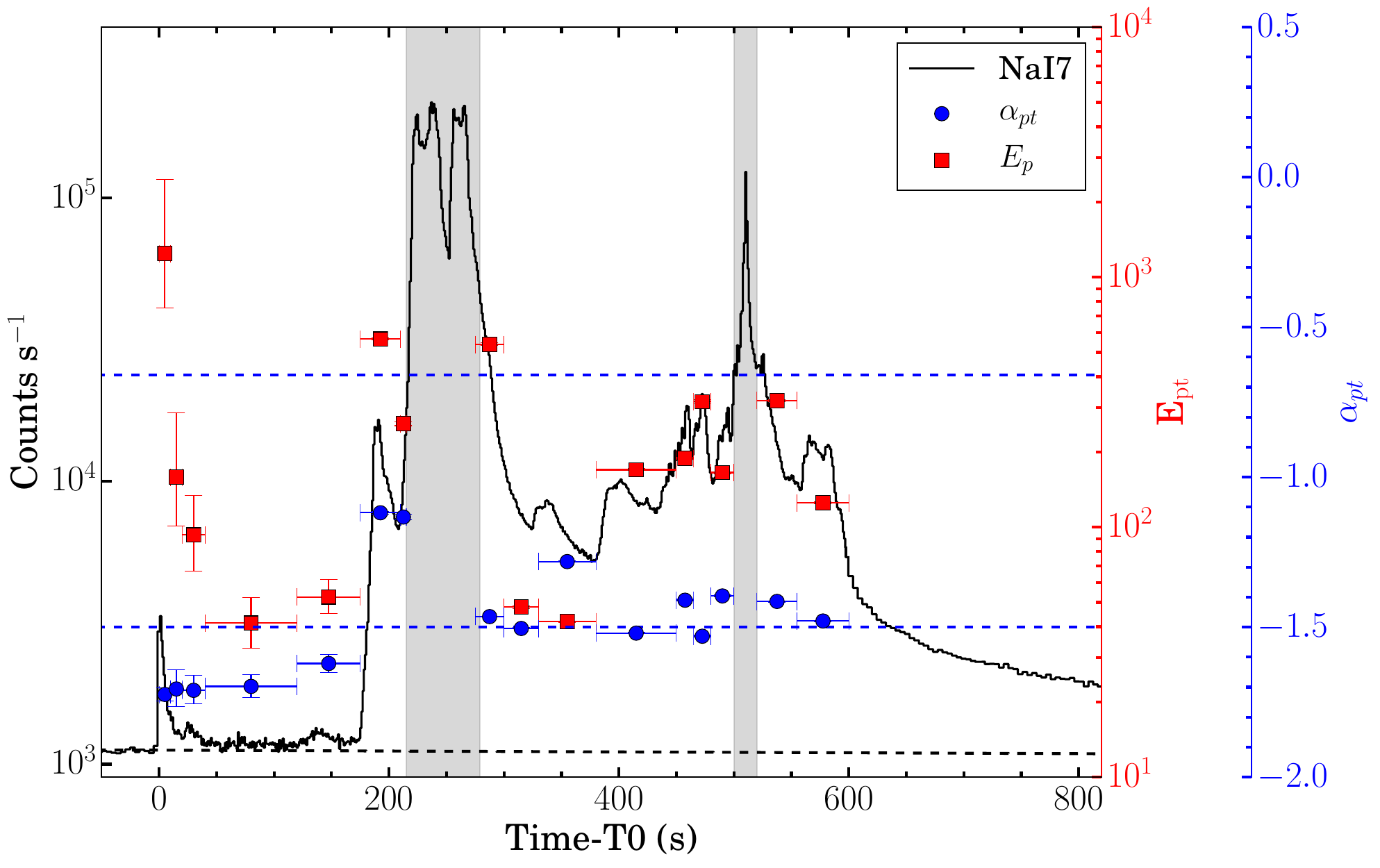}
\caption{Left: \fermi-GBM light curve of GRB 221009A. The shaded area represents the quiescent phase, a zoomed version of which is shown in the inset. Right: the evolution of prompt emission spectral parameters \Ep (red squares) and $\alpha_{\rm pt}$ (blue circles) along the GBM light curve (black) is shown. The blue dashed lines at [-3/2,-2/3] represent the synchrotron line of death for low energy spectral index $\alpha_{\rm pt}$. Black dashed lines depict the background counts. Shaded areas represent the bad time intervals [217-280 s] and [508-514 s] \cite{2023ApJ...952L..42L} omitted in the spectral fitting.}
\label{fig:gbm_fit_param}
\end{figure}

We acquired the \fermi-GBM data for GRB 221009A from the \fermi-GBM Burst Catalogue\footnote{\url{https://heasarc.gsfc.nasa.gov/W3Browse/fermi/fermigbrst.html}} and conducted temporal and spectral data analysis using the techniques outlined in \cite{2022MNRAS.511.1694G, 2023arXiv231216265G}. To perform the temporal and spectral analysis on the GBM data, we employed the Python-based package \sw{GBM-Tool} \citep{GbmDataTools}, focusing on the brightest sodium iodide detectors (NaI-7) as well as the brightest bismuth germanate detector (BGO-1). The multi-channel \fermi-GBM light curves of GRB 221009A are shown in Figures \ref{fig:multichannel_gbm} and \ref{fig:gbm_fit_param}). The \fermi-GBM light curve of GRB 221009A consists of a faint precursor emission followed by a main and extremely bright emission episode. We noted that the counts remain above the background or even consist of very faint and weaker emission in between the precursor and the main burst, started at $\sim$ \fermiT + 180 s (see inset in the left panel of Figure \ref{fig:gbm_fit_param}). This is important and proves that in the past, this has been missed for other bursts with precursors due to the combination of higher $z$ and sensitivity limits of detectors. Figure \ref{fig:multichannel_gbm} shows the \fermi count-rate light curve in different energy ranges and the evolution of the hardness ratio (HR). During the main and very long (more than 1000 s) emission phase of GRB 221009A, most of the GRB detecting instruments, including \fermi-GBM, were saturated, making the prompt emission analysis challenging. Following the analysis by \cite{2023ApJ...952L..42L, 2023ApJ...943L...2L, 2023ApJ...956L..21Z}, we omit the interval [217-280 s] and [508-514 s] to avoid pile up. 
To perform the spectral analysis of \fermi-GBM observations, we have utilized several empirical and physical models to fit the time-integrated and time-resolved spectrum (to observe the spectral parameter evolution) of GRB 221009A. Since this burst lasts longer than 1000 s, we have used \sw{CSPEC} files to represent the count-rate light curve and spectral analysis. The fit statistic \sw{PGstat} is used, and Bayesian Information Criteria (BIC) was applied to find the best-fit model. The time average spectrum (8 \keV to 40 MeV) in the temporal range \fermiT+170 s – \fermiT+600 s is best fit by the \sw{Band} function, showing the lowest BIC value of all the fits. 

The evolution of spectral parameters serves as a crucial tool for deciphering emission mechanisms during the prompt phase of GRBs \citep{2015AdAst2015E..22P}. The analysis results of \cite{1986ApJ...301..213N, 1983Natur.306..451G, 1985ApJ...290..728L} revealed that, within the \sw{Band} function, the peak energy (\Ep) exhibits four distinct types of evolution: (a) transitioning from hard to soft, (b) following flux variations, (c) shifting from soft to hard, and (d) displaying chaotic patterns. Conversely, the evolution of the low-energy spectral index ($\alpha_{\rm pt}$) is less predictable; however, some studies have noted flux-tracking patterns in $\alpha_{\rm pt}$, for example GRB 140102A \citep{2021MNRAS.505.4086G} and GRB 201216C \citep{2023ApJ...942...34R}.

In our analysis of time-resolved spectra, we divided the light curve into multiple time intervals of varying durations (see Table \ref{tab:Prompt}). Each interval's spectrum was fitted using different empirical functions, such as the \sw{Band} function and \sw{Cutoff power-law}, and subsequently re-evaluated by incorporating the thermal \sw{Blackbody} component. The spectral parameters obtained from the fitting are listed in Table \ref{tab:Prompt}. Figure \ref{fig:gbm_fit_param} illustrates the spectral parameter obtained from the best-fit model, showing that both \Ep and $\alpha_{\rm pt}$ seem to track the intensity for GRB 221009A. We observed that the spectrum created near the peak of the light curve is best described by a combination of the \sw{Band} function and \sw{Blackbody}, whereas between the peaks, a single \sw{Band} function provides the best fit for the spectra. This pattern has been observed in other extensively studied VHE-detected GRBs such as 180720B \citep{2021ApJ...920...53C} and GRB 201216C \citep{2023ApJ...942...34R}. Our prompt emission analysis of GRB 221009A is consistent with the results of \cite{2023ApJ...943L...2L, 2023ApJ...956L..21Z}.

In the precursor phase of GRB 221009A, we noticed a deviation of the low energy spectral index from the expected synchrotron fast and slow cooling range (-3/2, -2/3) (as illustrated in Figure \ref{fig:gbm_fit_param}), which poses a challenge to explanations involving the synchrotron mechanism \citep{2002ApJ...581.1248P}. The spectral fitting outcomes for the precursor emission are given in Table \ref{Precursor}. The \sw{Band} function provides the best fit for the precursor spectrum. However, adding a \sw{Blackbody} component to the \sw{Band} function slightly improves the spectral fitting, indicating the presence of a thermal component in the precursor emission spectrum \citep{2007MNRAS.380..621L}.

\subsection{Afterglow Analysis of GRB 221009A} 
\label{sec:afterglow_analysis}
In the following section, we provide details of the afterglow follow-up observations of GRB 221009A and analysis conducted with the 3.6m Devasthal optical telescope (DOT, \citealt{2016RMxAC..48...83P, 2018sn87.conf..149B}) in conjunction with publicly accessible afterglow data.

\subsubsection{DOT NIR follow-up observations and analysis of GRB 221009A}
\label{DOT Observations}

We conducted near-infrared (NIR) observations of GRB 221009A using the 3.6m DOT at ARIES. The burst was observed over five nights, from October 21 to 26, 2022. These observations were made in the J, H, and Ks filters of the TIRCAM2 instrument, which was mounted on the side port of the 3.6m DOT. We observed source frames at five different dither positions (D1-D5) to remove background noise. Initially, we carried out image pre-processing, for which dark and flat frames were observed separately on every observation night. Subsequently, the dither sets were combined to obtain sky frames after pre-processing. After sky subtraction, we aligned the source images and conducted PSF photometry on the resulting stacked image. Early on, we observed the optical afterglow of GRB 221009A on October 16, 2023, about seven days after the burst, using the 1.3m Devasthal Fast Optical Telescope at ARIES \citep{2022GCN.32811....1G}. We took multiple frames in the R-band and used the Image Reduction and Analysis Facility (IRAF) to process the optical data \citep{2022JApA...43...11G}. The magnitudes of the source obtained from our observations are listed in Table \ref{tab:DOT_obs}, and finding charts are shown in Figure \ref{fig:finding_chart}.

In addition to our observations from the 3.6m DOT of GRB 221009A, data points in various NIR-Optical bands have been obtained from the General Coordinates Network (GCN\footnote{\url{https://gcn.gsfc.nasa.gov/other/221009A.gcn3}}), as well as publications by \cite{2023arXiv230204388L}, \cite{2023arXiv230203829S}, \cite{2023arXiv230207906O}, and R. Sánchez-Ramírez et al. (2024, under review) to get a well-sampled light curve.
We obtained the \swift-XRT light curve at 10\,\keV from the official \swift-webpage\footnote{\url{https://www.swift.ac.uk/burst_analyser/01126853/}}. We employed the MCMC fitting technique to model the afterglow light curves. We used a smoothly joined broken power law to fit the optical R-band light curve (good coverage) to determine the break time T$_{\rm b,o}$. The smoothness parameter was fixed at 3. The fitting parameters obtained were $\alpha_{\rm o,1}$ = 0.57$_{-0.04}^{+0.04}$ and $\alpha_{\rm o,2}$ = 1.43$_{-0.02}^{+0.02}$, with a break time T$_{\rm b,o}$ = 24700$_{-2500}^{+2500}$. The late-time light curves in other optical bands were fitted using a simple power-law function, and the resulting decay indices were consistent within the error bar of the post-break decay index ($\alpha_{\rm o,2}$) in the R-band light curve. We also observed an achromatic behavior in the optical and NIR light curves, with temporal decay indices of $\alpha_{\rm J}$ $\sim$ 1.4 and $\alpha_{\rm K}$ $\sim$ 1.45 in the J and K bands, respectively. 

The light curve of GRB 221009A from \swift-XRT displays a smooth decay. We utilized a simple power-law function to fit it, yielding a decay index of $\alpha_{\rm x}$ = 1.66$_{-0.01}^{+0.01}$ with BIC = 35120. Additionally, we attempted fitting the \swift-XRT light curve using the smoothly joined broken power-law with smoothness parameter fixed at 3 and constrain $\alpha_{\rm x,1}$ = 1.57$_{-0.02}^{+0.02}$, $\alpha_{\rm x,2}$ = 1.86$_{-0.02}^{+0.02}$, and the break time T$_{\rm b,x}$ $\sim$ 84000$_{-6500}^{+4500}$ s with BIC = 34203. The smoothly joined broken power-law model provided a better fit with a lower BIC value ($\Delta$BIC $<$ 900) to the observed data.

\subsubsection{Spectral energy distribution analysis and Closure relations} 

To create the NIR-optical-X-ray SED, we have used the methodology described in \cite{2023ApJ...942...34R, 2023arXiv231216265G}. We have retrieved the \swift-XRT spectra for GRB 221009A from the \swift webpage\footnote{\url{https://www.swift.ac.uk/xrt_spectra/01126853/}} at two epochs (corresponding to which we have enough NIR/optical observations). To model the XRT spectrum using \sw{XSPEC}, we employed a power-law function combined with Galactic (\sw{phabs} fixed at NH$_{\rm Gal}$ = 5.38 $\times$ 10$^{21}$ cm$^{-2}$) and intrinsic (\sw{zphabs}) absorption components. To constrain the intrinsic hydrogen column density (NH$_{\rm z}$), we fitted the late time spectra and found NH$_{\rm z}$ = 1.40 $\times$ 10$^{22}$ cm$^{-2}$. We fixed NH$_{\rm z}$ for the spectral analysis and found X-ray spectral indices $\beta_{\rm x}$ = 0.80$^{+0.04}_{-0.04}$. The $\beta_{\rm x}$ obtained from our analysis are consistent with the results of \cite{2023ApJ...946L..24W}. We fit the NIR-optical observations with a simple power-law to constrain the optical spectral indices ($\beta_{\rm o}$).

GRB afterglows are assumed to follow certain closure relations that help to constrain the spectral regime of the afterglow emission as well as the circumburst medium surrounding the burst \citep{Sari_1998, 2013NewAR..57..141G}. For GRB 221009A, we have studied several cases and found that the spectral and temporal indices obtained for different spectral regimes are not fully consistent with the existing closure relations. For $p$ $>$ 2 case and in the spectral regime $\nu_{m}$ $<$ $\nu_{o}$ $<$ $\nu_{x}$ $<$ $\nu_{c}$, we have $p$ = 2$\beta$ + 1 $\sim$ 2.6. The $p$ value obtained in this spectral regime for a WIND-like medium only explains the decay $\alpha$ = (3$p$-1)/4 = 1.7 of the XRT light curve $\alpha_{x}$ = 1.66 $\pm$ 0.01 but not the optical light curve $\alpha_{o,2}$ = 1.43 $\pm$ 0.02 in 5$\sigma$ range. In an ISM-like medium, the $\alpha$ = 3($p$-1)/4 = 1.2 is again too shallow to explain the decay in optical and X-ray light curves. In the spectral regime $\nu_{m}$ $<$ $\nu_{c}$ $<$ $\nu_{o}$ $<$ $\nu_{x}$, $p$ = 2.6 in both ISM and WIND-like media is able to explain the decay of optical light with decay index $\alpha$ = (3$p$-2)/4 = 1.45 within 1$\sigma$, but this is sallower than the decay of X-ray light curve. If we consider an early jet break, then for $p$ = 2.6, the decay indices obtained for the ISM and the WIND-like circumburst medium are much steeper than the observed values.

For $p$ $<$ 2 cases, $p$ = 2$\beta$ $\sim$ 1.6, without considering an early jet break. In the spectral regime $\nu_{m}$ $<$ $\nu_{o}$ $<$ $\nu_{x}$ $<$ $\nu_{c}$, the optical and XRT decay indices are not consistent with the closure relations $\alpha$ = $\frac{3(p+2)}{16}$ = 0.68 and $\alpha$ = $\frac{(p+8)}{8}$ = 1.2 for ISM and WIND-like circumburst medium, respectively. Similarly, for the spectral regime, $\nu_{m}$ $<$ $\nu_{c}$ $<$ $\nu_{o}$ $<$ $\nu_{x}$, $\alpha$ = $\frac{(3p+10)}{16}$ = 0.92 (ISM) and $\alpha$ = $\frac{(p+6)}{8}$ = 0.95 (wind) are again too shallow compared to optical and X-ray decay indices.

For $p$ $<$ 2 with an early jet break, the optical decay indices are consistent with the relation $\alpha$ = $\frac{3(p+6)}{16}$ $\sim$ 1.43 in the ISM-like circumburst medium for the spectral regime $\nu_{m}$ $<$ $\nu_{o}$ $<$ $\nu_{c}$ $<$ $\nu_{x}$. The X-ray decay index within the same spectral regime is consistent with the relation $\alpha$ = $\frac{3p+22}{16}$ = 1.67 within 1$\sigma$.

Thus, if we consider a break in the \swift-XRT light curve, then these relations are feasible to explain the decay of the optical and XRT light curves; such a break and ISM-like medium are also consistent with the analysis of \cite{2023arXiv230207761L, 2023arXiv230207906O}.

\subsection{ Calculation of bore time (t$_{b}$) for three different scenarios} \label{tb_calculation}
Utilizing the equation \ref{eqn:collapsar} with L$_{\rm \gamma, iso}$ = 9.91 $\times$ 10$^{53}$ erg s$^{-1}$ and $\theta_{j}$ = 0.1 rad (5.7$^{\circ}$) \citep{2023arXiv230207761L, 2023arXiv230207906O, 2023arXiv230206225K}, we have calculated the value of t$_{b}$ for GRB 221009A for three different cases:
\begin{enumerate}
    \item R and M fixed: For R = 10$^{11}$ cm and M = 15\,M$_\odot$, we obtain T$_{90}$/t$_{b}$ = 4270.76.
    
    \item R fixed M changed: For R fixed at 10$^{11}$ cm but M increased to 30\,M$_\odot$, we obtain T$_{90}$/t$_{b}$ = 3390.494, i.e there is 20\% decrement in T$_{90}$/t$_{b}$.
    
    \item R and M, both changed: Further, by varying the radius according to relation R = 1.33M$^{0.55}$ corresponding to M = 30\,M$_\odot$. We obtain the ratio T$_{90}$/t$_{b}$ = 2635.142, i.e 38\% decrement to the initial value. These results are consistent with our analysis given in section \ref{sec:collapsar}.
    
\end{enumerate}

\begin{figure*}[!ht]
\centering
\includegraphics[scale=0.43]{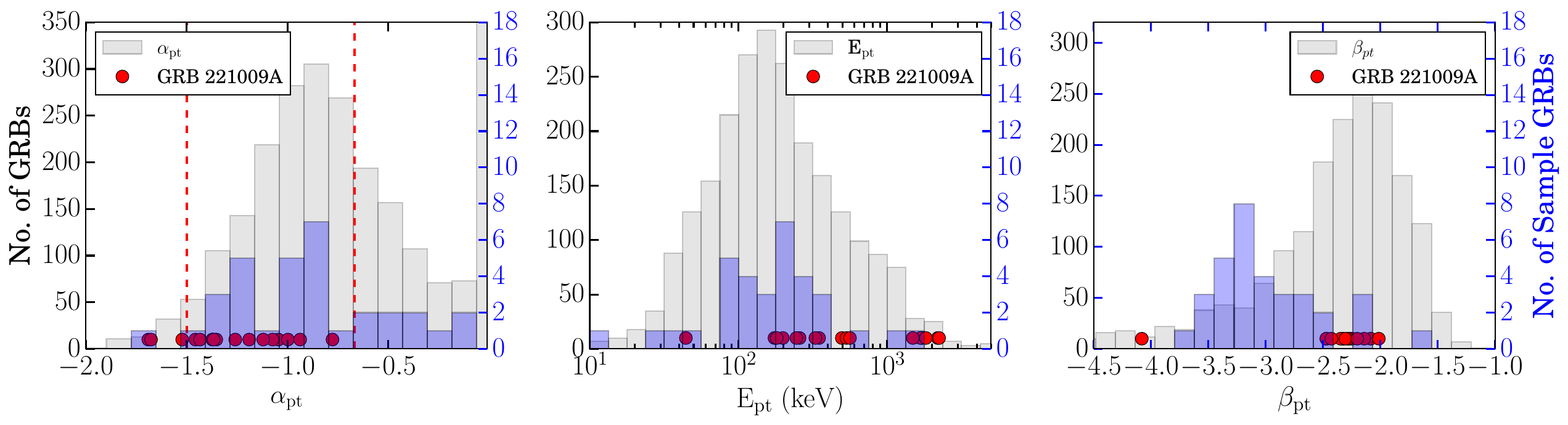} 
\caption{Left panel: distribution of $\alpha_{\rm pt}$ of the complete sample \fermi-detected GRBs (gray) along with those GRBs with \fermi-GBM detection (blue) in our selected sample. Similarly, the middle and right panels show the distribution of \Ep and $\beta_{\rm pt}$, respectively. The red dots represent the distribution of parameters obtained from the time-resolved analysis of \fermi-GBM observation of GRB 221009A.}
\label{fig:prompt_param_ULGRBs}
\end{figure*}

\begin{figure}[!ht]
\centering
\includegraphics[scale=0.282]{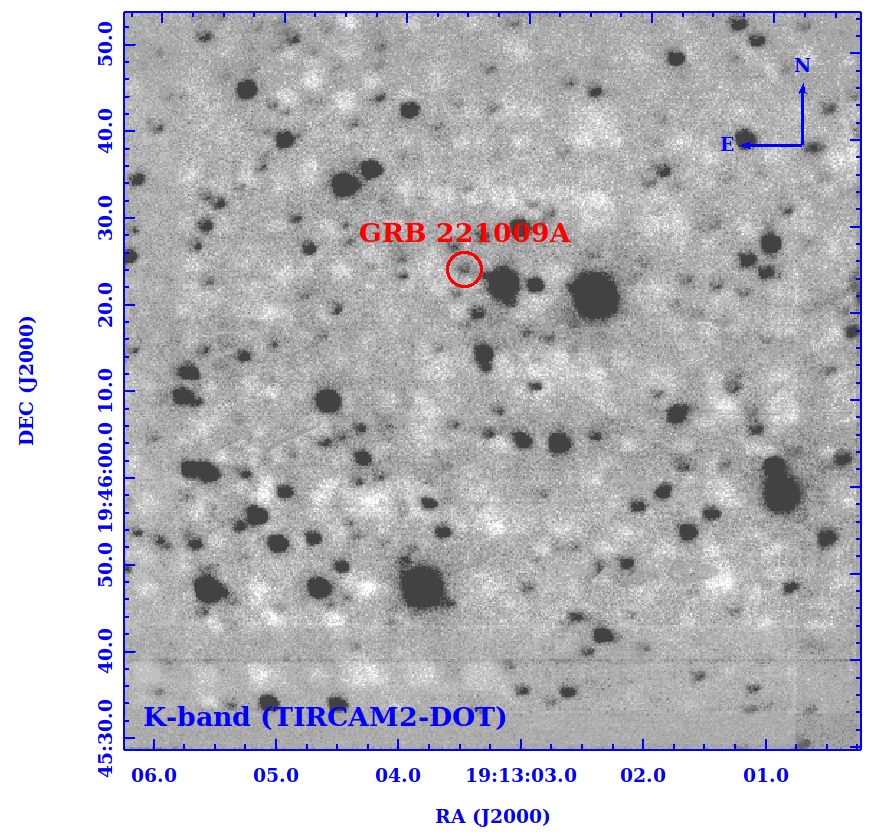} 
\includegraphics[scale=0.3]{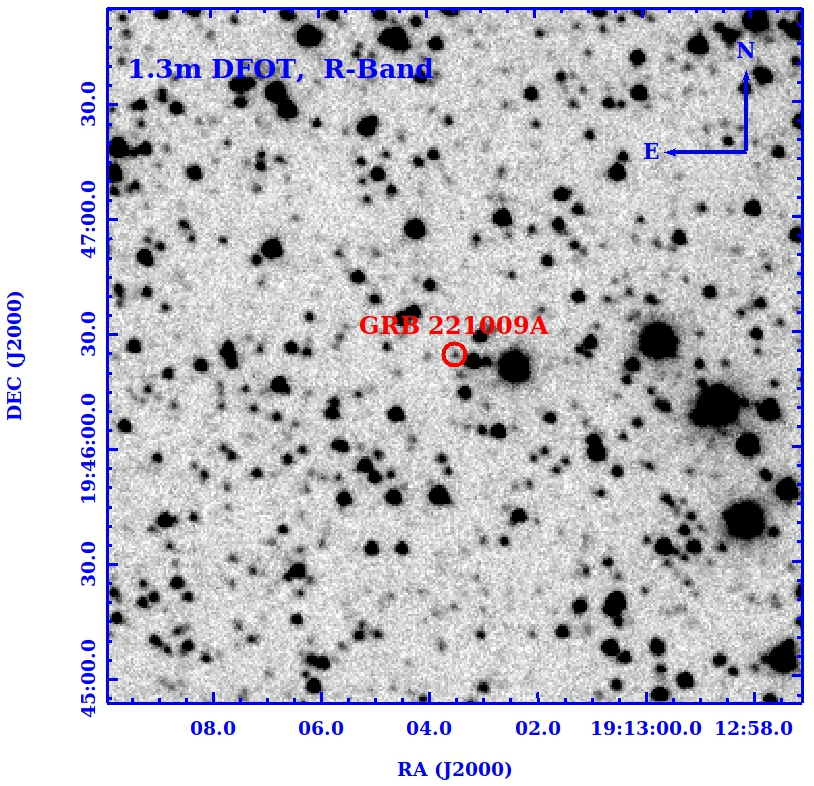} 
\caption{Left, the K-band finding chart for GRB 221009A from observations made with the TIRCAM2 instrument mounted on the side port of the 3.6m DOT. Right, the R-band finding chart is generated from observations made with the 1.3m DFOT.}
\label{fig:finding_chart}
\end{figure}

\FloatBarrier
\addtolength{\tabcolsep}{-3pt}
\begin{longtable}{|c|c|c|c|c|c|c|c|c|c|} 
\caption{The characteristics of the ULGRBs in our sample that met the selection criteria given in Section \ref{sec:selection_criteria}.}
\label{tab:collapsar}\\ \hline

\bf GRB &\bf z & \boldmath T$_{90}$ & \bf \boldmath E$_{\rm \gamma,iso}$ & \bf \boldmath L$_{\rm \gamma,iso}$ & \bf \boldmath E$_{\rm K,iso}$ & \bf \boldmath E$_{\rm X,iso}$ & \bf \boldmath $\frac{\rm T_{90,z}}{\rm t_{b}}$ & \textbf{Non-Collapsar} \\

 & & \bf \boldmath (s) & \bf \boldmath ($\times$10$^{52}$ erg) & \bf \boldmath ($\times$10$^{51}$ erg s$^{-1}$) & \bf \boldmath ($\times$10$^{52}$ erg) & \bf \boldmath ($\times$10$^{50}$ erg) &  & \textbf{Probability} \\ \hline

\endfirsthead
\hline 
\multicolumn{9}{|r|}{{Continued on next page}} \\ \hline
\endfoot
\hline
\bf GRB &\bf z & \boldmath T$_{90}$ & \bf \boldmath E$_{\rm \gamma,iso}$ & \bf \boldmath L$_{\rm \gamma,iso}$ & \bf \boldmath E$_{\rm K,iso}$ & \bf \boldmath E$_{\rm X,iso}$ & \bf \boldmath $\frac{\rm T_{90,z}}{\rm t_{b}}$ & \textbf{Non-Collapsar} \\

 & & \bf \boldmath (s) & \bf \boldmath ($\times$10$^{52}$ erg) & \bf \boldmath ($\times$10$^{51}$ erg s$^{-1}$) & \bf \boldmath ($\times$10$^{52}$ erg) & \bf \boldmath ($\times$10$^{50}$ erg) &  &\textbf{Probability} \\ \hline
 
\endhead
\hline
\endlastfoot

\multicolumn{9}{|c|}{\large {\textbf{Bronze Sample}}} \\ [2ex] \hline
050315 &1.95&95.4& 14.95 $\pm$ 0.66 & 5.28 $\pm$ 0.64 &  &  &30.88&6.21 $\times$ 10$^{-8}$\\ 
050319 &3.2425&152& 15.12 $\pm$ 1.65 & 10.12 $\pm$ 1.60 &  &  &34.53&6.43 $\times$ 10$^{-9}$\\ 
050505 &4.2748&58.9& 44.16 $\pm$ 3.28 & 29.62 $\pm$ 4.43 &  &  &15.37&5.94 $\times$ 10$^{-7}$\\ 
050724 &0.257&98.7& 0.08 $\pm$ 0.01 & 0.19 $\pm$ 0.02 &  &  &12.42&5.27 $\times$ 10$^{-8}$\\ 
050730 &3.9693&155& 37.23 $\pm$ 2.45 & 7.26 $\pm$ 1.99 &  & 273.00 $\pm$ 11.00 &79.9&5.84 $\times$ 10$^{-9}$\\ 
050803 &3.5&88.1& 27.83 $\pm$ 1.74 & 10.50 $\pm$ 1.42 &  &  &17.64&9.08 $\times$ 10$^{-8}$\\ 
050814 &5.3&143& 47.16 $\pm$ 5.19 & 11.93 $\pm$ 4.88 &  &  &46.48&8.69 $\times$ 10$^{-9}$\\ 
050915A &2.5273&53.4& 6.24 $\pm$ 0.66 & 5.47 $\pm$ 0.99 &  &  &14&9.26 $\times$ 10$^{-7}$\\ 
051111 &1.54948&64& 13.17 $\pm$ 0.48 & 5 $\pm$ 0.57 &  &  &21.03&4.06 $\times$ 10$^{-7}$\\ 
060115 &3.5328&139& 23.94 $\pm$ 1.70 & 9.42 $\pm$ 1.47 &  &  &61.36&9.98 $\times$ 10$^{-9}$\\ 
060116 &6.6&105& 81.71 $\pm$ 8.89 & 31.11 $\pm$ 8.19 &  &  &29.4&3.91 $\times$ 10$^{-8}$\\ 
060418 &1.49&109& 24.06 $\pm$ 0.72 & 14.29 $\pm$ 0.77 &  &  &45.23&3.27 $\times$ 10$^{-8}$\\ 
060522 &5.11&69.1& 26.40 $\pm$ 2.57 & 7.97 $\pm$ 2.66 &  &  &18.23&2.85 $\times$ 10$^{-7}$\\ 
060602A &0.787&74.8& 1.31 $\pm$ 0.13 & 0.30 $\pm$ 0.15 &  &  &7.56&1.97 $\times$ 10$^{-7}$\\ 
060604 &2.1357&96& 2.09 $\pm$ 0.55 &0.04& 30.27 $\pm$ 15.10 & 3.86 $\pm$ 1.09 &3.42&6.02 $\times$ 10$^{-8}$\\ 
060605 &3.773&79.8& 10.34 $\pm$ 1.33 & 6.88 $\pm$ 1.67 & 154.67 $\pm$ 65.15 & 60.00 $\pm$ 3.90 &35.15&1.45 $\times$ 10$^{-7}$\\ 
060607A &3.0749&103& 26.76 $\pm$ 1.20 & 13.33 $\pm$ 1.30 &  &  &71.64&4.29 $\times$ 10$^{-8}$\\ 
060614 &0.1254&109& 0.37 $\pm$ 0.01 & 0.15 $\pm$ 0.01 & 11.20 $\pm$ 2.21 & 1.91 $\pm$ 0.24 &22.84&3.27 $\times$ 10$^{-8}$\\ 
060707 &3.424&66.6& 21.14 $\pm$ 1.74 & 9.44 $\pm$ 2.17 & 11.51 $\pm$ 2.42 & 5.75 $\pm$ 0.44 &22.12&3.38 $\times$ 10$^{-7}$\\ 
060714 &2.7108&116& 24.25 $\pm$ 1.43 & 8.16 $\pm$ 0.92 & 35.51 $\pm$ 6.16 & 49.80 $\pm$ 3.80 &25.05&2.42 $\times$ 10$^{-8}$\\ 
060719 &1.532&66.9& 4.62 $\pm$ 0.29 & 4.50 $\pm$ 0.41 &  &  &14.54&3.31 $\times$ 10$^{-7}$\\ 
060729 &0.5428&113& 0.98 $\pm$ 0.08 & 0.29 $\pm$ 0.04 & 19.46 $\pm$ 1.68 & 0.711 $\pm$ 0.017 &9.47&2.74 $\times$ 10$^{-8}$\\ 
060814 &1.9229&145& 68.06 $\pm$ 1.15 & 27.87 $\pm$ 1.13 & 19.39 $\pm$ 15.33 & 3.18 $\pm$ 0.23 &157.07&8.11 $\times$ 10$^{-9}$\\ 
060906 &3.6856&44.6& 32.08 $\pm$ 1.96 & 17.09 $\pm$ 2.73 & 68.13 $\pm$ 41.64 & 3.17 $\pm$ 1.31 &31.52&2.06 $\times$ 10$^{-6}$\\ 
061007 &1.2622&75.7& 93.75 $\pm$ 1.04 & 32.50 $\pm$ 0.83 &  &  &39.97&1.86 $\times$ 10$^{-7}$\\ 
061021 &0.3463&47.8& 0.44 $\pm$ 0.02 & 0.80 $\pm$ 0.04 &  &  &5.03&1.52 $\times$ 10$^{-6}$\\ 
061110B &3.4344&135& 16.87 $\pm$ 1.56 & 5.40 $\pm$ 1.32 &  &  &24.98&1.15 $\times$ 10$^{-8}$\\ 
061121 &1.3145&81.2& 31.33 $\pm$ 0.45 & 44.17 $\pm$ 0.90 & 28.80 $\pm$ 9.43 & 37.00 $\pm$ 3.30 &36.7&1.34 $\times$ 10$^{-7}$\\ 
061210 &0.4095&85.2& 0.23 $\pm$ 0.04 & 0.61 $\pm$ 0.06 &  &  &12.04&1.06 $\times$ 10$^{-7}$\\ 
061222A &2.088&100& 45.40 $\pm$ 1.07 & 39.69 $\pm$ 1.28 &  &  &88.5&4.95 $\times$ 10$^{-8}$\\ 
070110 &2.3521&88.4& 10.74 $\pm$ 0.72 & 3.12 $\pm$ 0.70 &  &  &11.45&8.93 $\times$ 10$^{-8}$\\ 
070208 &1.165&64& 0.92 $\pm$ 0.21 & 0.61 $\pm$ 0.28 &  &  &11.51&4.06 $\times$ 10$^{-7}$\\ 
070318 &0.8397&130& 2.43 $\pm$ 0.13 & 1.22 $\pm$ 0.13 &  &  &28.42&1.39 $\times$ 10$^{-8}$\\ 
070411 &2.9538&116& 26.19 $\pm$ 1.41 & 6.90 $\pm$ 1.09 &  &  &24.21&2.42 $\times$ 10$^{-8}$\\ 
070419A &0.9705&160& 0.83 $\pm$ 0.13 &0.02&  &  &6.56&5.00 $\times$ 10$^{-9}$\\ 
070529 &2.4996&109& 18.16 $\pm$ 1.79 & 8.13 $\pm$ 2.22 & 30.54 $\pm$ 17.65 & 8.87 $\pm$ 0.92 &28.68&3.27 $\times$ 10$^{-8}$\\ 
070714B &0.9225&65.6& 0.83 $\pm$ 0.10 & 3.04 $\pm$ 0.19 & 758.30 $\pm$ 40.09 & 3.85 $\pm$ 0.45 &57.41&3.62 $\times$ 10$^{-7}$\\ 
071003 &1.60435&148& 27.49 $\pm$ 0.94 & 20.91 $\pm$ 1.33 & 9.91 $\pm$ 1.15 & 6.33 $\pm$ 0.49 &60.93&7.34 $\times$ 10$^{-9}$\\ 
071028B &0.94&51.2&  &  &  &  &  &1.12 $\times$ 10$^{-6}$\\ 
071122 &1.14&71.4& 1.05 $\pm$ 0.21 & 0.38 $\pm$ 0.21 &  &  &34.51&2.44 $\times$ 10$^{-7}$\\ 
080210A &2.641&45&14.31&1.06&  &  &14.06&1.98 $\times$ 10$^{-6}$\\ 
080330 &1.5119&60.4& 0.87 $\pm$ 0.22 & 1.39 $\pm$ 0.37 &  &  &  &5.29 $\times$ 10$^{-7}$\\ 
080411 &1.0301&56.3& 37.02 $\pm$ 0.49 & 42.47 $\pm$ 0.70 &  &  &23.63&7.29 $\times$ 10$^{-7}$\\ 
080413A &2.433&46.4& 24.60 $\pm$ 0.70 & 34.31 $\pm$ 1.47 &  &  &27.31&1.73 $\times$ 10$^{-6}$\\ 
080603B &2.6892&59.1& 21.43 $\pm$ 0.71 & 26.49 $\pm$ 1.41 &  &  &55.21&5.85 $\times$ 10$^{-7}$\\ 
080604 &1.4171&77.6& 2.05 $\pm$ 0.23 & 0.82 $\pm$ 0.26 &  &  &12.75&1.66 $\times$ 10$^{-7}$\\ 
080607 &3.0368&79& 251.89 $\pm$ 4.59 & 241.69 $\pm$ 11.73 & 177.31 $\pm$ 15.39 & 53.30 $\pm$ 4.50 &80.52&1.52 $\times$ 10$^{-7}$\\ 
080710 &0.8454&143& 1.44 $\pm$ 0.21 & 0.57 $\pm$ 0.15 &  &  &19.05&8.69 $\times$ 10$^{-9}$\\ 
080805 &1.5042&107& 7.73 $\pm$ 0.36 & 3.35 $\pm$ 0.39 &  &  &20.44&3.57 $\times$ 10$^{-8}$\\ 
080810 &3.3604&108& 55.95 $\pm$ 1.93 & 20.38 $\pm$ 2.17 &  &  &39.88&3.42 $\times$ 10$^{-8}$\\ 
080905B &2.3739&121& 12.12 $\pm$ 1.07 & 8.44 $\pm$ 1.44 & 53.34 $\pm$ 16.35 & 109.00 $\pm$ 6.00 &72.85&1.97 $\times$ 10$^{-8}$\\ 
080906 &2.13&148& 20.18 $\pm$ 1.02 & 4.49 $\pm$ 0.80 & 33.94 $\pm$ 3.23 & 6.40 $\pm$ 1.53 &30.68&7.34 $\times$ 10$^{-9}$\\ 
080916A &0.6887&61.3& 2.67 $\pm$ 0.07 & 1.61 $\pm$ 0.10 &  &  &27.32&4.95 $\times$ 10$^{-7}$\\ 
081118 &2.58&53.4& 9.13 $\pm$ 0.85 & 3.38 $\pm$ 1.07 &  &  &9.21&9.26 $\times$ 10$^{-7}$\\ 
090313 &3.3736&83& 18.32 $\pm$ 3.16 & 7.03 $\pm$ 3.17 &  &  &26.88&1.20 $\times$ 10$^{-7}$\\ 
090418A &1.608&56.3& 15.41 $\pm$ 0.58 & 5.22 $\pm$ 0.91 & 14.16 $\pm$ 3.49 & 31.30 $\pm$ 2.70 &55.78&7.29 $\times$ 10$^{-7}$\\ 
090424 &0.544&49.5& 8.37 $\pm$ 0.15 & 22.27 $\pm$ 0.61 &  &  &35.02&1.30 $\times$ 10$^{-6}$\\ 
090519 &3.85&58& 17.73 $\pm$ 1.95 & 10.31 $\pm$ 2.62 &  &  &27.93&6.37 $\times$ 10$^{-7}$\\ 
090530A &1.266&48&2.39&0.58&  &  &10.22&1.49 $\times$ 10$^{-6}$\\ 
090618 &0.54&113& 40.66 $\pm$ 0.37 & 11.68 $\pm$ 0.21 & 11.97 $\pm$ 1.88 & 11.20 $\pm$ 1.30 &69.82&2.74 $\times$ 10$^{-8}$\\ 
090726 &2.71&56.7& 6.61 $\pm$ 0.74 & 4.47 $\pm$ 1.06 &  &  &17.28&7.06 $\times$ 10$^{-7}$\\ 
090812 &2.452&74.5& 40.59 $\pm$ 1.13 & 24.93 $\pm$ 1.56 &  &  &37.45&2.00 $\times$ 10$^{-7}$\\ 
090814A &0.696&78.1& 0.79 $\pm$ 0.10 & 0.27 $\pm$ 0.11 &  &  &18.14&1.61 $\times$ 10$^{-7}$\\ 
090926B &1.24&99.3&15.18& 5.45 $\pm$ 0.58 &  &  &59.25&5.12 $\times$ 10$^{-8}$\\ 
091109A &3.076&48& 17.09 $\pm$ 2.29 & 10.31 $\pm$ 3.12 &  &  &11.05&1.49 $\times$ 10$^{-6}$\\ 
100424A &2.465&104& 10.72 $\pm$ 0.86 & 1.64 $\pm$ 0.77 & 57.01 $\pm$ 10.42 & 69.30 $\pm$ 2.80 &31.09&4.10 $\times$ 10$^{-8}$\\ 
100513A &4.772&83.5& 30.09 $\pm$ 2.52 & 10.04 $\pm$ 2.72 &  &  &17.76&1.17 $\times$ 10$^{-7}$\\ 
100621A &0.542&63.6& 7.74 $\pm$ 0.11 & 3.63 $\pm$ 0.10 & 55.34 $\pm$ 32.18 & 10.50 $\pm$ 0.60 &37.94&4.18 $\times$ 10$^{-7}$\\ 
100906A &1.727&114.4&45.78&3.09& 348.78 $\pm$ 121.14 & 44.70 $\pm$ 2.50 &75.82&2.58 $\times$ 10$^{-8}$\\ 
110213A &1.4607&48& 16.47 $\pm$ 1.09 & 1.97 $\pm$ 0.86 & 172.24 $\pm$ 22.11 & 4740 $\pm$ 1260 &26.96&1.49 $\times$ 10$^{-6}$\\ 
110808A &1.348&48&0.81&0.09& 1.74 $\pm$ 0.64 & 4.66 $\pm$ 0.63 &4.03&1.49 $\times$ 10$^{-6}$\\ 
110818A &3.36&103& 48.10 $\pm$ 3.01 & 17 $\pm$ 3.01 &  &  &41.68&4.29 $\times$ 10$^{-8}$\\ 
111008A &4.99005&62.8& 117.16 $\pm$ 5.86 & 121.90 $\pm$ 14.53 & 45.70 $\pm$ 16.25 & 102.00 $\pm$ 5.00 &55.83&4.43 $\times$ 10$^{-7}$\\ 
111225A &0.297&106& 0.13 $\pm$ 0.01 & 0.05 $\pm$ 0.01 &  &  &9.35&3.74 $\times$ 10$^{-8}$\\ 
111228A &0.71627&101& 5.53 $\pm$ 0.15 & 5.39 $\pm$ 0.23 & 19.80 $\pm$ 5.43 & 5.31 $\pm$ 0.26 &91.72&4.72 $\times$ 10$^{-8}$\\ 
120119A &1.728&68& 64.27 $\pm$ 0.95 & 33.23 $\pm$ 1.08 &  &  &99.32&3.07 $\times$ 10$^{-7}$\\ 
120326A &1.798&69.5& 10.86 $\pm$ 0.41 & 12.61 $\pm$ 0.61 & 15.18 $\pm$ 6.76 & 46.00 $\pm$ 1.50 &37.96&2.77 $\times$ 10$^{-7}$\\ 
120327A &2.8145&63.5& 32.04 $\pm$ 1.12 & 29.43 $\pm$ 1.66 & 6.83 $\pm$ 4.94 & 36.20 $\pm$ 2.40 &89.48&4.21 $\times$ 10$^{-7}$\\ 
120714B &0.3984&157& 0.25 $\pm$ 0.03 & 0.04 $\pm$ 0.02 &  &  &  &5.49 $\times$ 10$^{-9}$\\ 
120724A &1.48&77.9& 2.26 $\pm$ 0.35 & 1.24 $\pm$ 0.39 & 2.08 $\pm$ 0.41 & 0.77 $\pm$ 0.15 &24.93&1.63 $\times$ 10$^{-7}$\\ 
120729A &0.8&93.9& 2.11 $\pm$ 0.10 & 1.73 $\pm$ 0.13 &  &  &57.18&6.70 $\times$ 10$^{-8}$\\ 
120802A &3.796&50.3& 25.56 $\pm$ 1.62 & 31.59 $\pm$ 2.50 & 68.98 $\pm$ 12.18 & 28.20 $\pm$ 9.40 &37.53&1.21 $\times$ 10$^{-6}$\\ 
120805A &3.1&48& 8.88 $\pm$ 1.42 &1.29& 0.51 $\pm$ 0.23 &  &8.09&1.49 $\times$ 10$^{-6}$\\ 
121024A &2.298&68& 6.96 $\pm$ 0.95 & 7.90 $\pm$ 1.45 &  &  &31.33&3.07 $\times$ 10$^{-7}$\\ 
121201A &3.385&85&9.38&0.71&  &  &8.53&1.08 $\times$ 10$^{-7}$\\ 
130215A &0.597&66.2& 2.48 $\pm$ 0.21 & 0.85 $\pm$ 0.23 &  &  &  &3.47 $\times$ 10$^{-7}$\\ 
130420A &1.297&121& 16.67 $\pm$ 0.43 & 5.05 $\pm$ 0.35 &  &  &69.08&1.97 $\times$ 10$^{-8}$\\ 
130505A &2.27&89.3& 21.48 $\pm$ 1.73 &  &  &  &  &8.51 $\times$ 10$^{-8}$\\ 
130610A &2.092&47.7& 13.54 $\pm$ 0.62 & 7.98 $\pm$ 0.88 &  &  &16.07&1.53 $\times$ 10$^{-6}$\\ 
131105A &1.686&112&  &  & 41.22 $\pm$ 19.38 & 20.00 $\pm$ 1.40 &  &2.86 $\times$ 10$^{-8}$\\ 
140114A &3&140& 31.98 $\pm$ 1.50 & 5.66 $\pm$ 1.13 & 41.63 $\pm$ 10.79 &  &65.73&9.64 $\times$ 10$^{-9}$\\ 
140206A &2.73&94.2& 143.16 $\pm$ 2.56 & 148.27 $\pm$ 3.41 & 44.34 $\pm$ 11.71 & 728.00 $\pm$ 9.00 &119.42&6.60 $\times$ 10$^{-8}$\\ 
140213A &1.2076&59.9& 22.94 $\pm$ 0.38 & 31.16 $\pm$ 0.96 &  &  &42.52&5.50 $\times$ 10$^{-7}$\\ 
140311A &4.954&70.5& 46.80 $\pm$ 6.80 & 19.68 $\pm$ 7.44 &  &  &18.46&2.59 $\times$ 10$^{-7}$\\ 
140419A &3.956&80.1& 246.36 $\pm$ 4.71 & 65.59 $\pm$ 3.30 &  &  &33.81&1.43 $\times$ 10$^{-7}$\\ 
140423A &3.26&134& 107.49 $\pm$ 3.44 & 21.34 $\pm$ 2.70 &  &  &98.59&1.19 $\times$ 10$^{-8}$\\ 
140506A &0.889&111& 2.77 $\pm$ 0.27 & 9.17 $\pm$ 0.76 &  &  &28.29&2.99 $\times$ 10$^{-8}$\\ 
140512A &0.725&154& 8.92 $\pm$ 0.17 & 3.52 $\pm$ 0.15 & 51.44 $\pm$ 14.15 & 35.80 $\pm$ 0.70 &97.76&6.03 $\times$ 10$^{-9}$\\ 
140518A &4.707&60.5& 24.10 $\pm$ 1.85 & 14.56 $\pm$ 2.45 &  &  &46.87&5.25 $\times$ 10$^{-7}$\\ 
140703A &3.14&68.6& 41.83 $\pm$ 3.23 & 26.09 $\pm$ 5.18 & 1260.01 $\pm$ 581.01 & 75.40 $\pm$ 4.20 &42.06&2.94 $\times$ 10$^{-7}$\\ 
140907A &1.21&80& 8.65 $\pm$ 0.36 & 2.48 $\pm$ 0.35 &  &  &49&1.43 $\times$ 10$^{-7}$\\ 
141026A &3.35&139& 15.48 $\pm$ 1.62 & 3.65 $\pm$ 1.55 &  &  &46.76&9.98 $\times$ 10$^{-9}$\\ 
150206A &2.087&75& 78.89 $\pm$ 2.13 & 47.92 $\pm$ 2.03 &  &  &24.11&1.94 $\times$ 10$^{-7}$\\ 
150323A &0.593&150& 2.54 $\pm$ 0.11 & 1.37 $\pm$ 0.07 & 6.23 $\pm$ 1.12 &  &83.78&6.87 $\times$ 10$^{-9}$\\ 
150424A &0.3&81.1& 0.16 $\pm$ 0.01 & 1.34 $\pm$ 0.05 &  &  &23.46&1.34 $\times$ 10$^{-7}$\\ 
150727A &0.313&88& 0.43 $\pm$ 0.02 & 0.12 $\pm$ 0.02 &  &  &8.9&9.13 $\times$ 10$^{-8}$\\ 
150818A &0.282&143& 0.41 $\pm$ 0.02 & 0.14 $\pm$ 0.02 &  &  &8.48&8.69 $\times$ 10$^{-9}$\\ 
150910A &1.36&112& 11.48 $\pm$ 0.85 & 1.99 $\pm$ 0.86 & 875.01 $\pm$ 210.01 & 15.50 $\pm$ 0.30 &54.09&2.86 $\times$ 10$^{-8}$\\ 
150915A &1.968&160& 3.54 $\pm$ 0.86 &0.48& 1.75 $\pm$ 1.43 & 45.20 $\pm$ 3.60 &22.14&5.00 $\times$ 10$^{-9}$\\ 
151021A &2.33&110& 177.54 $\pm$ 3.24 & 49.63 $\pm$ 4.89 &  &  &61.39&3.13 $\times$ 10$^{-8}$\\ 
151027A &0.81&130& 6.27 $\pm$ 0.19 & 5.01 $\pm$ 0.41 & 28.71 $\pm$ 2.04 & 3.40 $\pm$ 0.05 &40.46&1.39 $\times$ 10$^{-8}$\\ 
151027B &4.063&80& 24.06 $\pm$ 4.17 &2.46& 43.55 $\pm$ 10.63 & 50.70 $\pm$ 3.70 &19.08&1.43 $\times$ 10$^{-7}$\\ 
151111A &3.5&76&  &  &  &  &  &1.83 $\times$ 10$^{-7}$\\ 
160703A &1.5&45& 26.02 $\pm$ 0.48 & 12.56 $\pm$ 0.57 &  &  &22.1&1.98 $\times$ 10$^{-6}$\\ 
160804A &0.736&153& 8 $\pm$ 0.21 & 1.42 $\pm$ 0.18 &  &  &90.89&6.23 $\times$ 10$^{-9}$\\ 
161108A &0.5&116& 0.35 $\pm$ 0.04 & 0.11 $\pm$ 0.03 & 1.72 $\pm$ 0.32 & 1.44 $\pm$ 0.22 &203.58&2.42 $\times$ 10$^{-8}$\\ 
161117A &1.549&126&64.73& 14.18 $\pm$ 0.59 & 42.15 $\pm$ 11.73 & 16.40 $\pm$ 0.90 &148.12&1.61 $\times$ 10$^{-8}$\\ 
170202A &3.645&46.2&44.79&5.55& 47.71 $\pm$ 6.36 & 21.10 $\pm$ 1.50 &11.36&1.77 $\times$ 10$^{-6}$\\ 
170405A &3.51&165& 50.23 $\pm$ 3.56 &  &  &  &  &4.30 $\times$ 10$^{-9}$\\ 
180314A &1.445&50.5& 29.73 $\pm$ 1.08 & 16.02 $\pm$ 1.22 &  &  &20.15&1.19 $\times$ 10$^{-6}$\\ 
180325A &2.25&92.8& 38.93 $\pm$ 1.04 & 55.75 $\pm$ 2.10 &  & 156.00 $\pm$ 7.00 &98.26&7.09 $\times$ 10$^{-8}$\\ 
180510B &1.305&134& 4.71 $\pm$ 0.50 & 1.92 $\pm$ 0.45 &  &  &22.96&1.19 $\times$ 10$^{-8}$\\ 
180618A &1.2&47.4& 1.27 $\pm$ 0.20 & 3.85 $\pm$ 0.42 &  &  &11.22&1.58 $\times$ 10$^{-6}$\\ 
180720B &0.654&108& 46.83 $\pm$ 0.67 & 35.01 $\pm$ 1.44 &  &  &135.15&3.42 $\times$ 10$^{-8}$\\ 
181110A &1.505&138& 28.67 $\pm$ 0.77 & 6.98 $\pm$ 0.55 &  &  &139.49&1.03 $\times$ 10$^{-8}$\\ 
190106A &1.859&78.1& 25.34 $\pm$ 0.71 & 15.31 $\pm$ 0.84 &  & 72.90 $\pm$ 5.40 &24.67&1.61 $\times$ 10$^{-7}$\\ 
190114A &3.3765&67.1& 9.76 $\pm$ 1.40 & 5.01 $\pm$ 1.91 &  & 13.60 $\pm$ 3.20 &20.89&3.26 $\times$ 10$^{-7}$\\ 
190829A &0.0785&56.9& 0.04 $\pm$ 0 & 0.05 $\pm$ 0.01 & 649.85 $\pm$ 56.32 & 1.89 $\pm$ 0.03 &4.34&6.95 $\times$ 10$^{-7}$\\ 
191019A &0.248&64.3& 0.76 $\pm$ 0.02 & 0.27 $\pm$ 0.02 &  &  &19.61&3.97 $\times$ 10$^{-7}$\\ 
191221B &1.148&48& 33.24 $\pm$ 0.87 & 5.68 $\pm$ 0.94 &  &  &24.03&1.49 $\times$ 10$^{-6}$\\ 
201216C &1.1&48& 71.77 $\pm$ 1.19 & 19.57 $\pm$ 1.27 &  &  &61.83&1.49 $\times$ 10$^{-6}$\\ 
201221A &5.7&44.3& 52.88 $\pm$ 4.82 & 21.59 $\pm$ 6.73 & 9.91 $\pm$ 2.61 & 24.80 $\pm$ 2.90 &27.54&2.12 $\times$ 10$^{-6}$\\ 
210420B &1.4&158& 3.84 $\pm$ 0.48 & 0.75 $\pm$ 0.33 &  &  &30.99&5.32 $\times$ 10$^{-9}$\\ 
210504A &2.077&143& 14.48 $\pm$ 1.19 & 3.37 $\pm$ 1 &  &  &62.71&8.69 $\times$ 10$^{-9}$\\ 
210610B &1.13&69.4& 60.75 $\pm$ 0.84 & 24.16 $\pm$ 1.43 &  &  &64.67&2.79 $\times$ 10$^{-7}$\\ 
210619B &1.937&60.9& 442.24 $\pm$ 5.13 & 512.61 $\pm$ 9.32 &  &  &44.59&5.10 $\times$ 10$^{-7}$\\ 
220101A &4.61&162& 472.79 $\pm$ 7.98 & 124.28 $\pm$ 5.69 &  & 611.32 $\pm$ 11.27 &90.05&4.70 $\times$ 10$^{-9}$\\ 
220117A &4.961&50.6& 37.29 $\pm$ 4.47 & 29.92 $\pm$ 6.81 &  &  &40.87&1.18 $\times$ 10$^{-6}$\\ 
220611A &2.3608&57& 4.42 $\pm$ 0.88 & 2.48 $\pm$ 1.42 & 73.84 $\pm$ 21.81 &  &10.93&6.89 $\times$ 10$^{-7}$\\ 
230818A &2.42&64& 14.89 $\pm$ 1.32 & 1.77 $\pm$ 1.66 &  & 29.27 $\pm$ 2.96 &20.18&4.06 $\times$ 10$^{-7}$\\ 
\hline
\multicolumn{9}{|c|}{\large{\textbf{Silver Sample}}}\\[2ex]\hline 
050820A &2.6147&241& 30.31 $\pm$ 1.94 & 17.97 $\pm$ 2.06 & 70.00 $\pm$ 20.00 & 100.98 $\pm$ 11.31 & 80.67& 6.65 $\times$ 10$^{-10}$\\ 
050904 &6.295&182& 166.20 $\pm$ 6.38 & 17.16 $\pm$ 4.88 &  &  &46.45&2.65 $\times$ 10$^{-9}$\\ 
051001 &2.4296&190& 12.27 $\pm$ 1.01 & 2.47 $\pm$ 0.65 &  &  &34.74&2.14 $\times$ 10$^{-9}$\\ 
060202 &0.783&193& 1.78 $\pm$ 0.13 & 0.34 $\pm$ 0.11 &  &  &24.4&1.98 $\times$ 10$^{-9}$\\ 
060210 &3.9122&288& 118.40 $\pm$ 6.32 & 18.19 $\pm$ 4.03 &  &  &75.11&2.80 $\times$ 10$^{-10}$\\ 
060510B &4.941&263& 88.33 $\pm$ 3.66 & 10.01 $\pm$ 2.27 &  &  &51.6&4.34 $\times$ 10$^{-10}$\\ 
060526 &3.2213&298& 14.51 $\pm$ 1.86 & 15.52 $\pm$ 1.80 &  &  &177.27&2.37 $\times$ 10$^{-10}$\\ 
060904B &0.7029&190& 1.10 $\pm$ 0.10 & 1.35 $\pm$ 0.13 &  &  &38.27&2.14 $\times$ 10$^{-9}$\\ 
070129 &2.3384&460& 19.56 $\pm$ 1.73 & 2.31 $\pm$ 0.63 & 42.92 $\pm$ 12.30 & 0.816 $\pm$ 0.064 &67.94&3.16 $\times$ 10$^{-11}$\\ 
070306 &1.49594&209& 15.76 $\pm$ 0.82 & 8.74 $\pm$ 0.48 &  &  &138.05&1.34 $\times$ 10$^{-9}$\\ 
070612A &0.617&365& 5.31 $\pm$ 0.30 & 0.73 $\pm$ 0.16 &  &  &  &9.06 $\times$ 10$^{-11}$\\ 
070721B &3.6298&337& 48.66 $\pm$ 3.08 & 21.46 $\pm$ 3.42 &  &  &294.46&1.32 $\times$ 10$^{-10}$\\ 
071021 &2.452&229& 9.63 $\pm$ 1.20 & 3.65 $\pm$ 0.83 &  &  &48.67&8.55 $\times$ 10$^{-10}$\\ 
071025 &5.2&241& 175.60 $\pm$ 5.40 & 31.19 $\pm$ 3.96 &  &  &88.99&6.65 $\times$ 10$^{-10}$\\ 
071031 &2.6918&181& 7.47 $\pm$ 1.07 & 2.68 $\pm$ 0.95 &  &  &35.75&2.72 $\times$ 10$^{-9}$\\ 
080207 &2.0858&292& 34.08 $\pm$ 1.14 & 4.61 $\pm$ 1.36 &  &  &105.35&2.62 $\times$ 10$^{-10}$\\ 
080310 &2.42743&363& 16.08 $\pm$ 1.36 & 5.86 $\pm$ 0.92 & 19.67 $\pm$ 9.71 & 28.20 $\pm$ 1.80 &215.88&9.30 $\times$ 10$^{-11}$\\ 
080928 &1.6919&234& 8.83 $\pm$ 0.60 & 5.74 $\pm$ 0.47 &  &  &183.06&7.69 $\times$ 10$^{-10}$\\ 
081008 &1.9683&188& 20.39 $\pm$ 0.84 & 5.04 $\pm$ 0.64 &  &  &105.64&2.26 $\times$ 10$^{-9}$\\ 
081028A &3.038&284& 40.31 $\pm$ 1.94 & 4.17 $\pm$ 1.07 &  &  &49.46&2.99 $\times$ 10$^{-10}$\\ 
081029 &3.8479&275& 32.13 $\pm$ 3.68 & 4.59 $\pm$ 2.01 & 49.36 $\pm$ 3.46 & 32.70 $\pm$ 1.90 &109.74&3.50 $\times$ 10$^{-10}$\\ 
081109A &0.9787&221& 5.05 $\pm$ 0.25 & 1.27 $\pm$ 0.17 &  &  &97.96&1.02 $\times$ 10$^{-9}$\\ 
081203A &2.05&223& 40.42 $\pm$ 1.34 & 12.70 $\pm$ 0.98 &  &  &93.85&9.74 $\times$ 10$^{-10}$\\ 
090407 &1.4485&315& 3.09 $\pm$ 0.56 & 1.69 $\pm$ 0.38 & 690.01 $\pm$ 349.01 & 9.24 $\pm$ 0.49 &78.52&1.82 $\times$ 10$^{-10}$\\ 
090516A &4.109&181& 153.48 $\pm$ 11.25 & 19.83 $\pm$ 7.93 & 74.03 $\pm$ 25.59 & 66.60 $\pm$ 3.30 &135.15&2.72 $\times$ 10$^{-9}$\\ 
090715B &3&266& 56.46 $\pm$ 2.05 & 31.08 $\pm$ 1.70 &  &  &80.06&4.11 $\times$ 10$^{-10}$\\ 
100413A &3.9&191&93.89&1.04&  &  &79.08&2.09 $\times$ 10$^{-9}$\\ 
100704A &3.6&197.5&79.81&5.38&  &  &43.57&1.77 $\times$ 10$^{-9}$\\ 
100814A &1.44&177& 23.76 $\pm$ 0.54 & 7.46 $\pm$ 0.46 & 2366.10 $\pm$ 162.31 & 46.10 $\pm$ 1.30 &78.1&3.04 $\times$ 10$^{-9}$\\ 
100901A &1.4084&437& 5.02 $\pm$ 0.67 & 1.34 $\pm$ 0.46 & 7.97 $\pm$ 3.01 & 11.50 $\pm$ 0.60 &151.32&3.97 $\times$ 10$^{-11}$\\ 
100902A &4.5&428.8&61.01&1.32&  &  &42.06&4.32 $\times$ 10$^{-11}$\\ 
110205A &2.22&249& 94.01 $\pm$ 2.08 & 15.11 $\pm$ 0.89 &  &  &115.59&5.67 $\times$ 10$^{-10}$\\ 
110801A &1.858&385& 19.70 $\pm$ 1.14 & 2.84 $\pm$ 0.55 &  &  &225.22&7.07 $\times$ 10$^{-11}$\\ 
111123A &3.1516&290& 77.02 $\pm$ 2.77 & 7.95 $\pm$ 1.34 &  &  &136.34&2.71 $\times$ 10$^{-10}$\\ 
120624B &2.1974&180& 172.96 $\pm$ 2.34 & 30.56 $\pm$ 1.52 &  &  &  &2.80 $\times$ 10$^{-9}$\\ 
120909A &3.93&221& 116.46 $\pm$ 6.99 & 26.24 $\pm$ 3.88 &  &  &94.41&1.02 $\times$ 10$^{-9}$\\ 
120922A &3.1&168& 58.26 $\pm$ 2.53 & 11.19 $\pm$ 1.34 & 78.44 $\pm$ 38.26 & 17.90 $\pm$ 1.60 &42.11&3.93 $\times$ 10$^{-9}$\\ 
121211A &1.023&183& 1.75 $\pm$ 0.26 & 1.36 $\pm$ 0.39 &  &  &64.84&2.58 $\times$ 10$^{-9}$\\ 
130418A &1.218&275& 3.44 $\pm$ 0.36 & 0.65 $\pm$ 0.26 &  &  &94.02&3.50 $\times$ 10$^{-10}$\\ 
130427A &0.3399&244& 52.91 $\pm$ 0.57 & 53.62 $\pm$ 0.64 &  &  &135.43&6.26 $\times$ 10$^{-10}$\\ 
130514A &3.6&214& 121.40 $\pm$ 3.30 & 30.08 $\pm$ 2.90 &  &  &90.65&1.19 $\times$ 10$^{-9}$\\ 
130528A &1.25&640& 11.61 $\pm$ 0.65 & 0.33 $\pm$ 0.25 & 8.20 $\pm$ 2.40 & 4.63 $\pm$ 0.46 &94.67&7.95 $\times$ 10$^{-12}$\\ 
130606A &5.9134&277& 80.28 $\pm$ 6.11 & 68.07 $\pm$ 5.38 &  &  &214.69&3.38 $\times$ 10$^{-10}$\\ 
130907A &1.238&364& 272.88 $\pm$ 2.01 & 48.96 $\pm$ 0.90 &  &  &938.97&9.18 $\times$ 10$^{-11}$\\ 
140331A &4.65&210& 14.41 $\pm$ 2.85 & 3.64 $\pm$ 2.05 &  53.00 $\pm$ 18.00 & 33.41 $\pm$ 3.74  &52.70&1.31 $\times$ 10$^{-9}$\\ 
140430A &1.6&174& 3.70 $\pm$ 0.51 & 6.52 $\pm$ 0.54 & 17.31 $\pm$ 4.79 & 10.30 $\pm$ 0.70 &88.26&3.31 $\times$ 10$^{-9}$\\ 
141109A &2.993&200& 66.29 $\pm$ 2.49 & 20.90 $\pm$ 1.69 &  &  &49.81&1.66 $\times$ 10$^{-9}$\\ 
150413A &3.139&244& 46.01 $\pm$ 3.83 & 14.12 $\pm$ 2.59 &  &  &  &6.26 $\times$ 10$^{-10}$\\ 
150821A &0.755&169& 0.63 $\pm$ 0.15 &  &  &  &  &3.82 $\times$ 10$^{-9}$\\ 
160131A &0.972&328& 25.01 $\pm$ 0.62 & 7.22 $\pm$ 0.45 &  &  &196.96&1.50 $\times$ 10$^{-10}$\\ 
160227A &2.38&316& 22.46 $\pm$ 1.48 & 3.85 $\pm$ 0.69 & 12.76 $\pm$ 2.62 & 37.40 $\pm$ 2.40 &56.64&1.79 $\times$ 10$^{-10}$\\ 
160425A &0.555&305& 0.79 $\pm$ 0.08 & 0.83 $\pm$ 0.08 &  &  &42.27&2.12 $\times$ 10$^{-10}$\\ 
161017A &2.013&217& 26.83 $\pm$ 0.90 & 12.44 $\pm$ 0.82 &  &  &135.28&1.11 $\times$ 10$^{-9}$\\ 
170519A &0.818&220& 0.98 $\pm$ 0.15 & 0.48 $\pm$ 0.11 & 31.34 $\pm$ 7.51 & 2.19 $\pm$ 0.13 &83.05&1.04 $\times$ 10$^{-9}$\\ 
170531B &2.366&170& 12.78 $\pm$ 1.16 & 3.43 $\pm$ 0.97 &  &  &51.36&3.71 $\times$ 10$^{-9}$\\ 
170607A &0.557&320& 3.04 $\pm$ 0.13 & 0.76 $\pm$ 0.08 &  4.40 $\pm$ 1.20 & 2.59 $\pm$ 0.11  &49.29&1.69 $\times$ 10$^{-10}$\\ 
170705A &2.01&223& 45.94 $\pm$ 1.30 & 57.30 $\pm$ 1.99 & 7.79 $\pm$ 1.31 & 64.80 $\pm$ 2.50 &240.34&9.74 $\times$ 10$^{-10}$\\ 
171205A &0.037&190& 0.01 $\pm$ 0 & 0 $\pm$ 0 & 1.10 $\pm$ 0.21 & (1.65$\pm$0.11)$\times$10$^{-3}$ &  &2.14 $\times$ 10$^{-9}$\\ 
171222A &2.409&174& 12.91 $\pm$ 1.34 & 2.69 $\pm$ 1.06 & 27.18 $\pm$ 5.89 &  &64.93&3.31 $\times$ 10$^{-9}$\\ 
180329B &1.998&214& 17.10 $\pm$ 1.40 & 4.48 $\pm$ 1.65 & 0.30 $\pm$ 0.12 & 9.74 $\pm$ 0.72 &140.13&1.19 $\times$ 10$^{-9}$\\ 
180620B &1.1175&224& 16.74 $\pm$ 0.40 & 5.40 $\pm$ 0.35 & 68.00 $\pm$ 22.00 & 30.38 $\pm$ 3.01 &75.70&9.53 $\times$ 10$^{-10}$\\ 
180624A &2.855&467& 50.75 $\pm$ 2.80 & 8.24 $\pm$ 1.65 &  &  &152.25&2.95 $\times$ 10$^{-11}$\\ 
181020A &2.938&238& 77.96 $\pm$ 2.65 & 73.33 $\pm$ 3.38 &  & 292.00 $\pm$ 3.00 &318.97&7.08 $\times$ 10$^{-10}$\\ 
190114C &0.42&361& 18.13 $\pm$ 0.21 & 21.90 $\pm$ 0.35 &  &  &240.52&9.54 $\times$ 10$^{-11}$\\ 
190719C &2.469&186& 36.49 $\pm$ 1.90 & 31.40 $\pm$ 2.01 & 84.00 $\pm$ 23.00 & 75.91 $\pm$ 7.91 &128.23&2.38 $\times$ 10$^{-9}$\\ 
191004B &3.503&300& 38.84 $\pm$ 3.48 & 57.55 $\pm$ 2.97 &  &  &190.11&2.30 $\times$ 10$^{-10}$\\ 
200205B &1.465&454& 19.75 $\pm$ 0.78 & 4.38 $\pm$ 0.39 &  &  &281.76&3.35 $\times$ 10$^{-11}$\\ 
210702A &1.1757&219& 17.87 $\pm$ 0.97 &  &  &  &  &1.06 $\times$ 10$^{-9}$\\ 
210822A &1.736&186& 75.11 $\pm$ 1.53 & 99.52 $\pm$ 2.67 &  &  &401.21&2.38 $\times$ 10$^{-9}$\\ 
211024B &1.1137&600& 11.22 $\pm$ 0.56 & 1.23 $\pm$ 0.24 &  &  &268.44&1.03 $\times$ 10$^{-11}$\\ 
\hline
\multicolumn{9}{|c|}{\large{\textbf{Gold Sample}}}\\[2ex]\hline
060124A &2.29&750& 4.36 $\pm$ 0.79 & 6.73 $\pm$ 2.59 &  &  &281.16&4.42 $\times$ 10$^{-12}$\\ 
060218A &0.034&2100& 0.003 $\pm$ 0.0003 & 0.47 $\pm$ 0.96 & 0.47 $\pm$ 0.096 & 2.57 $\pm$ 0.01 &210.6&1.09 $\times$ 10$^{-12}$\\ 
070419B &1.9588&4930& 57.61 $\pm$ 2.38 & 11.38 $\pm$ 2.69 &  56.00 $\pm$ 14.00 & 17.19 $\pm$ 0.85  &3018.07&4.43 $\times$ 10$^{-10}$\\ 
070518A &1.16&57900& 0.40 $\pm$ 0.09 & 0.98 $\pm$ 0.35 &  &  &6413.03& \\ 
080319B &0.938&1340& 238.60 $\pm$ 4.05 & 68.57 $\pm$ 2.43 &  &  &2119.36&1.02 $\times$ 10$^{-12}$\\ 
090404A &2.87&44700& 35.90 $\pm$ 1.90 & 18.27 $\pm$ 2.78 &  &  &21417.06& \\ 
090417B &0.345&2130& 0.48 $\pm$ 0.06 & 0.04 $\pm$ 0.03 &  &  &117.79&1.12 $\times$ 10$^{-12}$\\ 
091024A &1.09&1300& 20.00 $\pm$ 1.64 & 6.60 $\pm$ 1.64 & 21.00 &  &714.02&1.06 $\times$ 10$^{-12}$\\ 
091127A &0.49&5398& 3.53 $\pm$ 0.20 & 13.47 $\pm$ 1.16 &  &  &4372.42&1.74 $\times$ 10$^{-9}$\\ 
100316D &0.059&1300& 0.004 $\pm$ 0.001 & 0.003 $\pm$ 0.0004 &0.004& 0.002 & 36.43 & 1.06 $\times$ 10$^{-12}$\\ 
100728A &1.57&1460& 267.78 $\pm$ 5.53 & 51.85 $\pm$ 4.58 &  &  &1039.18&9.34 $\times$ 10$^{-13}$\\ 
101225A &0.85&6420& 2.94 $\pm$ 0.98 &  &64.5&  &  & \\ 
111209A &0.677&18200& 38.79 $\pm$ 0.95 & 0.72 $\pm$ 0.44 & 96.00 & 50.80 $\pm$ 0.60 &2707.35& \\ 
111215A &2.1&1120& 24.36 $\pm$ 5.36 & 4.08 $\pm$ 3.34 &  &  &214.15&1.39 $\times$ 10$^{-12}$\\ 
121027A &1.773&5730& 12.96 $\pm$ 1.26 & 5.81 $\pm$ 1.48 & 90.64 $\pm$ 22.97 & 207.00 $\pm$ 8.00 &2104.28&4.77 $\times$ 10$^{-9}$\\ 
121217A &3.1&778& 112.93 $\pm$ 9.45 & 25.44 $\pm$ 4.06 &  &  &627.18&3.90 $\times$ 10$^{-12}$\\ 
130925A & 0.348 & 4500 & 6.01 $\pm$ 0.19 & 0.99 $\pm$ 0.14& 0.50 & & 1206.36 & 1.35 $\times$ 10$^{-10}$\\ 
140614A &4.233&720& 19.42 $\pm$ 3.80 & 10.33 $\pm$ 14.65 & 13.78 $\pm$ 4.75 & 10.00 $\pm$ 1.30 &408.2&5.11 $\times$ 10$^{-12}$\\ 
141121A & 1.47 & 1410 & 12.06 $\pm$ 2.67 & 2.36 $\pm$ 1.35 & 100.00 & 13.80 $\pm$ 0.86 & 343.57 & 1.03 $\times$ 10$^{-12}$\\ 
170714A &0.793&1000& 3.78 $\pm$ 0.67 & 0.24 $\pm$ 0.48 &  &  &114.28&1.82 $\times$ 10$^{-12}$\\ 
210905A &6.32&778& 241.61 $\pm$ 51.34 & 140.22 $\pm$ 45.56 &  &  &619.7&3.90 $\times$ 10$^{-12}$\\ 
221009A &0.151&1100& 1000 $\pm$ 7.00 & 991 $\pm$ 6.00 & 541.00 $\pm$ 157.97 &  &4270.76&1.45 $\times$ 10$^{-12}$\\  

\hline
\end{longtable}
\FloatBarrier
\begin{table*}
\caption{Characteristics of a sample of well-studied GRBs included in our diamond sub-sample obtained from the various published papers with corresponding references in the last column.}
\label{tab:lit_ulGRBs}
\centering
\begin{tabular}{|c|c|c|c|c|c|c|c|} \hline
\bf GRB & \bf $z$ & \bf T$_{90}^{*}\footnote{Durations given in references}$ (s) & \bf E$_{\rm \gamma,iso}$ (erg) & \bf E$_{\rm K,iso}$ (erg) & \bf E$_{\rm X,iso}$ (erg) & \bf E$_{\rm pt}$ (KeV) & References \\ \hline
090309A & & 5276 &  &  &  &  & \cite{2016ApJ...829....7L}\\ 
101024A & & 4883 &  &  &  & & \cite{2016ApJ...829....7L}\\ 
110709B & & 900 &  &  &  & & \cite{2013ApJ...778...54V}\\ 
220627A & 3.08 & 1092 & 4.81 $\pm$ 0.02 $\times$ 10$^{54}$ & 9.01 $\pm$ 7.41 $\times$ 10$^{53}$ & 2.00 $\times$ 10$^{52}$ & 205 $\pm$ 109 & \cite{2023arXiv230710339D} \\ \hline
\end{tabular}
\end{table*}

\vspace{-1cm}
\begin{table*}
\centering
\caption{Parameters obtained from the spectral fitting of prompt emission of GRB 221009A with Band and Band + {Blackbody} function.}
\label{tab:Prompt}
\begin{center}
\begin{tabular}{|c|c|c|c|c|c|c|c|c|}\hline
\boldmath T$_{\rm start}$ & \boldmath T$_{\rm end}$ &\boldmath $\alpha_{\rm pt}$ & \boldmath E$_{\rm pt}$ &\boldmath $\beta_{\rm pt}$ & \bf kT & \bf Photon flux & \bf Energy Flux & \bf BIC \\ \hline
\bf (s) &\bf (s) & - & \bf (keV) & - &\bf (KeV)& \bf (photon s$^{-1}$ cm$^{-2}$) & \bf (erg s$^{-1}$ cm$^{-2}$) & - \\ \hline
0 & 10 & $-1.724_{-0.023}^{+0.023}$ & $1241.579_{-488.208}^{+1213.568}$ & $-7.422_{-0.000}^{+5.380}$ & - & 11.502 & 1.90 $\times$ $10^{-6}$ & 641.84 \\
10 & 20 & $-1.706_{-0.059}^{+0.064}$ & $158.330_{-57.429}^{+129.004}$ & $-6.115_{-0.000}^{+0.000}$ & - & 3.244 & 2.76 $\times$ $10^{-7}$ & 325.05 \\
20 & 40 & $-1.711_{-0.046}^{+0.050}$ & $93.057_{-26.391}^{+40.965}$ & $-10.000_{-0.000}^{+0.000}$ & - & 2.656 & 1.88 $\times$ $10^{-7}$ & 422.71 \\
40 & 120 & $-1.697_{-0.037}^{+0.040}$ & $41.305_{-8.489}^{+11.102}$ & $-6.100_{-0.000}^{+3.743}$ & - & 1.218 & 6.51 $\times$ $10^{-8}$ & 623.04 \\
120 & 175 & $-1.621_{-0.030}^{+0.031}$ & $52.615_{-7.467}^{+9.056}$ & $-7.359_{-0.000}^{+4.865}$ & - & 2.062 & 1.17 $\times$ $10^{-7}$ & 688.46 \\
175 & 210 & $-1.119_{-0.005}^{+0.005}$ & $565.048_{-5.555}^{+5.546}$ & $-2.337_{-0.033}^{+0.030}$ & - & 86.41 & 2.58 $\times$ $10^{-5}$ & 17020.85 \\
210 & 215 & $-1.133_{-0.009}^{+0.009}$ & $259.168_{-4.439}^{+4.513}$ & $-2.255_{-0.044}^{+0.044}$ & - & 109.696 & 2.20 $\times$ $10^{-5}$ & 4624.26 \\
280 & 300 & $-1.465_{-0.002}^{+0.002}$ & $538.138_{-3.667}^{+8.768}$ & $-2.209_{-0.015}^{+0.015}$ & - & 384.494 & 7.88 $\times$ $10^{-5}$ & 79848.27 \\
300 & 330 & $-1.504_{-0.002}^{+0.003}$ & $47.966_{-0.699}^{+0.412}$ & $-2.024_{-0.006}^{+0.006}$ & - & 111.441 & 1.33 $\times$ $10^{-5}$ & 34106.81 \\
330 & 380 & $-1.281_{-0.002}^{+0.002}$ & $41.923_{-0.252}^{+0.243}$ & $-2.224_{-0.008}^{+0.008}$ & - & 88.713 & 7.86 $\times$ $10^{-6}$ & 50251.11 \\
380 & 450 & $-1.520_{-0.002}^{+0.002}$ & $169.723_{-1.525}^{+1.591}$ & $-2.220_{-0.014}^{+0.015}$ & - & 115.398 & 1.50 $\times$ $10^{-5}$ & 92334.54 \\
450 & 465 & $-1.410_{-0.003}^{+0.003}$ & $188.551_{-2.294}^{+2.228}$ & $-2.229_{-0.020}^{+0.020}$ & - & 194.609 & 2.80 $\times$ $10^{-5}$ & 35179.5 \\
465 & 480 & $-1.530_{-0.003}^{+0.003}$ & $317.729_{-3.869}^{+6.432}$ & $-3.386_{-0.040}^{+1.109}$ & - & 212.591 & 2.63 $\times$ $10^{-5}$ & 42367.11 \\
480 & 500 & $-1.396_{-0.003}^{+0.003}$ & $165.436_{-1.741}^{+1.529}$ & $-2.247_{-0.018}^{+0.017}$ & - & 187.32 & 2.55 $\times$ $10^{-5}$ & 48050.02 \\
520 & 555 & $-1.414_{-0.002}^{+0.002}$ & $321.032_{-2.945}^{+2.879}$ & $-2.150_{-0.013}^{+0.013}$ & - & 217.701 & 4.05 $\times$ $10^{-5}$ & 99834.32 \\
555 & 600 & $-1.479_{-0.002}^{+0.002}$ & $125.181_{-1.109}^{+0.854}$ & $-2.279_{-0.012}^{+0.015}$ & - & 154.187 & 1.76 $\times$ $10^{-5}$ & 111863.65 \\
170 & 600 & $-1.287_{-0.000}^{+0.000}$ & $2367.750_{-6.056}^{+7.242}$ & $-2.423_{-0.006}^{+0.007}$ & - & 314.243 & 1.52 $\times$ $10^{-4}$ & 1397389.35 \\
\hline

\boldmath T$_{\rm start}$ & \boldmath T$_{\rm end}$ &\boldmath $\alpha_{\rm pt}$ &\boldmath E$_{\rm pt}$ & \boldmath $\beta_{\rm pt}$ & \bf kT & \bf Photon flux & \bf Energy Flux & \bf BIC \\ \hline
\bf (s) &\bf (s) & - & \bf (keV) & - &\bf (KeV)& \bf (photon s$^{-1}$ cm$^{-2}$) & \bf (erg s$^{-1}$ cm$^{-2}$) & -\\ \hline
0 & 10 & $-1.724_{-0.027}^{+0.027}$ & $1240.536_{-540.508}^{+1579.592}$ & $-7.957_{-0.000}^{+0.000}$ & $11.309_{-0.000}^{+11.309}$ & 11.502 & 1.90 $\times$ $10^{-6}$ & 653.65 \\
10 & 20 & $-1.839_{-0.000}^{+0.085}$ & $316.852_{-231.345}^{+19402.886}$ & $-5.199_{-0.000}^{+0.000}$ & $29.476_{-0.000}^{+29.476}$ & 3.28 & 3.39 $\times$ $10^{-7}$ & 337.78 \\
20 & 40 & $-1.900_{-0.000}^{+0.056}$ & $68.317_{-36.847}^{+269.097}$ & $-9.614_{-0.000}^{+0.000}$ & $22.339_{-5.423}^{+22.339}$ & 2.66 & 1.95 $\times$ $10^{-7}$ & 433.12 \\
40 & 120 & $-1.699_{-0.043}^{+0.047}$ & $41.067_{-9.605}^{+13.423}$ & $-10.000_{-0.000}^{+7.725}$ & $7.090_{-0.000}^{+7.090}$ & 1.218 & 6.49 $\times$ $10^{-8}$ & 634.84 \\
120 & 175 & $-0.516_{-0.000}^{+0.000}$ & $499.964_{-0.000}^{+0.000}$ & $-2.500_{-0.000}^{+0.000}$ & $29.418_{-2.820}^{+29.418}$ & 0.693 & 9.00 $\times$ $10^{-8}$ & 1359.53 \\
175 & 210 & $-1.239_{-0.005}^{+0.005}$ & $950.288_{-17.305}^{+17.382}$ & $-2.679_{-0.097}^{+0.085}$ & $38.465_{-0.444}^{+38.465}$ & 86.881 & 2.63 $\times$ $10^{-5}$ & 16741.2 \\
210 & 215 & $-1.134_{-0.010}^{+0.010}$ & $259.616_{-5.128}^{+5.397}$ & $-2.251_{-0.056}^{+0.048}$ & $17.538_{-0.000}^{+17.538}$ & 109.654 & 2.21 $\times$ $10^{-5}$ & 4636.08 \\
280 & 300 & $-1.471_{-0.003}^{+0.002}$ & $563.222_{-6.983}^{+7.250}$ & $-2.228_{-0.019}^{+0.019}$ & $25.975_{-0.000}^{+25.975}$ & 384.667 & 7.88 $\times$ $10^{-5}$ & 79852.25 \\
300 & 330 & $-1.570_{-0.004}^{+0.004}$ & $125.862_{-3.415}^{+2.032}$ & $-2.079_{-0.018}^{+0.017}$ & $5.672_{-0.040}^{+5.672}$ & 110.641 & 1.27 $\times$ $10^{-5}$ & 33905.29 \\
330 & 380 & $-1.151_{-0.003}^{+0.003}$ & $67.959_{-0.546}^{+0.448}$ & $-2.318_{-0.016}^{+0.015}$ & $4.510_{-0.022}^{+4.510}$ & 88.57 & 7.43 $\times$ $10^{-6}$ & 49888.79 \\
380 & 450 & $-1.520_{-0.002}^{+0.002}$ & $170.098_{-1.823}^{+1.832}$ & $-2.222_{-0.017}^{+0.018}$ & $25.866_{-0.000}^{+25.866}$ & 115.403 & 1.49 $\times$ $10^{-5}$ & 92346.29 \\
450 & 465 & $-1.420_{-0.004}^{+0.004}$ & $195.499_{-3.520}^{+2.237}$ & $-2.244_{-0.028}^{+0.023}$ & $25.368_{-0.000}^{+25.368}$ & 194.662 & 2.79 $\times$ $10^{-5}$ & 35194.32 \\
465 & 480 & $-1.207_{-0.005}^{+0.005}$ & $214.797_{-2.206}^{+2.319}$ & $-2.239_{-0.023}^{+0.022}$ & $6.342_{-0.040}^{+6.342}$ & 209.6 & 3.11 $\times$ $10^{-5}$ & 41367.58 \\
480 & 500 & $-1.090_{-0.005}^{+0.005}$ & $166.160_{-1.315}^{+1.363}$ & $-2.262_{-0.018}^{+0.019}$ & $5.969_{-0.033}^{+5.969}$ & 185.27 & 2.53 $\times$ $10^{-5}$ & 47310.49 \\
520 & 555 & $1.287_{-0.004}^{+0.004}$ & $110.254_{-0.158}^{+0.151}$ & $-2.000_{-0.001}^{+0.000}$ & $5.941_{-0.009}^{+5.941}$ & 214.325 & 4.29 $\times$ $10^{-5}$ & 101515.36 \\
555 & 600 & $-1.135_{-0.003}^{+0.003}$ & $133.009_{-0.805}^{+0.811}$ & $-2.303_{-0.014}^{+0.015}$ & $5.301_{-0.019}^{+5.301}$ & 152.857 & 1.74 $\times$ $10^{-5}$ & 110518.52 \\
170 & 600 & $-1.216_{-0.001}^{+0.001}$ & $1975.222_{-4.828}^{+6.313}$ & $-2.342_{-0.006}^{+0.005}$ & $6.910_{-0.015}^{+6.910}$ & 313.455 & 1.49 $\times$ $10^{-4}$ & 1387598.57 \\
\hline
\end{tabular}
\end{center}
\end{table*}

\begin{table}
\centering
\caption{Parameters obtained from the spectral fitting of non-saturated precursor emission of GRB 221009A with \sw{Band}, \sw{Band} + \sw{Blackbody}, physical \sw{synchrotron} model.}
\label{Precursor}
\begin{center}
\begin{tabular}{|c|c|c|c|c|c|c|c|}\hline
\multicolumn{8}{|c|}{\sw{Band}}\\[2ex]\hline
\boldmath T$_{\rm start} (s)$ & \boldmath T$_{\rm end} (s)$ & \boldmath $\alpha_{\rm pt}$ &\bf \boldmath E$_{\rm pt}$ (keV) & \boldmath $\beta_{\rm pt}$ & - & \bf Flux \bf (erg s$^{-1}$ cm$^{-2}$) & \bf BIC \\ \hline 
0 & 10 & -1.661$_{-0.016}^{+0.016}$ & 1531.055$_{-284.783}^{+292.657}$ & -3.093$_{-0.501}^{+0.502}$ & - & 2.48$_{-0.13}^{+0.12}$ $\times$ $10^{-6}$ & 2697 \\
10 & 20 &-1.701$_{-0.071}^{+0.069}$ & 838.774$_{-476.816}^{+473.362}$ & -3.001$_{-0.558}^{+0.566}$ & - & 3.07$_{-0.58}^{+0.51}$ $\times$ $10^{-7}$ & 2617 \\ \hline 

\multicolumn{8}{|c|}{\sw{Band} + \sw{Blackbody}}\\ [2ex]\hline
\boldmath T$_{\rm start} (s)$ & \boldmath T$_{\rm end} (s)$ &\boldmath $\alpha_{\rm pt}$ &\bf \boldmath E$_{\rm pt}$ (keV) &\boldmath $\beta_{\rm pt}$ & \bf kT (KeV) & \bf Flux \bf (erg s$^{-1}$ cm$^{-2}$) & \bf BIC  \\ \hline 
0 & 10 & -1.555$_{-0.037}^{+0.039}$ & 1205.956$_{-250.925}^{+252.291}$ & -3.009$_{-0.414}^{+0.388}$ & 3.761$_{-0.589}^{+0.271}$ & 2.46$_{-0.23}^{+0.28}$ $\times$ $10^{-6}$ & 2685\\ 
10 & 20 &-1.758$_{-0.087}^{+0.089}$ & 931.427$_{-433.363}^{+437.622}$ & -2.878$_{-0.573}^{+0.539}$ & 15.072$_{-8.888}^{+8.323}$ & 2.98$_{-0.42}^{+0.49}$ $\times$ $10^{-7}$ & 2595  \\ \hline 

\multicolumn{8}{|c|}{\sw{Cutoff powerlaw}}\\ [2ex]\hline
\boldmath T$_{\rm start} (s)$ & \boldmath T$_{\rm end} (s)$ &\boldmath $\alpha_{\rm pt}$ &\bf \boldmath E$_{\rm pt}$ (KeV)& - & - & \bf Flux \bf (erg s$^{-1}$ cm$^{-2}$) & \bf BIC \\ \hline 
0 & 10 & -1.598$_{-0.017}^{0.0168}$ & 1309.932$_{-130.35}^{130.161}$ & - & - & 1.80$_{-0.22}^{0.22}$ $\times$ $10^{-6}$ & 2746\\ 
10 & 20 & -1.641$_{-0.098}^{0.098}$ & 572.897$_{-171.438}^{170.357}$ & - & - & 1.91$_{-2.48}^{0.89}$ $\times$ $10^{-7}$ & 2621\\ \hline 
\multicolumn{8}{|c|}{\sw{Cutoff powerlaw} + \sw{Blackbody}}\\ [2ex]\hline
\boldmath T$_{\rm start} (s)$ & \boldmath T$_{\rm end} (s)$ &\boldmath $\alpha_{\rm pt}$ & \bf \boldmath E$_{\rm pt}$ (KeV)& - &\bf kT (KeV)& \bf Flux \bf (erg s$^{-1}$ cm$^{-2}$) & \bf BIC \\ \hline
0 & 10 & -1.596$_{-0.016}^{0.016}$ & 1299.694$_{-125.262}^{125.674}$ & - & 10.804$_{-7.276}^{7.889}$ & 1.80$_{-0.21}^{0.25}$ $\times$ $10^{-6}$ & 2722\\ 
10 & 20 & -1.627$_{-0.098}^{0.097}$ & 564.421$_{-171.011}^{173.628}$ & - & 11.431$_{-7.734}^{8.051}$ & 1.99$_{-0.94}^{2.78}$ $\times$ $10^{-7}$ & 2599\\ \hline 

\multicolumn{8}{|c|}{\sw{Synchrotron}}\\ [2ex]\hline
\boldmath T$_{\rm start} (s)$ & \boldmath T$_{\rm end} (s)$ & \bf B (G) & \boldmath $p$ &\boldmath $\gamma_{cool}$ & - & \bf Flux \bf (erg s$^{-1}$ cm$^{-2}$) & \bf BIC\\  \hline 
0 & 10 & 0.343$_{-0.161}^{+0.169}$ & 1.681$_{-0.611}^{+0.629}$ & 34957576.163$_{-34813370.352}^{+48641616.967}$ & - & 2.66$_{-0.12}^{+0.13}$ $\times$ $10^{-6}$ & 2718 \\ 
10 & 20 &0.924$_{-0.902}^{+0.043}$ & 1.703$_{-0.479}^{+0.493}$ & 7294590.381$_{-7099455.667}^{+3587786.823}$ & - & 2.58$_{-0.47}^{+0.61}$ $\times$ $10^{-7}$ & 2626 \\ \hline 
\end{tabular}
\end{center}
\end{table}

\begin{table}
\centering
\caption{Our NIR observations of GRB 221009A using 3.6m DOT at ARIES. Magnitude is not corrected for Galactic extinction.}
\label{tab:DOT_obs}
\begin{tabular}{|c|c|c|c|c|c|c|} \hline
~~\bf T-T$_{0}$ (days) ~~~~ & ~~~~\bf Exposure (s) ~~~~ & ~~ \bf Filter ~~ & ~~~~\bf Magnitude ~~~~ &\bf Magnitude error & ~~~~ \bf Telescope ~~~~ \\ \hline
7.05882 & 200 $\times$ 10 & R & 21.3 & 0.04 & DFOT \\ 
13.5428 & 40 $\times$ 50 & J & 19.70 & 0.05 & DOT \\ 
14.5271 & 40 $\times$ 50 & J & 20.17 & 0.05 & DOT \\ 
14.5700 & 40 $\times$ 50 & H & 19.66 & 0.09 & DOT \\ 
16.5497 & 40 $\times$ 50 & H & 19.65 & 0.05 & DOT \\ 
13.6084 & 20 $\times$ 100 & K & 18.92 & 0.05 & DOT \\ 
14.6148 & 20 $\times$ 100 & K & 18.81 & 0.08 & DOT \\ 
15.6124 & 20 $\times$ 100 & K & 18.96 & 0.08 & DOT \\ 
16.6022 & 20 $\times$ 100 & K & 19.05 & 0.09 & DOT \\ \hline
\end{tabular}
\end{table}

\begin{table*}
\addtolength{\tabcolsep}{6pt}
\centering
\caption{ Prompt emission spectral characteristic of ULGRBs obtained from the fitting of \fermi-GBM observation.}
\label{tab:gbm_ulGRBs}
\begin{tabular}{|c|c|c|c|c|c|c|c|}\hline
\bf \fermi ID & \bf $z$ & \bf \boldmath T$_{90}$ & \bf LAT & \boldmath $\alpha_{\rm pt}$ &\bf \boldmath E$_{\rm pt}$ & \boldmath $\beta_{\rm pt}$ &\bf \boldmath E$_{\rm \gamma,iso}$ \\
 &  &  & \bf boresight &  &  & &  \\
 &  & \bf (s) & \bf (deg) &  & \bf \boldmath (\keV) & & \bf (erg) \\ \hline
080810549 & 3.36 & 107.67 & 60 &  {-0.851}$^{+0.089}_{-0.091}$  &  {305.672}$^{+36.751}_{-36.225}$  &  {-3.101}$^{+0.507}_{-0.527}$  &  {62.69}$^{+14.21}_{-11.17}$ $\times$ 10$^{52}$ \\ 
080905705 & 0.12 & 128 & 99 &  {-1.345}$^{+0.161}_{-0.166}$  &  {223.605}$^{+69.603}_{-68.397}$  &  {-2.856}$^{+0.639}_{-0.651}$  &  {0.07}$^{+0.01}_{-0.01}$ $\times$ 10$^{52}$ \\ 
080916406 & 0.69 & 61.35 & 76 &  {-0.981}$^{+0.074}_{-0.073}$  &  {120.847}$^{+9.085}_{-9.144}$  &  {-3.281}$^{+0.414}_{-0.432}$  &  {1.27}$^{+0.27}_{-0.21}$ $\times$ 10$^{52}$ \\ 
081008832 & 1.97 & 185.5 & - &  {-0.238}$^{+0.269}_{-0.264}$  &  {77.703}$^{+6.609}_{-6.749}$  &  {-3.353}$^{+0.406}_{-0.408}$  &  {4.33}$^{+3.66}_{-2.07}$ $\times$ 10$^{52}$ \\ 
090519881 & 3.9 & 64 & 47 &  {-0.361}$^{+0.313}_{-0.314}$  &  {208.829}$^{+43.924}_{-44.034}$  &  {-3.161}$^{+0.511}_{-0.517}$  &  {6.61}$^{+5.2}_{-2.86}$ $\times$ 10$^{52}$ \\ 
090618353 & 0.54 & 113.34 & 133 &  {-0.509}$^{+0.061}_{-0.061}$  &  {94.387}$^{+2.922}_{-2.931}$  &  {-2.123}$^{+0.015}_{-0.015}$  &  {25.06}$^{+3.73}_{-3.18}$ $\times$ 10$^{52}$ \\ 
090926914 & 2.11 & 81 & 100 &  {-0.131}$^{+0.121}_{-0.118}$  &  {89.572}$^{+3.921}_{-3.899}$  &  {-3.512}$^{+0.331}_{-0.332}$  &  {2.14}$^{+0.71}_{-0.52}$ $\times$ 10$^{52}$ \\ 
091024372 & 1.09 & 1020 & 98 &  {-0.842}$^{+0.077}_{-0.076}$  &  {337.521}$^{+40.266}_{-40.672}$  &  {-3.166}$^{+0.481}_{-0.493}$  &  {80.97}$^{+18.13}_{-13.92}$ $\times$ 10$^{52}$ \\ 
100413732 & 3.9 & 184.06 & 84 &  {-0.077}$^{+0.239}_{-0.241}$  &  {349.718}$^{+51.801}_{-51.415}$  &  {-2.917}$^{+0.571}_{-0.609}$  &  {76.52}$^{+38.93}_{-23.85}$ $\times$ 10$^{52}$ \\ 
100704149 & 3.6 & 197.5 & 64 &  {-0.489}$^{+0.144}_{-0.146}$  &  {175.796}$^{+16.955}_{-17.286}$  &  {-3.397}$^{+0.396}_{-0.401}$  &  {73.84}$^{+23.72}_{-17.3}$ $\times$ 10$^{52}$ \\ 
100814160 & 1.44 & 174.5 & 87 &  {-0.643}$^{+0.118}_{-0.118}$  &  {120.259}$^{+9.728}_{-9.751}$  &  {-3.486}$^{+0.358}_{-0.354}$  &  {7.77}$^{+2.32}_{-1.79}$ $\times$ 10$^{52}$ \\ 
110818860 & 3.36 & 103 & 95 &  {-1.213}$^{+0.091}_{-0.091}$  &  {200.447}$^{+31.341}_{-30.984}$  &  {-2.936}$^{+0.553}_{-0.586}$  &  {28.27}$^{+8.53}_{-5.86}$ $\times$ 10$^{52}$ \\ 
111228657 & 0.72 & 101.2 & 68 &  {-1.369}$^{+0.381}_{-0.331}$  &  {26.903}$^{+5.182}_{-5.341}$  &  {-2.633}$^{+0.328}_{-0.419}$  &  {1.45}$^{+3.76}_{-1.02}$ $\times$ 10$^{52}$ \\ 
120922939 & 3.11 & 173 & 85 &  {-0.821}$^{+0.419}_{-0.428}$  &  {201.682}$^{+50.354}_{-49.392}$  &  {-2.719}$^{+0.481}_{-0.491}$  &  {34.01}$^{+4.04}_{-3.85}$ $\times$ 10$^{52}$ \\ 
130420313 & 1.29 & 123.5 & 134 &  {-0.808}$^{+0.361}_{-0.353}$  &  {56.05}$^{+10.734}_{-10.652}$  &  {-3.233}$^{+0.456}_{-0.474}$  &  {7.91}$^{+4.77}_{-2.21}$ $\times$ 10$^{52}$ \\ 
131105087 & 1.68 & 112.3 & 36 &  {-1.201}$^{+0.041}_{-0.041}$  &  {203.115}$^{+17.389}_{-17.546}$  &  {-3.227}$^{+0.428}_{-0.453}$  &  {14.16}$^{+1.92}_{-1.54}$ $\times$ 10$^{52}$ \\ 
140506880 & 0.89 & 111.1 & 142 &  {-1.184}$^{+0.252}_{-0.251}$  &  {193.196}$^{+63.929}_{-62.518}$  &  {-2.969}$^{+0.584}_{-0.598}$  &  {1.69}$^{+1.22}_{-0.65}$ $\times$ 10$^{52}$ \\ 
140512814 & 0.72 & 154.8 & - &  {-1.261}$^{+0.037}_{-0.038}$  &  {392.697}$^{+48.378}_{-49.035}$  &  {-3.218}$^{+0.465}_{-0.483}$  &  {5.46}$^{+0.72}_{-0.61}$ $\times$ 10$^{52}$ \\ 
140703026 & 3.14 & 84.04 & 16 &  {-1.265}$^{+0.122}_{-0.121}$  &  {127.816}$^{+19.894}_{-19.811}$  &  {-3.164}$^{+0.488}_{-0.501}$  &  {12.18}$^{+4.24}_{-3.04}$ $\times$ 10$^{52}$ \\ 
150727793 & 0.31 & 49.409 & 46 &  {-0.334}$^{+0.235}_{-0.235}$  &  {170.106}$^{+20.908}_{-21.247}$  &  {-3.111}$^{+0.492}_{-0.509}$  &  {0.14}$^{+0.08}_{-0.05}$ $\times$ 10$^{52}$ \\ 
151027166 & 0.81 & 123.39 & 9 &  {-1.375}$^{+0.102}_{-0.102}$  &  {172.112}$^{+37.216}_{-37.495}$  &  {-2.825}$^{+0.538}_{-0.604}$  &  {2.96}$^{+1.06}_{-0.74}$ $\times$ 10$^{52}$ \\ 
160804065 & 0.74 & 131.58 & 88 &  {-1.036}$^{+0.217}_{-0.217}$  &  {81.913}$^{+11.164}_{-11.315}$  &  {-3.271}$^{+0.443}_{-0.448}$  &  {1.58}$^{+1.12}_{-0.65}$ $\times$ 10$^{52}$ \\ 
161117066 & 1.55 & 122.18 & 91 &  {-0.858}$^{+0.044}_{-0.043}$  &  {84.824}$^{+2.297}_{-2.283}$  &  {-3.597}$^{+0.282}_{-0.281}$  &  {14.77}$^{+1.76}_{-1.56}$ $\times$ 10$^{52}$ \\ 
170405777 & 3.51 & 78.6 & 52 &  {-0.662}$^{+0.037}_{-0.037}$  &  {257.887}$^{+14.149}_{-14.122}$  &  {-2.209}$^{+0.112}_{-0.109}$  &  {465.12}$^{+64.69}_{-60.03}$ $\times$ 10$^{52}$ \\ 
171222684 & 2.41 & 80.452 & 43 &  {-1.412}$^{+0.073}_{-0.073}$  &  {35.031}$^{+11.59}_{-11.59}$  &  {-1.601}$^{+0.501}_{-0.501}$  &  {3.41}$^{+1.82}_{-1.83}$ $\times$ 10$^{52}$ \\ 
180620660 & 1.12 & 46.797 & 136 &  {-0.861}$^{+0.183}_{-0.183}$  &  {102.944}$^{+13.173}_{-13.606}$  &  {-2.671}$^{+0.356}_{-0.379}$  &  {2.65}$^{+1.65}_{-0.96}$ $\times$ 10$^{52}$ \\ 
180720598 & 0.65 & 48.89 & 50 & {-1.023}$^{+0.011}_{-0.011}$ & {633.603}$^{+17.934}_{-18.142}$ & {-2.475}$^{+0.054}_{-0.054}$ & {31.82}$^{+1.08}_{-1.11}$ $\times$ 10$^{52}$ \\ 
190114873 & 0.42 & 361.5 & 68 &  {-1.111}$^{+0.004}_{-0.004}$  &  {1069.591}$^{+19.987}_{-20.043}$  &  {-3.712}$^{+0.181}_{-0.187}$  &  {28.42}$^{+1.55}_{-1.28}$ $\times$ 10$^{52}$ \\ 
190829830 & 0.08 & 10.37 & 33 &  {-0.924}$^{+0.298}_{-0.285}$  &  {10.687}$^{+0.689}_{-0.672}$  &  {-2.431}$^{+0.021}_{-0.022}$  &  {0.02}$^{+0.02}_{-0.02}$ $\times$ 10$^{52}$ \\ 
210610827 & 1.13 & 55.04 & 63 &  {-0.687}$^{+0.018}_{-0.018}$  &  {283.452}$^{+6.948}_{-6.883}$  &  {-3.441}$^{+0.355}_{-0.362}$  &  {17.45}$^{+1.19}_{-0.88}$ $\times$ 10$^{52}$ \\ 
210619999 & 1.94 & 54.785 & 108 &  {-0.913}$^{+0.013}_{-0.013}$  &  {226.805}$^{+6.234}_{-6.033}$  &  {-2.111}$^{+0.028}_{-0.028}$  &  {288.03}$^{+6.89}_{-7.13}$ $\times$ 10$^{52}$ \\ 
220101215 & 4.62 & 237 & 18 &  {-1.028}$^{+0.043}_{-0.043}$  &  {305.924}$^{+27.324}_{-27.224}$  &  {-3.318}$^{+0.426}_{-0.438}$  &  {274.99}$^{+37.86}_{-29.2}$ $\times$ 10$^{52}$ \\ 
221009553 & 0.15 &  1100  & 62 &  {-1.661}$^{+0.016}_{-0.016}$  &  {1531.27}$^{+292.65}_{-284.78}$  &  {-3.093}$^{+0.502}_{-0.501}$  &  {$\sim$ 1000} $\times$ 10$^{52}$ \\ \hline
\end{tabular}
\end{table*}

\begin{table*}
\vspace{-0.2cm}
\addtolength{\tabcolsep}{7pt}
\centering
\caption{Spectral characteristic of ULGRBs obtained from the analysis of \fermi-LAT observation.}
\label{tab:lat_ulGRBs}
\begin{tabular}{|c|c|c|c|c|c|c|}\hline
\fermi ID & MET & boresight & $\Gamma_{\rm LAT}$ & Energy flux & photon flux & TS \\
 & (s) & (deg) &  & $\times$ 10$^{-10}$ erg $s^{-1} cm^{-2}$ & $\times$ 10$^{-6}$ Ph $s^{-1} cm^{-2}$ &  \\ \hline
151027166 & 467611108.033 & 9  & -2.73 $\pm$ 0.62 & 0.72 $\pm$ 0.34 & 1.96 $\pm$ 1.04 & 17 \\
170405777 & 513110367.886 & 52 & -2.61 $\pm$ 0.36 & 0.46 $\pm$ 0.15 & 1.16 $\pm$ 0.36 & 33 \\
190114873 & 569192227.626 & 67 & -2.61 $\pm$ 0.36 & 0.46 $\pm$ 0.15 & 1.16 $\pm$ 0.36 & 16 \\
210619999 & 645839970.604 & 109& -1.42 $\pm$ 0.27 & 0.45 $\pm$ 0.22 & 0.24 $\pm$ 0.15 & 30 \\
220101215 & 662706616.734 & 18 & -2.68 $\pm$ 0.29 & 1.75 $\pm$ 4.42 & 4.64 $\pm$ 1.10 & 61 \\
220627890 & 678057665.086 & 27 & -2.19 $\pm$ 0.16 & 5.35 $\pm$ 1.12 & 9.14 $\pm$ 1.70 & 140 \\
221009553 & 687014224.988 & 62 & -1.74 $\pm$ 0.08 & 19.4 $\pm$ 2.28 & 17.9 $\pm$ 2.16 & 494 \\ \hline
\end{tabular}
\end{table*}

\begin{table*}
\centering
\caption{Summary of initial and final parameters of massive star models simulated using the \sw{MESA} code. Starting from the PMS, the models are evolved till they reach the stage of the onset of core collapse. Here, M$_{\rm ZAMS}$ is the mass of the model at ZAMS, and $\Omega/\Omega_{c}$ is the ratio of initial angular rotational velocity and critical angular rotational velocity. Further, the M$_{\rm final}$ is the mass, R$_{\rm final}$ is the radius, M$_{\rm Fe-core}$ is the Iron-core mass, T$_{\rm eff}$ is the effective temperature, and L is the corresponding Luminosity of the model at the stage of the onset of core collapse. Additionally, t$_{\rm ff}$ and t$_{\rm b}$ are the free-fall time of the star model and the bore time of the weak jet, respectively.}
\begin{tabular}{|c|c|c|c|c|c|c|c|c|} \hline
~~ \bf M$_{\rm ZAMS}$ ~~ & ~~ \boldmath $\Omega/\Omega_{c}$ ~~ & ~~ \bf \boldmath M$_{\rm final}$ ~~ & ~~ \bf \boldmath R$_{\rm final}$ ~~ & ~~ \bf \boldmath M$_{\rm Fe-core}$ ~~ & ~~ \bf \boldmath log(T$_{\rm eff}$) ~~ & ~~ \bf Log (L) ~~ & ~~~~~~ \bf \boldmath t$_{\rm ff} ~~~~~ $ & ~~~~ \bf t$_{\rm b}$ ~~~~ \\ 

\bf \boldmath (M$_{\odot}$) & & \bf \boldmath (M$_{\odot}$) & \bf \boldmath(R$_{\odot}$) & \bf \boldmath (M$_{\odot}$) & \bf \boldmath (K) & \bf \boldmath (L$_\odot$) & \bf (s) & \bf (s)\\ \hline

15 & 0.1 & 14.96 & 603.23 & 1.55 & 3.64 & 5.08 & 1438579.51 & 140.7\\
15 & 0.2 & 14.96 & 593.52 & 1.65 & 3.67 & 5.17 & 1403955.16 & 138.4\\
15 & 0.3 & 14.95 & 57.19 & 1.91 & 4.20 & 5.29 & 42011.37 & 13.3\\
15 & 0.4 & 13.62 & 11.11 & 2.05 & 4.70 & 5.85 & 3768.59 & 2.6\\
15 & 0.5 & 13.42 & 10.51 & 1.90 & 4.71 & 5.83 & 3494.82 & 2.4\\
15 & 0.6 & 13.26 & 9.55 & 1.93 & 4.72 & 5.81 & 3043.96 & 2.2\\
20 & 0.1 & 19.97 & 549.77 & 1.59 & 3.70 & 5.25 & 1083359.66 & 128.2\\
20 & 0.2 & 19.92 & 488.37 & 1.96 & 3.76 & 5.39 & 908098.89 & 113.9\\
20 & 0.3 & 18.22 & 12.93 & 2.05 & 4.70 & 5.98 & 4088.36 & 3.0\\
20 & 0.4 & 17.91 & 12.17 & 2.04 & 4.71 & 5.97 & 3769.02 & 2.8\\
20 & 0.5 & 17.60 & 12.43 & 1.94 & 4.70 & 5.96 & 3923.15 & 2.9\\
20 & 0.6 & 17.36 & 12.76 & 1.90 & 4.70 & 5.96 & 4110.22 & 3.0\\
25 & 0.1 & 24.91 & 489.57 & 1.98 & 3.79 & 5.50 & 815131.60 & 114.2\\
25 & 0.2 & 24.83 & 662.72 & 1.98 & 3.77 & 5.70 & 1285760.38 & 154.6\\
25 & 0.3 & 22.42 & 12.98 & 1.93 & 4.72 & 6.07 & 3707.46 & 3.0\\
25 & 0.4 & 22.10 & 14.16 & 1.77 & 4.70 & 6.06 & 4256.61 & 3.3\\
25 & 0.5 & 21.50 & 13.01 & 1.91 & 4.72 & 6.05 & 3800.50 & 3.0\\
25 & 0.6 & 21.17 & 12.37 & 1.88 & 4.73 & 6.04 & 3552.95 & 2.9 \\
30 & 0.1 & 29.89 & 77.90 & 1.91 & 4.23 & 5.65 & 47233.24 & 18.2\\
30 & 0.2 & 29.76 & 72.49 & 1.82 & 4.29 & 5.82 & 42492.10 & 16.9\\
30 & 0.3 & 26.62 & 14.31 & 1.86 & 4.72 & 6.15 & 3939.29 & 3.3\\
30 & 0.4 & 26.11 & 15.63 & 1.78 & 4.70 & 6.14 & 4540.96 & 3.6\\
30 & 0.5 & 25.44 & 14.31 & 1.79 & 4.72 & 6.13 & 4029.45 & 3.3\\
30 & 0.6 & 25.09 & 13.75 & 1.93 & 4.72 & 6.12 & 3820.44 & 3.2\\

\hline
\end{tabular}
\label{tab:mesa_table}
\end{table*}

\end{document}